\documentclass[journal]{IEEEtran}
\usepackage{mathtools}

\usepackage{amsmath}
\usepackage{caption}
\usepackage{multicol}                        
\usepackage{graphicx}
\usepackage{algorithm}
\usepackage{adjustbox}
\usepackage{algpseudocode}
 \usepackage{multirow}
 \usepackage{array}
\usepackage{subcaption}
\makeatletter
\renewcommand{\theALG@line}{\arabic{ALG@line}}
\makeatother

\newcommand{\Input}{\State \textbf{Input:}~}
\newcommand{\Output}{\State \textbf{Output:}~}

\usepackage{amssymb}

\pdfoutput=1

\ifCLASSINFOpdf
\else
\fi
\begin{document}
%
\title{SI-ChainFL: Shapley-Incentivized Secure Federated Learning for High-Speed Rail Data Sharing}
%
%
%

\author{Mingjie~Zhao, \textit{Student member, IEEE}, Cheng Dai, \textit{Member, IEEE}, Fei Chen, Kui Ye, Xin Chen, Kaoru Ota, \textit{Member, IEEE}, Mianxiong Dong, \textit{Senior member, IEEE}, and Bing Guo* 
\thanks{This work was supported in part by the National Natural Science Foundation of China under Grant No. U2268204 and 62172061; National Key R\&D Program of China under Grant No. 2023YFB3308300; the Science and Technology Project of Sichuan Province under Grant No. 2024ZDZX0012, 2023ZHCG0011, 2021YFG0152. Additionally, this work was supported in part by China Scholarship Council (CSC) under Grant 202506240164.(Corresponding
authors: Bing Guo)}
\thanks{Mingjie Zhao, Cheng Dai, Bing Guo, Fei Chen, Kui Ye, and Xin Chen are with
the College of Computer Scicnce, Sichuan University, Chengdu 610065, China (e-mail: zhaomingjie@stu.scu.edu.cn;
daicheng@scu.edu.cn; guobing@scu.edu.cn;  chenfei@stu.scu.edu.cn; 2021323040029@stu.scu.edu.cn; 2024323045015@stu.scu.edu.cn;). Mingjie Zhao is also a visiting student at Muroran Institute of Technology (e-mail:25061115p@muroran-it.ac.jp).}
\thanks{Kaoru Ota is with the Graduate School of Information Sciences, To-hoku University, Sendai 980-8579, Japan.(e-mail: k.ota@tohoku.ac.jp)}
\thanks{Mianxiong Dong is with the Department of Sciences and Informatics, Muroran Institute of Technology, Muroran 0508585, Japan. (e-mail: mxdong@muroran-it.ac.jp)}}

%
%

\markboth{IEEE TRANSACTIONS ON MOBILE COMPUTING, VOL. XX, NO. XX, XXX 2026}%
{Shell \MakeLowercase{\textit{et al.}}: Bare Demo of IEEEtran.cls for IEEE Journals}
%



\maketitle

\begin{abstract}
In high-speed rail (HSR) systems, federated learning (FL) enables cross-departmental flow prediction without sharing raw data. However, existing schemes suffer from two key limitations: (1) insufficient incentives, leading to free-riding and model poisoning; and (2) centralized aggregation, which introduces a single point of failure. We propose a secure and efficient framework SI-ChainFL that addresses these issues by combining contribution-aware incentives with decentralized aggregation. First, we quantify client contributions using a Shapley value metric that jointly considers rare-event utility, data diversity, data quality, and timeliness. To reduce computational overhead, we further develop a rare positive driven client clustering strategy to accelerate Shapley estimation. Moreover, we design a blockchain-based consensus protocol for decentralized aggregation, where aggregation eligibility is tied to Shapley incentives. This design motivates clients to submit high-quality updates and enables efficient and secure global aggregation. Experiments on MNIST, CIFAR 10 and CIFAR 100, and a HSR flow dataset show that SI ChainFL remains effective under 90\% malicious clients in PA attacks, achieving 14.12\% higher accuracy than RAGA. Theoretical analysis further guarantees an upper bound on performance degradation caused by malicious participants.
\end{abstract}

\begin{IEEEkeywords}
Federated learning, Incentive mechanism, Shapley value, Data sharing, Blockchain.
\end{IEEEkeywords}

%
\IEEEpeerreviewmaketitle

\section{Introduction}
\subsection{Background}
%
%
%
%
\IEEEPARstart 
{C}{hina’s} high-speed rail network has grown from scratch to the world’s largest in the past decade \cite{lasserre2020emergence} By 2025, China's high-speed rail passenger volume is expected to reach nearly 4.255 billion trips, including approximately 513 million trips during the 40-day Spring Festival travel rush. Hub stations will maintain high capacity operations for extended periods. Extreme weather or unforeseen events may lead to capacity constraints and short-term passenger surges, causing station congestion and resource shortages \cite{huang2025metasignal}. Therefore, passenger flow forecasting for high-speed rail scenarios is crucial for traffic capacity scheduling and congestion early warning.

\begin{figure*}[t]
  \centering
  \includegraphics[width=\textwidth]{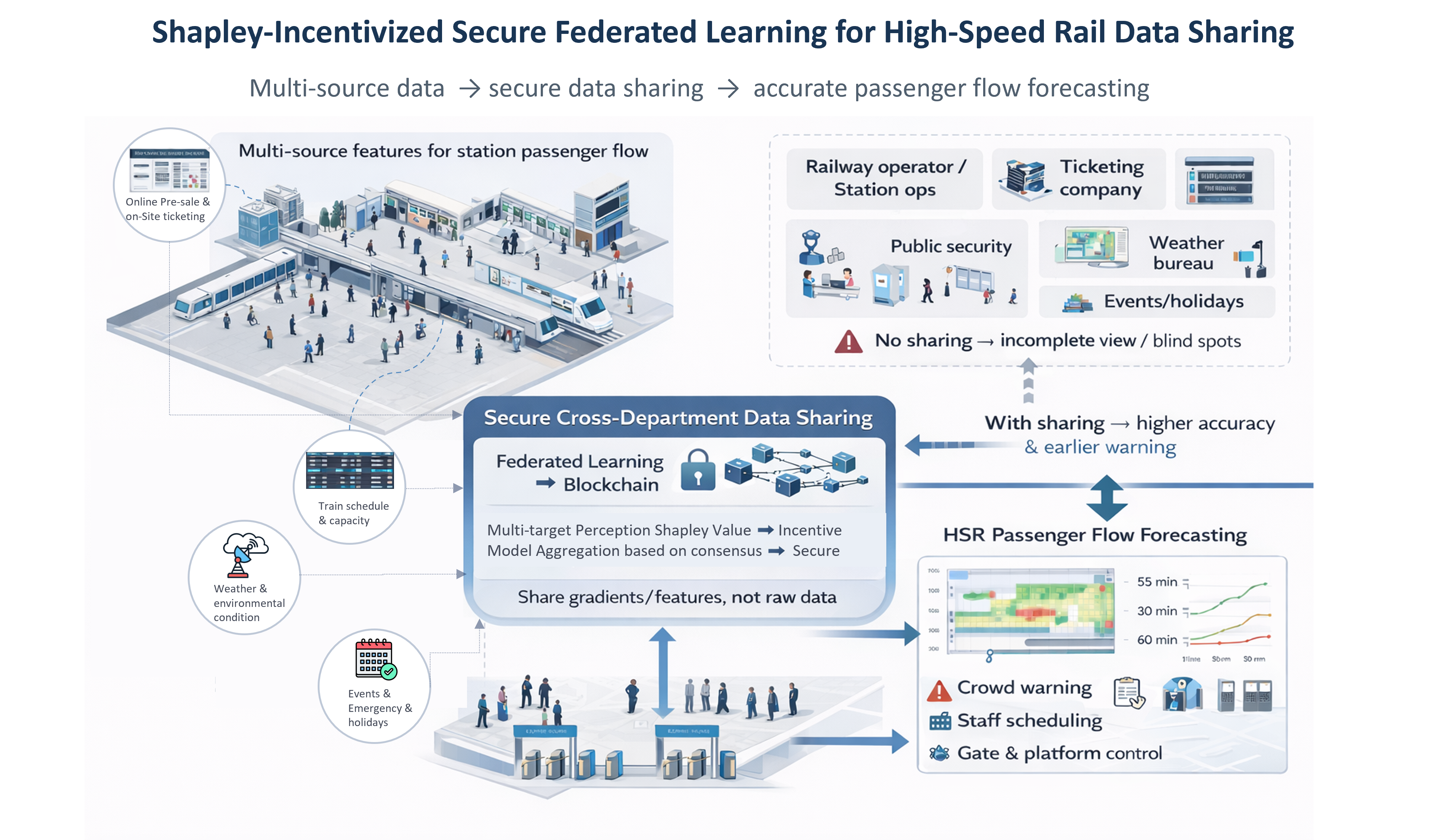}
  \caption{High-speed rail data sharing application scenarios}
  \label{fig:scenios}
\end{figure*}

As shown in Fig. \ref{fig:scenios}, passenger flow forecasting typically relies on cross-departmental heterogeneous data sharing, but privacy risks and compliance requirements (such as the EU's General Data Protection Regulation, GDPR) hinder data owners from sharing raw data. To address real-world problems and verify the practicality of the method, this paper constructs a real high-speed rail dataset based on multi-source data from station operations, ticket pre-sales, and meteorology. This dataset exhibits typical characteristics such as multi-source heterogeneity, non-independent and identically distributed data, and time sensitivity, effectively reflecting the key challenges of cross-departmental data sharing and collaborative modeling. Federated learning allows multiple parties to collaboratively train locally without exchanging raw data \cite{hu2023data}  , providing a feasible solution for transportation data sharing \cite{he2025fedvkd}.

However, in large-scale collaborative training, the lack of a fair incentive mechanism that matches actual contributions can easily lead to insufficient client participation, unstable training quality, and free-riding behavior, thereby reducing the convergence speed of the global model. Therefore, recent research has introduced blockchain \cite{wu2025blockchain} and incentive mechanisms \cite{tian2025pricing} to verify upload updates and constrain training behavior, thereby encouraging high-quality data contributions and improving the robustness and efficiency of training.

\subsection{Motivation}

Existing solutions often use sample size and gradient alignment for incentive allocation \cite{shi2022fedfaim}, \cite{feng2019joint}, \cite{jiao2020toward}. On the one hand, this design underestimates the value of rare and highly informative data. In high-speed rail scenarios, representative rare samples often bring higher marginal utility than large amounts of redundant data. On the other hand, gradient alignment focuses on the consistency between local update directions and global gradients \cite{xu2021gradient}, while ignoring gradient magnitude and potential adversarial behavior. Malicious nodes may even construct updates with directions consistent with the global gradient but numerically harmful. Allocating rewards solely based on sample size and gradient alignment is inherently unfair.

In federated learning environments with limited computing power and highly heterogeneous data, the challenge lies in how to quantify the actual contributions of each client and rationally allocate incentives to maximize the global model's gains. Therefore, it is necessary to construct a fair and incentive compatible mechanism to encourage high-quality data contributions and stable training.

\begin{enumerate}

\item[i)] In federated learning, the central coordinator strives to maximize the global model utility under budget and resource constraints, while clients aim to obtain the most incentives with minimal data and training costs. Therefore, the incentive mechanism needs to simultaneously satisfy fairness, robustness, and budget constraints.

\item[ii)] Client utility is influenced by various factors such as gradient information, data diversity, data quality, and timeliness, non-i.i.d. data, unstable training state, and dynamic network conditions further interfere with the calculation and statistics of multidimensional contribution estimation.

\item[iii)] Client selection is discrete, while contribution evaluation and incentive allocation rely on a continuous, nonlinear utility function derived from Shapley values via a nonlinear mapping. The coupling of discrete selection and continuous nonlinear allocation forms a MINLP problem, which is often difficult to solve precisely. This prompts us to seek scalable approximation methods, such as stochastic permutation Shapley estimation \cite{mann1960values}.

\end{enumerate}

\subsection{Contributions}

This paper investigates how to achieve secure and efficient multi-party data sharing in federated learning. The design of contribution evaluation, incentive allocation, and secure aggregation must be integrated to avoid free-rider attacks, poisoning attacks, and single points of failure. Unlike most existing works (i) employing heuristic reward rules based on sample size or loss reduction, and (ii) focusing only on blockchain-based secure aggregation while neglecting fine-grained contribution modeling, SI-ChainFL couples a multi-dimensional Shapley valuation mechanism with a blockchain-supported aggregation layer within a unified framework.

The main contributions of this paper are summarized below:

\begin{itemize}

\item We propose a multi-objective Shapley value data contribution evaluation method that comprehensively considers the marginal contribution of rare positive example predictions, data diversity, data quality, and timeliness. We propose a rare positive example clustering client to quickly calculate Shapley values, reducing computational complexity from exponential to near-linear.

\item We designed SI-ChainFL to embed Shapley scores into the incentive mechanism and aggregation process, binding participation in model aggregation and access to the global model. This mechanism encourages clients to actively participate in training and contribute high-quality data, thereby achieving decentralized and verifiable aggregation.

\item We validated SI-ChainFL on MNIST, CIFAR-10, CIFAR-100, and a real-world High-Speed Rail dataset (HSR). Experimental results show that, compared with the baseline, this model achieves better performance under malicious client attacks.

\end{itemize}

Sections~II and III survey the relevant literature and underlying technologies. 
Section~IV presents the system model, threat model, and design goals. 
Section~V details the proposed SI-ChainFL framework,  
Section~VI provides a comprehensive security analysis of SI-ChainFL. 
Section~VII reports the experimental results and corresponding analysis. 
Finally, Section~VIII summarizes the paper and discusses potential directions for future research.

\section{RELATED WORK}
This work focuses on designing contribution-aware incentive mechanisms and using blockchain to secure aggregate model for federated learning (FL) in multi-party data sharing scenarios. To provide a more detailed overview of the work presented in this paper, in this section we mainly review the lines of relevant research: (i) Incentive mechanisms for FL, and (ii) blockchain-enabled secure aggregation in FL.

\subsection{Incentive mechanisms for FL}
Federated learning (FL) was first formalized by McMahan et al. via the FedAvg algorithm, which enables collaborative training by aggregating local updates instead of exchanging raw data \cite{mcmahan2017communication}. Despite its privacy benefits, practical FL often suffers from low participation, making incentive design critical for large-scale training. Some researchers \cite{yang2025fairness_tmc, ding2025incentive_tmc} develop incentives under a Stackelberg game formulation and treat privacy-preserving performance as a key evaluation criterion. Huang et al. \cite{huang2024hierarchical} model participant interactions through a Stackelberg framework, derive Nash and Stackelberg equilibria, and address upper–lower level coupling. Guo et al. \cite{guo2025lcefl_tmc} assess participants’ contributions to assign trust levels, which are then used as aggregation weights for local models. Kang et al. \cite{kang2019reliablefl} take reputation as the primary basis for worker selection and employ a consortium blockchain to manage clients’ reputations. Auction schemes have also been studied to reduce social costs in the FL service market and improve social welfare \cite{jiao2021auctionfl, pang2023auctionfl}. Ding et al. \cite{ding2025dynamicpricing} jointly design participation incentives and network pricing to balance the benefits of servers and users, while leveraging deep reinforcement learning to derive optimal training and pricing strategies for participants \cite{zhan2020learningbased}.
 However, most of these studies use a single, coarse metric, such as reputation and sample size, without explicitly breaking down the participants' contributions into multiple metrics. For fine-grained valuation, Ghorbani and Zou \cite{ghorbani2019datashapley} introduce data Shapley as a principled measure of each training datum’s marginal utility to model performance. Building on this line, Sun et al. \cite{sun2023shapleyfl} propose an adaptive weighting strategy inspired by inter-client Shapley disparities to enhance FL robustness. Singhal et al. \cite{singhal2024greedyshapley} adopt fast approximate Shapley computation to identify the most influential clients in each communication round, and Fan et al. \cite{fan2024verfedsv} present an efficient Shapley-based valuation method for vertical FL to evaluate the contributions of heterogeneous data sources. To further reduce computational overhead, Lei et al. \cite{lei2025feddsv} propose a Monte Carlo variant sampling approach with flexible weighting, accommodating dynamic settings where participants may join or leave intermittently. Unlike these studies, we develop a multi-objective Shapley valuation method that jointly accounts for rare event utility, data diversity, data quality, and timeliness, thereby providing a more faithful assessment of data value and enabling fairer incentives for participation in FL.

\subsection{Blockchain-enabled secure aggregation in FL}

The immutability and traceability of blockchain enable accountable coordination and provide mechanisms for defending against attacks. Previous work mainly applied blockchain to three stages: model update uploading, training node selection, and consensus. At the update submission stage, blockchain enables update traceable accountability. Lu et al. \cite{lu2021commEffBCfl} incorporate blockchain into federated learning to strengthen communication security and privacy protection. In \cite{zhao2021privacybcfl}, to deter malicious manipulation during model update uploading, each participant signs its model update so that related activities become auditable and traceable. At the node selection stage, validators verify updates and select participants to improve the security of the training. Qu et al. \cite{qu2020blockchainedfl} employ blockchain to support model verification and participant selection, which in turn can improve model convergence. Zhang et al. \cite{zhang2024bitfl} integrate a randomized incentive mechanism by designing selection probabilities for participants, improving training participation while preserving the privacy of participants’ cost information. At the consensus stage, Dou et al. \cite{dou2025bcfl6g} record transactions on-chain and further propose a time-sensitive proof-of-stake consensus mechanism motivated by Newton’s law of cooling to select high-stake miners for block production, thereby boosting participants’ willingness to contribute to training. Chen et al. \cite{chen2025verifiableppfl} leverage cryptographic techniques to mitigate attacks during model submission and introduce a verification procedure that safeguards the aggregation integrity.

However, most existing blockchain consensus designs are driven primarily by participants’ historical reputation, and seldom quantify their multidimensional marginal utility to global model training. This limitation becomes particularly salient in complex settings where contribution evaluation must be timely and responsive. In contrast, we develop a lightweight blockchain consensus protocol that uses contribution-aware, multi-objective Shapley values to perform randomized yet verifiable selection of aggregation and validation nodes. This design eliminates reliance on a centralized server while enabling secure model aggregation that remains aligned with incentive objectives.

\section{PRELIMINARIES}
In this section, we review the relevant background knowledge and summarize the core techniques underlying our method.

\subsection{Federated Learning}

Federated Learning (FL) is a distributed machine learning paradigm that enables multiple participants to collaboratively train a model while keeping their own data locally and exchanging only model updates instead of raw data. The system consists of multiple clients and a central server, where the set of clients is denoted as  $\mathcal{P}=\{1,2,\ldots,P\}$. Client $i\in\mathcal{P}$ holds a local dataset $\mathcal{D}_i$. Formally, FL aims to minimize a global loss of the form

\begin{equation}\min_{w} F(w)= \sum_{i=1}^{P} p_i \, \mathcal{L}_i\bigl(w; \mathcal{D}_i\bigr),\end{equation}

where $\mathcal{L}_i$ is the local loss of client $i$, and  $p_i$ reflects its relative weight, with $\sum_{i=1}^{P} p_i=1$.

A typical federated learning algorithm such as FedAvg \cite{mcmahan2017communication} works as follows: in each round $(t)$ of communication, the server broadcasts the current global model $w^{(t)}$ to selected clients. In each global communication round, each selected client performs several local gradient update steps on its private dataset to minimize its local objective, and then uploads the updated model parameters $w_i^{(t+1)}$ to the server for weighted averaging. The server performs weighted average aggregation to obtain a global model:

\begin{equation}
w^{(t+1)} = \sum_{i \in \mathcal{S}^{(t)}}
\frac{n_i^{(t)}}{\sum_{j \in \mathcal{S}^{(t)}} n_j^{(t)}} \, w_i^{(t+1)},
\end{equation}

where $\mathcal{S}^{(t)}\subseteq\mathcal{P}$ is the set of clients participating in training in round $t$, $n_i^{(t)}$ is the number of samples used by client $i$ in round $t$, and $\sum_{j \in \mathcal{S}^{(t)}} n_j^{(t)}$ is the total number of samples over all participating clients in round $t$

FedSGD \cite{mcmahan2017communication} can be viewed as a special case of FedAvg where each selected client performs only a single local gradient step in each round and directly uploads its gradient to the central server for weighted averaging, the overall procedure is essentially equivalent to synchronous distributed SGD. Compared with FedAvg, FedSGD mitigates client drift caused by multiple local epochs and is therefore closer to the convergence behavior of standard SGD, but it requires more frequent communication rounds and thus incurs a higher communication cost.

FL offers practical benefits, including reduced raw data transfer, improved use of heterogeneous data, and lower wide area bandwidth consumption. However, FL also faces many challenges in practice. Non IID and imbalanced client data can hinder convergence and accuracy. Network and device heterogeneity increases coordination overhead. Weak incentives may encourage free riding or low quality participation. Moreover, model updates may leak sensitive information and remain vulnerable for poisoning or other adversarial attacks.

\subsection{Shapley Value}
The concept of Shapley Value was first proposed by Lloyd S. Shapley in cooperative game theory \cite{shapley1953stochastic}, which is mainly used to quantify the marginal contribution of individuals to the overall income of the alliance. Shapley Value is the contribution value $\phi_i(v) $ assigned to each participant  $i \in \mathcal{P} $ defined as follows:
\begin{equation}\phi_i(v) = \sum_{S \subseteq \mathcal{P} \setminus \{i\}} \frac{|S|!(|\mathcal{P}|-|S|-1)!}{|\mathcal{P}|!} [v(S \cup \{i\}) - v(S)]\end{equation}

where $S$ is the participant $i$ to join the previous consortium subset,  $v(S)$ is the value function $v(\cdot)$ calculated to represent the total revenue generated by the consortium $S$, $v: 2^{N} \to \mathbb{R}$.

Participants who contribute the same get the same Shapley value, while those who don't generate any marginal gain have a Shapley value of 0, so the Shapley Value satisfies symmetry and Dummy.

In a data sharing setting, higher rewards can be provided to participants with greater contributions based on the calculated Shapley Value of participants to increase participants' enthusiasm for participating in model training, effectively preventing free rider attacks. Additionally, if the Shapley Value consists of multiple independent components, the contributions of participants in each section can be linearly combined, which facilitates comprehensive multi-metric evaluation in complex scenarios. However, calculating Shapley Value requires traversing all subsets of participants, with high complexity $(O(2^n))$, so it is very important to design an accurate approximation algorithm to reduce the computational cost of the system.

\section{PROBLEM STATEMENT}
This section first describes the system model, the threat model, and then articulates the design objectives.

\subsection{System Model}

\begin{figure}[t]
  \centering
  \includegraphics[width=0.5\textwidth]{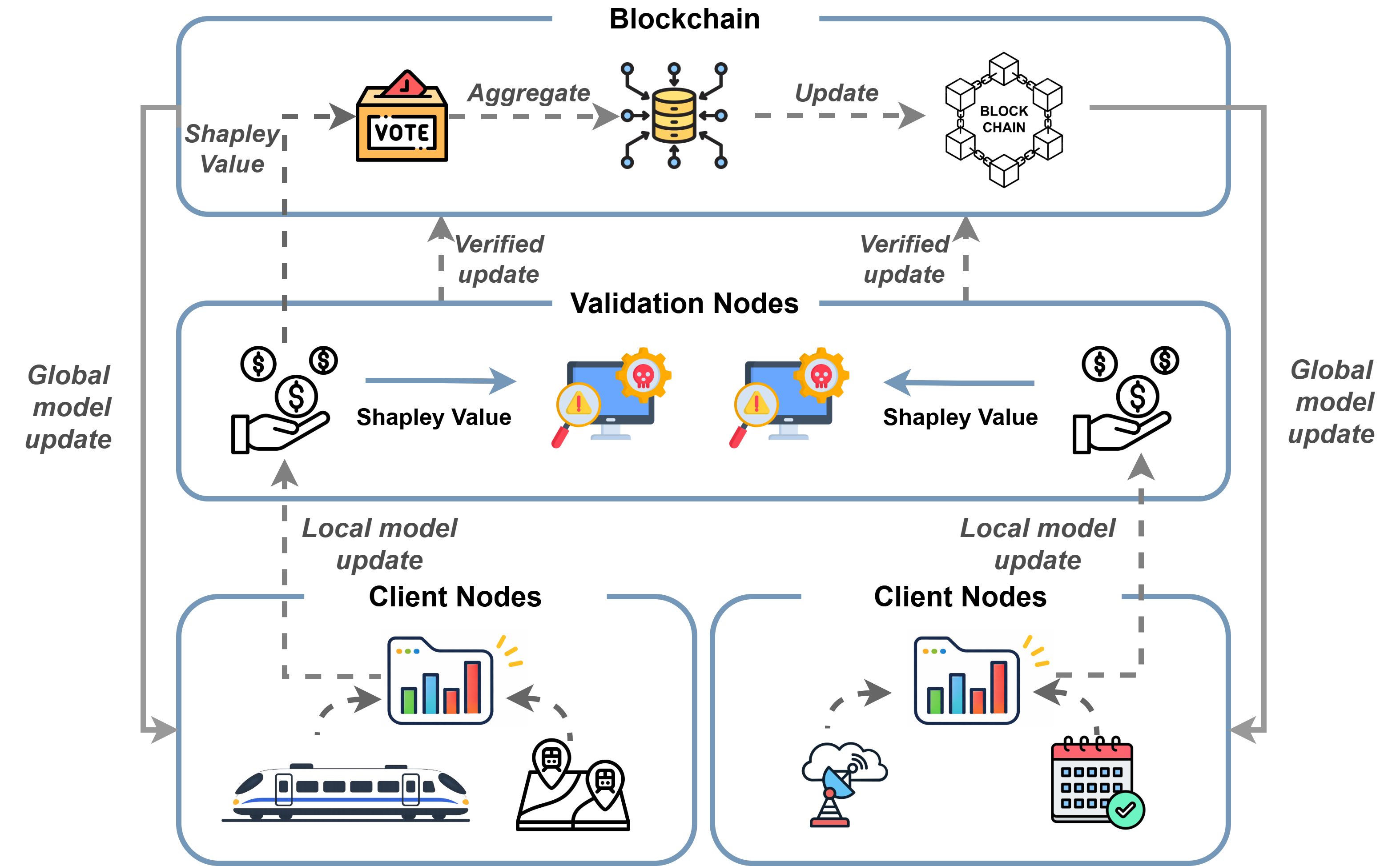}
  \caption{System model}
  \label{fig:System model}
\end{figure}

As shown in Fig. \ref{fig:System model}, the data sharing system consists of three types of entities: $n$ edge nodes, $m$ validation nodes, and a blockchain network. The functions of each component are described below.

\begin{itemize}
\item \textit{Client Nodes:} Client nodes are local edge devices deployed at different train stations and weather stations. They work together to build a high-speed rail passenger flow prediction model. Edge nodes at high-speed rail stations collect data on the numbers of trains, passengers, and ticket pre-sales, while weather stations collect data on meteorological features including weather conditions, temperature, and wind speed. Each edge node independently trains its model on its private data and only uploads model updates to the blockchain network.

\item \textit{Validation Nodes:} The validator nodes quantify the marginal benefits of each edge node based on the improvement of global model performance, assess the data quality and diversity provided by edge nodes, and identify potential malicious nodes. These assessment provide the basis for subsequent client selection and incentive allocation.

\item \textit{Blockchain Network:} The blockchain network serves as the infrastructure for global model aggregation. Validator nodes use the consensus protocol to select client updates for global aggregation and broadcast the new global model to eligible clients. This process enhances the security, auditability, and transparency of FL.

\end{itemize}

\subsection{Threat Model}

In our system model, we assume that participating edge nodes are heterogeneous and can be grouped into four categories: (i) honest nodes with high-quality data, (ii) honest nodes with low-quality data, (iii) passive nodes that are unwilling to contribute computing resources, and (iv) malicious nodes with attack behavior.

\begin{itemize}
\item \textit{Honest nodes with high quality data:} These nodes collect data have high feature diversity, and have highly reliable labels, and the data distribution is closely related to the training target task. Moreover, these nodes will actively participate local training.

\item \textit{Honest nodes with low quality data:} The data collected by these nodes has serious redundancy, noise, or distribution deviations, which have little effect on improving the performance of the global model.

\item \textit{Passive nodes:} These nodes skip effective training, upload arbitrary or outdated model updates and still obtain global model updates, thereby free-riding on the system.

\item \textit{Malicious Nodes:} These nodes intentionally upload biased local model updates to  poison the model training and may exploit received model updates to infer other to infer the private information of other participants.

\end{itemize}

\subsection{Design Goals}

Our goal is to build a data sharing framework to achieve efficient and safe model collaborative training, the framework is designed to focus on the following four aspects:

\begin{itemize}
\item \textit{Effectiveness:}
Across various datasets and attack scenarios, our approach consistently outperforms baselines in global model performance.

\item \textit{Efficiency:}
Our proposed framework accelerates model convergence compared to the baseline method.

\item \textit{Privacy Protection:}
By evaluating each node’s marginal utility, the system can filter out malicious clients from global model aggregation. This mitigates poisoning and free-riding behavior and lowers the risk that model updates are exploited to infer other participants’ private data.

\item \textit{Robustness:}
We use a blockchain to aggregate global model, remove the risk of single points of failure, while a consensus protocol selects participating clients, thereby ensuring transparent and secure global model updates.

\end{itemize}

\section{Proposed Approach}

In this section, we first show the design of our SI-ChainFL and, then provide the working mechanisms at each stage.

\subsection{Overall Design}
\begin{figure*}[t]
  \centering
  \includegraphics[width=0.9\textwidth]{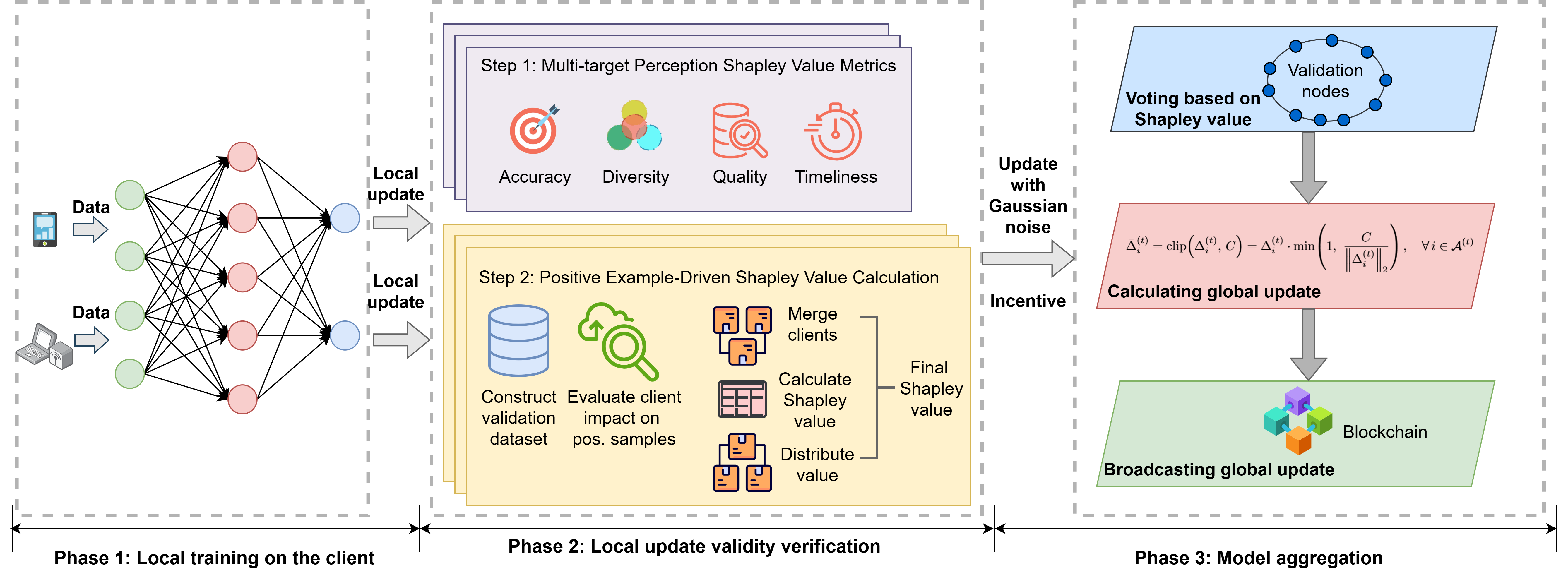}
  \caption{System workflow}
  \label{fig:System_workflow}
\end{figure*}

\begin{algorithm}[t]
\caption{SI-ChainFL Main Procedure}
\label{alg:sichainfl_main_concise}
\begin{algorithmic}[1]
  \Input Clients $N$, rounds $T$, init model $\theta^{(0)}$,
         datasets $\{\mathcal{D}_i\}$, val sets $\{\mathcal{D}^{(r)}_{\mathrm{val}}\}_{r=1}^R$,
         weights $(\lambda_{\mathrm{Acc}},\lambda_{\mathrm{Div}},\lambda_{\mathrm{Qua}})$,
         top-$K$, ratio $\rho$, similarity $\kappa$, time decay $\gamma$,
         clip bound $C$, activation $\psi(\cdot)$, consensus threshold $\tau$
  \Output Final model $\theta^{(T)}$, Shapley scores $\{\widehat{\Phi}_i\}$

  \State Initialize $t\gets 0$, $\widehat{\Phi}_i\gets 0$ for all $i\in N$

  \While{$t < T$}

    \State \textbf{Stage 1: Contribution Quantification}
    \For{each client $i\in\mathcal{P}^{(t)}$}
      \State $\Delta_i^{(t)} \gets \mathrm{LocalTrain}(\theta^{(t)},\mathcal{D}_i)$
    \EndFor
    \State Define Shapley value $\nu^{(t)}(S)$ using $(\lambda_{\mathrm{Acc}},\lambda_{\mathrm{Div}},\lambda_{\mathrm{Qua}})$

    \State \textbf{Stage 2: Approximate Shapley}
    \State $\{\widetilde{\mathcal{D}}^{(r)}_{\mathrm{val}}\}\gets \mathrm{StratifyVal}(\{\mathcal{D}^{(r)}_{\mathrm{val}}\},\rho)$
    \State $\eta_i^{(t)} \gets \mathrm{PosImpact}(\theta^{(t)},\{\Delta_i^{(t)}\},\{\widetilde{\mathcal{D}}^{(r)}_{\mathrm{val}}\})$
    \State $\mathcal{H}^{(t)} \gets \mathrm{TopK}(\{\eta_i^{(t)}\},K)$
    \State $\mathcal{G}^{(t)} \gets \mathrm{Merge}(\mathcal{P}^{(t)}\!\setminus\!\mathcal{H}^{(t)},\kappa)$
    \State $\{\hat{\phi}_G^{(t)}\}\gets \mathrm{ShapleyPerm}(\nu^{(t)},\ \mathcal{G}^{(t)}\cup\{\mathcal{H}^{(t)}\})$
    \State $\hat{\phi}_i^{(t)} \gets \mathrm{Distribute}(\{\hat{\phi}_G^{(t)}\},\{\eta_i^{(t)}\})$
    \State $\widehat{\Phi}_i \gets \gamma\,\widehat{\Phi}_i + (1-\gamma)\,\hat{\phi}_i^{(t)}$

    \State \textbf{Stage 3: Secure Aggregation}
    \State $\mathcal{A}^{(t)}\gets \mathrm{Filter}(\mathcal{P}^{(t)},\widehat{\Phi})$
    \State $\mathrm{Consensus}(t,\mathcal{A}^{(t)},\{\Delta_i^{(t)}\},\widehat{\Phi},\tau)$
    \For{each $i\in\mathcal{A}^{(t)}$}
      \State $w_i^{(t)} \gets \mathrm{Normalize}\big(\psi(\max(\widehat{\Phi}_i,0))\big)$
      \State $\bar{\Delta}_i^{(t)} \gets \mathrm{clip}(\Delta_i^{(t)},C)$
    \EndFor
    \State $\Delta^{(t)} \gets \sum_{i\in\mathcal{A}^{(t)}} w_i^{(t)}\,\bar{\Delta}_i^{(t)}$
    \State $\theta^{(t+1)} \gets \theta^{(t)} \oplus \eta\,\Delta^{(t)}$

    \State $t\gets t+1$
  \EndWhile

  \State \Return $\theta^{(T)}$, $\{\widehat{\Phi}_i\}$
\end{algorithmic}
\end{algorithm}

As shown in Fig.~\ref{fig:System_workflow}, SI-ChainFL consists of three parts: multi-objective Shapley value modeling, efficient Shapley value approximation, and blockchain-based global model aggregation. We quantify each client’s contribution using multi-objective metrics (e.g., rare-event utility, data diversity, data quality, and timeliness), approximate the resulting Shapley values via rare event aware sampling, and feed these scores into a consensus mechanism to achieve decentralized global model updates. The detailed workflow of SI-ChainFL is given in Alg.~\ref{alg:sichainfl_main_concise} and is summarized as follows:

\begin{itemize}

\item \textit{Stage 1: Client Contribution Quantification.}
We construct a client contribution metric based on Shapley values. In each round, clients’ rare-event prediction utility, data diversity, data quality, and timeliness are jointly incorporated into a multi-objective contribution function (see Sec.~V-B).

\item \textit{Stage 2: Approximate Shapley Value Computation.}
We selectively compute Shapley values based on each client’s contribution to rare positive examples. In each validation set, all positive samples and a fixed proportion of negatives are retained, reducing the number of Shapley evaluations and variance across splits (see Sec.~V-C).

\item \textit{Stage 3: Secure and fair global model aggregation.}
We designed a blockchain-based consensus algorithm that embeds the Shapley value into the incentive mechanism (see Sec.~V-D). 
\end{itemize}
\subsection{Multi-target Perception Shapley Value}
In this phase, we aim to accurately and fairly quantify each participant’s data and training contribution while accounting for the scarcity and timeliness of high-speed rail flow data. To this end, we employ a multi-objective contribution function that jointly considers (i) rare-event prediction utility, (ii) data diversity, and (iii) data quality. In addition, we apply an exponential time-discounting scheme across training rounds to capture the temporal relevance of contributions.

Let $N$ denote the set of clients, and let global training proceed over $T$ rounds. For a client $i \in N$, let $\theta_S$ be the model obtained at round $t$ by the coalition $S$ containing $i$, and let $p_{\theta_S}(x)$ denote the predicted event probability under $\theta_S$ on the validation set $D_{\mathrm{val}}$. Based on a multi-objective value function $\nu^{(t)}(\cdot)$, we derive a per-round Shapley value $\phi_i^{(t)}$ for each client and then aggregate these values with time decay weights over $t$ to obtain the Shapley score $\Phi_i$.

Specifically, the function consists of the following parts:

\noindent\textit{Rare-Event Prediction Utility:}
To emphasize rare flow events, we evaluate coalition $S$ with model $\theta_S$ on a validation set $\mathcal{D}_{\mathrm{val}}$ partitioned into $R$ subsets $\{\mathcal{D}_{\mathrm{val}}^{(r)}\}_{r=1}^R$. Let $\pi_r$ be the fraction of positives in $\mathcal{D}_{\mathrm{val}}^{(r)}$. We normalize the area under the precision--recall curve (AUPRC) as
\begin{equation}
\widetilde{A}_r(S)=
\frac{\mathrm{AUPRC}\!\left(\theta_S;\mathcal{D}_{\mathrm{val}}^{(r)}\right)-\pi_r}
{1-\pi_r},
\end{equation}

To constrain false alarms, we use the Matthews correlation coefficient (MCC) with an FPR budget:
\begin{equation}
\begin{aligned}
\mathrm{MCC}^{\star}_r(S)
&=\max_{\tau\in\mathcal{T}}
\ \mathrm{MCC}\!\left(\theta_S,\tau;\mathcal{D}_{\mathrm{val}}^{(r)}\right) \\
&\text{s.t.}\quad
\mathrm{FPR}\!\left(\theta_S,\tau;\mathcal{D}_{\mathrm{val}}^{(r)}\right)\le \rho ,
\end{aligned}
\end{equation}
where $\mathcal{T}$ is the set of thresholds and $\rho\in(0,1)$ is the maximum false-positive rate. 
We normalize $\mathrm{MCC}^{\star}_r(S)\in[0,1]$ via
\begin{equation}
\widetilde{M}_r(S)=\frac{\mathrm{MCC}^{\star}_r(S)+1}{2}.
\end{equation}

The rare-event utility on subset $r$ is then
\begin{equation}
v_r(S)=\big(\widetilde{A}_r(S)\big)^{\alpha}\big(\widetilde{M}_r(S)\big)^{1-\alpha},
\quad \alpha\in[0,1],
\end{equation}
and we aggregate across subsets by a weighted mean
\begin{equation}
v_{\mathrm{Acc}}(S)
=\exp\!\left(\sum_{r=1}^{R}\beta_r\log\bigl(v_r(S)+\varepsilon\bigr)\right),
\end{equation}
where $\beta_r\ge0$ and $\sum_{r}\beta_r=1$, and $\varepsilon>0$ is a small constant for numerical stability.

\textit{Data Diversity Metric:}
For regions with small datasets and sparse sample coverage, even highly unique data that improve model generalization may have their marginal utility underestimated if only sample size is considered, which is unfair. To address this issue, we introduce a data diversity measure based on a feature-representation similarity matrix. Specifically, we use cosine similarity to construct $K_{mn} \in [0,1]$:
\begin{equation}
  K_{mn} = \frac{1 + \cos(z_m, z_n)}{2},
\end{equation}
where $K_{mn}$ measures the similarity between the feature summaries of clients $m$ and $n$, and $z_m,z_n \in \mathbb{R}^d$ are locally computed summary vectors.

For any coalition $S \subseteq \mathcal{N}$ with $|S| < 2$, we set the data diversity to zero, i.e., $v_{\mathrm{Div}}(S) = 0$. For $S \subseteq \mathcal{N}$ with $|S| \geq 2$, we define the average similarity and diversity as
\begin{equation}
\begin{aligned}
  \overline{K}(S)
  = \frac{2}{|S|(|S|-1)}
    \sum_{\substack{m<n\\ m,n\in S}} K_{mn},
\\  v_{\mathrm{Div}}(S) = 1 - \overline{K}(S).
\end{aligned}
\end{equation}

\textit{Data Quality Metrics:} Low-quality data may introduce noise, negatively impacting model training. By introducing data quality as one of the data contribution evaluation metrics, we can better filter out nodes providing low-quality or malicious data, thus improving the model's generalization ability across different scenarios. We will evaluate data quality from two aspects: data cleanliness and the credibility of data labels.

We define data cleanliness as follows:
\begin{equation}
C_i=\exp\!\left(-\gamma_1 r_i^{\mathrm{miss}}-\gamma_2 r_i^{\mathrm{out}}-\gamma_3 r_i^{\mathrm{sync}}\right)\in(0,1].
\end{equation}
Where $r_i^{\mathrm{miss}},\, r_i^{\mathrm{out}},\, r_i^{\mathrm{sync}} \in [0,1]$ are the data missing rate, outlier rate, and time alignment error rate, respectively. 

We define label credibility as follows:
\begin{equation}
\ell_i=\frac{1}{|\mathcal{D}_i|}\sum_{(x,y)\in \mathcal{D}_i}\bigl(y-p_{\theta}(x)\bigr)^2,
L_i=\exp\!\left(-\gamma_4\,\ell_i\right)\in(0,1].
\end{equation}
If the label's prediction differs significantly from the global model's prediction, the label's reliability will decrease.

Combining the above two factors into a data quality metric:
\begin{equation}
\mathrm{DQ}(S)
= 1 - \prod_{i\in S}\bigl(1-C_i L_i\bigr),
\end{equation}

In each round, we will weight and fuse the above three parts to form a multi-objective function $v^{(t)}(S)$:
\begin{equation}
\nu^{(t)}(S)=
\lambda_{\mathrm{Acc}}\,\nu_{\mathrm{Acc}}^{(t)}(S)
+\lambda_{\mathrm{Div}}\,\nu_{\mathrm{Div}}^{(t)}(S)
+\lambda_{\mathrm{Qua}}\,\nu_{\mathrm{Qua}}^{(t)}(S),
\end{equation}
Where $\lambda_{\mathrm{Acc}},\,\lambda_{\mathrm{Div}},\,\lambda_{\mathrm{Qua}}\ge 0$ and $\lambda_{\mathrm{Acc}},\,\lambda_{\mathrm{Div}},\,\lambda_{\mathrm{Qua}} = 1$, substituting this into the Shapley value formula yields:
\begin{equation}\phi_i^{(t)}=\sum_{S\subseteq N\setminus\{i\}}\frac{|S|!\,\bigl(|N|-|S|-1\bigr)!}{|N|!}\left(\nu^{(t)}(S\cup\{i\})-\nu^{(t)}(S)\right),\end{equation}

Because surge event prediction is highly sensitive with time, the contribution of earlier training rounds should decay over time. We therefore introduce a time decay weight and compute a weighted sum of each round Shapley values:
\begin{equation}
\Phi_i=\sum_{t=1}^{T}\omega_t\,\phi_i^{(t)},\qquad
\omega_t \propto \exp\!\bigl(-\lambda (T-t)\bigr),
\end{equation}
where $\omega_t$ is the time-decay weight (normalized such that $\sum_{t=1}^{T}\omega_t=1$) and $\lambda>0$ is the time-decay rate, with larger $\lambda$ assigning more weight to recent rounds.

\subsection{Positive Example Driven Shapley Computation}
Computing Shapley values requires enumerating all coalitions $S \subseteq N$, which incurs an exponential complexity of $O(2^{|N|})$ and leads to prohibitive training overhead. Moreover, rare event positives are typically concentrated in a few specific periods or scenarios. If random sampling is applied, the number of positive examples can vary significantly across batches, which causes large variance in Shapley values.
To address this issue, we propose a fast Shapley value computation method driven by rare positive examples. The core idea is to partition the validation set into subsets $\{\mathcal{D}_{\mathrm{val}}^{(r)}\}_{r=1}^{R}$, retain all positive samples and only a fixed proportion of negative samples in each subset, and merge a large number of clients whose contributions to rare positives are negligible into a single virtual client. Shapley values are then computed only for the $K$ clients with substantial impact on rare positives together with the virtual client (a total of $K+1$ clients). Finally, the Shapley value assigned to the virtual client is redistributed to the merged clients according to their impact on rare positives. The description of this procedure is given in Alg.~\ref{alg:pos_shapley_concise}.

\begin{algorithm}[t]
\caption{Positive Example-Driven Approximate Shapley}
\label{alg:pos_shapley_concise}
\begin{algorithmic}[1]
  \Input Global model $\theta^{(t)}$, client updates $\{\Delta_i^{(t)}\}$, scenario-wise validation sets $\{\mathcal{D}^{(r)}_{\mathrm{val}}\}$, previous scores $\{\widehat{\Phi}_i\}$
  \Output Per-round Shapley values $\{\hat{\phi}_i^{(t)}\}$, updated scores $\{\widehat{\Phi}_i\}$

  \State \textbf{Step 1: Construct validation set}
  \State $\{\widetilde{\mathcal{D}}^{(r)}_{\mathrm{val}}\} \gets \mathrm{KeepPosSampleNeg}(\{\mathcal{D}^{(r)}_{\mathrm{val}}\})$

  \State \textbf{Step 2: Positive-impact scoring}
  \For{each client $i$}
    \State $\eta_i^{(t)} \gets \mathrm{PosImpact}(\theta^{(t)}, \Delta_i^{(t)}, \{\widetilde{\mathcal{D}}^{(r)}_{\mathrm{val}}\})$
  \EndFor

  \State \textbf{Step 3: Select and merge}
  \State $\mathcal{H}^{(t)} \gets \mathrm{TopK}(\{\eta_i^{(t)}\})$
  \State $\mathcal{G}^{(t)} \gets \mathrm{MergeClients}(\mathcal{H}^{(t)}, \{\eta_i^{(t)}\})$
  \State $\widetilde{\mathcal{G}}^{(t)} \gets \mathcal{G}^{(t)} \cup \{\mathcal{H}^{(t)}\}$

  \State \textbf{Step 4: Group-level Shapley}
  \State $\{\hat{\phi}_{G}^{(t)}\}_{G\in\widetilde{\mathcal{G}}^{(t)}} \gets \mathrm{ShapleyPerm}(\widetilde{\mathcal{G}}^{(t)})$

  \State \textbf{Step 5: Redistribute to clients}
  \State $\{\hat{\phi}_i^{(t)}\} \gets \mathrm{Distribute}(\{\hat{\phi}_{G}^{(t)}\}, \{\eta_i^{(t)}\}, \mathcal{G}^{(t)}, \mathcal{H}^{(t)})$

  \State \textbf{Step 6: Time-decayed accumulation}
  \For{each client $i$}
    \State $\widehat{\Phi}_i \gets \gamma\,\widehat{\Phi}_i + (1-\gamma)\,\hat{\phi}_i^{(t)}$
  \EndFor

  \State \Return $\{\hat{\phi}_i^{(t)}\},\ \{\widehat{\Phi}_i\}$
\end{algorithmic}
\end{algorithm}

Specifically, the method consists of the following steps:

\textit{1. Construct a validation subset.}
For each scenario $r$, let
$\mathcal{P}^{(r)}=\{(x,y)\in\mathcal{D}_{\mathrm{val}}^{(r)}:y=1\}$ 
and 
$\mathcal{N}^{(r)}=\{(x,y)\in\mathcal{D}_{\mathrm{val}}^{(r)}:y=0\}$ 
denote the positive and negative sets, respectively.
We retain all positives and only a fixed ratio $\rho$ of negatives by
\[
\widetilde{\mathcal{D}}_{\mathrm{val}}^{(r)}
=\mathcal{P}^{(r)}\cup\widetilde{\mathcal{N}}^{(r)},\qquad
\bigl|\widetilde{\mathcal{N}}^{(r)}\bigr|
=\min\!\left(\rho\,\bigl|\mathcal{P}^{(r)}\bigr|,\ \bigl|\mathcal{N}^{(r)}\bigr|\right).
\]

\textit{2. Evaluate each client’s influence on positives.}
For round $t$, let $\mathcal{P}^{(t)}$ be the set of participating clients.
We first compute the baseline prediction and the change caused by client $i$:
\begin{equation}
\begin{aligned}
s_0^{(t)}(x) &= p_{\theta^{(t)}}(x),\\
\Delta s_i^{(t)}(x) &=
p_{\theta^{(t)}\oplus \Delta_i^{(t)}}(x)-p_{\theta^{(t)}}(x),
\quad \forall\, i\in\mathcal{P}^{(t)}.
\end{aligned}
\end{equation}
For any coalition $S\subseteq \mathcal{P}^{(t)}$, we approximate
\begin{equation}
s_S^{(t)}(x)\approx
\Big[s_0^{(t)}(x)+\sum_{i\in S}\Delta s_i^{(t)}(x)\Big]_{0}^{1},
\end{equation}
where $[\cdot]_{0}^{1}$ clips the value to $[0,1]$.

We select hard positives and critical negatives as the bottom-$M$ and top-$H$ samples ranked by $s_0^{(t)}(x)$ in 
$\mathcal{P}^{(r)}$ and $\widetilde{\mathcal{N}}^{(r)}$, and set the decision threshold
$\tau_r^{\delta}$ as the $(1-\delta)$-quantile of $\{s_0^{(t)}(x): x\in\widetilde{\mathcal{N}}^{(r)}\}$.
The influence of client $i$ in scenario $r$ and round $t$ is
\begin{equation}
\begin{aligned}
\mathrm{Imp}_i^{(r,t)}
&= \frac{1}{\left|\mathcal{P}_{\mathrm{hard}}^{(r)}\right|}
   \sum_{x\in \mathcal{P}_{\mathrm{hard}}^{(r)}}
   \left[ s_{\{i\}}^{(t)}(x) - \tau_{r}^{\delta} \right]_+ \\
&\quad - \lambda_{\mathrm{fp}}\,
   \frac{1}{\left|\mathcal{N}_{\mathrm{crit}}^{(r)}\right|}
   \sum_{x\in \mathcal{N}_{\mathrm{crit}}^{(r)}}
   \left[ s_{\{i\}}^{(t)}(x) - \tau_{r}^{\delta} \right]_+ ,
\end{aligned}
\end{equation}
where $[z]_+ = \max(z,0)$ and $\lambda_{\mathrm{fp}}>0$ is the false-positive penalty.

\textit{3. Merge similar clients.}
Let $\mathcal{U}^{(t)}$ be the set of clients in round $t$, which is partitioned into
$\mathcal{G}^{(t)}=\{G_1,\ldots,G_M\}$ as a collection of disjoint groups whose union is $\mathcal{U}^{(t)}$.
Clients are clustered according to their impact vectors $\mathbf{v}_i^{(t)}$ via cosine similarity
\begin{equation}
\mathrm{sim}(i,j)
=\frac{\left\langle \mathbf{v}_i^{(t)},\mathbf{v}_j^{(t)}\right\rangle}
{\left\|\mathbf{v}_i^{(t)}\right\|\ \left\|\mathbf{v}_j^{(t)}\right\|},
\end{equation}
and merged when $\mathrm{sim}(i,j)\ge \kappa$.

\textit{4. Compute group Shapley values.}
For any group coalition $S\subseteq \mathcal{G}^{(t)}$, we define the scenario-specific utility
\begin{equation}
\begin{aligned}
u_r^{(t)}(S)
&=\frac{1}{\left|\mathcal{P}_{\mathrm{hard}}^{(r)}\right|}
  \sum_{x\in \mathcal{P}_{\mathrm{hard}}^{(r)}}
  \left[s_S^{(t)}(x)-\tau_r^{\delta}\right]_+ \\
&\quad -\lambda_{\mathrm{fp}}\,
  \frac{1}{\left|\mathcal{N}_{\mathrm{crit}}^{(r)}\right|}
  \sum_{x\in \mathcal{N}_{\mathrm{crit}}^{(r)}}
  \left[s_S^{(t)}(x)-\tau_r^{\delta}\right]_+ ,
\end{aligned}
\end{equation}
and aggregate across scenarios by
\begin{equation}
v^{(t)}(S)=\sum_{r=1}^{R}\omega_r\,u_r^{(t)}(S),\qquad \sum_{r=1}^{R}\omega_r=1.
\end{equation}
Using $K$ random permutations, the Shapley value of group $G_m$ is approximated by
\begin{equation}
\hat{\phi}_{G_m}^{(t)}
=\frac{1}{K}\sum_{k=1}^{K}
\Big(v^{(t)}\!\big(S_{k,m}\cup\{G_m\}\big)-v^{(t)}\!\big(S_{k,m}\big)\Big),
\end{equation}
where $S_{k,m}$ is the set of groups preceding $G_m$ in the $k$-th permutation.

\textit{5. Redistribute group Shapley values.}
We aggregate each client’s influence as
\begin{equation}
\begin{aligned}
\eta_i^{(t)}
&=\sum_{r=1}^{R}\omega_r\,\bigl[\mathrm{Imp}_i^{(r,t)}\bigr]_+,\\
\alpha_i^{(t)}
&=\frac{\eta_i^{(t)}}{\sum_{j\in G_m}\eta_j^{(t)}+\varepsilon},
\quad i\in G_m,
\end{aligned}
\end{equation}
and assign client Shapley values by
\begin{equation}
\hat{\phi}_i^{(t)}=\alpha_i^{(t)}\,\hat{\phi}_{G_m}^{(t)},\qquad i\in G_m.
\end{equation}

\textit{6. Final Shapley value calculation.}
We introduce a time-decay factor $\gamma\in(0,1]$ and accumulate values as
\begin{equation}
\widehat{\Phi}_i=\sum_{t=1}^{T}\gamma^{\,T-t}\,\hat{\phi}_i^{(t)},
\qquad \gamma\in(0,1].
\end{equation}

This method reduces the coalition from $n$ clients to $M$ groups and changes the complexity from exponential in $n$ to approximately linear in $M$.

\subsection{A Secure Model Aggregation Method Based on Consensus}

\begin{algorithm}[t]
\caption{Secure Model Aggregation}
\label{alg:secure_agg}
\begin{algorithmic}[1]
  \Input Round $t$, model $\theta^{(t)}$, client updates $\{\Delta_i^{(t)}\}_{i\in\mathcal{P}^{(t)}}$, Shapley scores $\{\Phi_i^{(t)}\}$, clip bound $C$, lr $\eta$, activation $\psi(\cdot)$, small constant $\varepsilon$, consensus threshold $\tau$, public key $PK_{\Theta}$
  \Output Next model $\theta^{(t+1)}$, block record $(msg_t,\sigma_t,\theta^{(t+1)})$

  \State \textbf{Step 1: Construct candidate aggregation set}
  \State $\mathcal{A}^{(t)} \gets \mathrm{Filter}(\mathcal{P}^{(t)},\{\Phi_i^{(t)}\})$
  \State $msg_t \gets H\!\bigl(t \,\|\, \mathcal{A}^{(t)} \,\|\, \mathrm{digest}(\{\Delta_i^{(t)}\}_{i\in\mathcal{A}^{(t)}})\bigr)$

  \State \textbf{Step 2: Voting \& threshold signature}
  \State Validators run stake voting on $msg_t$ and aggregate signatures into $\sigma_t$ if total affirmative weight $\ge \tau$
  \State Ensure $\mathrm{Verify}(PK_{\Theta}, msg_t, \sigma_t)=1$

  \State \textbf{Step 3: Aggregation based on Shapley value}
  \For{each $i\in\mathcal{A}^{(t)}$}
    \State $\tilde{\Phi}_i^{(t)} \gets \max(\Phi_i^{(t)},0)$
  \EndFor
  \State $Z \gets \sum_{j\in\mathcal{A}^{(t)}} \psi(\tilde{\Phi}_j^{(t)}) + \varepsilon$
  \State $\Delta^{(t)} \gets \sum_{i\in\mathcal{A}^{(t)}} \frac{\psi(\tilde{\Phi}_i^{(t)})}{Z}\,\mathrm{clip}(\Delta_i^{(t)},C)$
  \State $\theta^{(t+1)} \gets \theta^{(t)} \oplus \eta\,\Delta^{(t)}$

  \State \textbf{Step 4: Block packing and broadcast}
  \State Broadcast block $(msg_t,\sigma_t,\mathcal{A}^{(t)},\theta^{(t+1)})$

  \State \Return $\theta^{(t+1)}$, $(msg_t,\sigma_t,\theta^{(t+1)})$
\end{algorithmic}
\end{algorithm}

This stage embeds the aggregation step of the update uploaded by the client with added Gaussian noise into the consensus process of the blockchain. The consensus result is the only trusted global update, thereby avoiding the risk of single point attacks. The detailed procedure is given in Alg.~\ref{alg:secure_agg}.

We treat each client’s Shapley value as its contribution equity. Only clients that receive sufficient votes in a consensus round are admitted to the aggregatable set $\mathcal{A}^{(t)}$ and allowed to join model aggregation and receive global updates. Because eligibility and rewards depend on these Shapley scores, clients are incentivized to improve local training quality.

Specifically, in round $t$, let the validator committee be $\mathcal{C}^{(t)}$ with voting weights $s_v^{(t)} > 0$ for $v\in\mathcal{C}^{(t)}$ and total stake
$W_t \triangleq \sum_{v\in \mathcal{C}^{(t)}} s_v^{(t)}$.
For each client update $\Delta_i^{(t)}$, validator $v$ returns a binary decision $b_{v,i}^{(t)}\in\{0,1\}$.
The weighted vote received by client $i$ is
\begin{equation}
B_i^{(t)} \triangleq \sum_{v\in \mathcal{C}^{(t)}} s_v^{(t)}\, b_{v,i}^{(t)}\, \psi\!\left(\tilde{\Phi}_i^{(t)}\right),
\qquad
\tilde{\Phi}_i^{(t)}=\max\!\left(\Phi_i^{(t)},0\right),
\end{equation}
where $\psi(\cdot)\ge 0$.

A client is admitted to the candidate aggregation set iff
\begin{equation}
\mathcal{A}^{(t)} \triangleq \left\{ i:\ B_i^{(t)} \ge \zeta W_t \right\}.
\end{equation}

The proposer then packages $\mathcal{A}^{(t)}$ and its updates into
\begin{equation}
msg_t = H\!\left(t \,\|\, \mathcal{A}^{(t)} \,\|\, \mathrm{digest}\!\left(\{\Delta_i^{(t)}\}_{i\in\mathcal{A}^{(t)}}\right)\right),
\end{equation}
and the committee gives a threshold signature $\sigma_t$ such that
$\mathrm{Verify}\!\left(PK_{\Theta},\, msg_t,\, \sigma_t\right)=1$.

We perform aggregation by normalizing nonnegative scores:
\begin{equation}
\begin{aligned}
w_i^{(t)} \; &=\;
\frac{\psi\!\left(\tilde{\Phi}_i^{(t)}\right)}
{\sum_{j\in\mathcal{A}^{(t)}} \psi\!\left(\tilde{\Phi}_j^{(t)}\right)+\varepsilon},\\[2pt]
\Delta^{(t)} \; &=\;
\sum_{i\in\mathcal{A}^{(t)}} w_i^{(t)}\,\mathrm{clip}\!\left(\Delta_i^{(t)},C\right),\\[2pt]
\theta^{(t+1)} \; &=\;
\theta^{(t)} \oplus \eta\,\Delta^{(t)} .
\end{aligned}
\end{equation}

The updated model $\theta^{(t+1)}$ is then broadcast only to clients in $\mathcal{A}^{(t)}$. In this way, the validation committee excludes malicious clients from $\mathcal{A}^{(t)}$, and 
aggregation based on Shapley values further ensures the security of the global model.

\section{SECURITY ANALYSIS}
In this section, we theoretically analyze the proposed SI-ChainFL method and its overall dependability.

\subsection{Basic Assumptions}

\textit{Assumption 1 (Clipped Update Bound):}
For any round $t$ and any client $i\in\mathcal{A}^{(t)}$, the uploaded update is $\ell_2$–clipped with a fixed bound $C>0$, i.e.,
\begin{equation}
\|\bar{\Delta}_i^{(t)}\|_2 \le C,\qquad \forall\, i\in\mathcal{A}^{(t)}.
\end{equation}

\textit{Assumption 2 (Shapley Weight Fidelity):}
The estimated Shapley score $\Phi_i^{(t)}$ is positively aligned with the true marginal gain $\Delta v_i^{(t)}$ in expectation, so that larger $\Phi_i^{(t)}$ statistically corresponds to larger $\Delta v_i^{(t)}$ over clients in $\mathcal{A}^{(t)}$.

\textit{Assumption 3 (Stake-Weighted Committee Honesty):}
In each consensus round, the stake share of Byzantine validators in $\mathcal{C}^{(t)}$ is bounded by a constant margin below $1/3$, i.e.,
\begin{equation}
\frac{\sum_{v\in \mathcal{C}^{(t)}} b_v^{(t)} s_v^{(t)}}{\sum_{v\in \mathcal{C}^{(t)}} s_v^{(t)}}
\le \frac{1}{3}-\xi ,
\end{equation}
where $b_v^{(t)}\in\{0,1\}$ indicates whether validator $v$ is Byzantine and $\xi>0$ is a constant slack.

\vspace{0.4em}
\subsection{Robustness and Convergence}

Validator nodes first form a candidate set $\mathcal{A}^{(t)}$ of acceptable updates, then perform Shapley-based weighted aggregation:
\begin{equation}
\begin{aligned}
w_i^{(t)} &=
\frac{\psi\big(\max(\Phi_i^{(t)},0)\big)}
{\sum_{j\in\mathcal{A}^{(t)}} \psi\big(\max(\Phi_j^{(t)},0)\big) + \varepsilon},\\
\Delta^{(t)} &=
\sum_{i\in\mathcal{A}^{(t)}} w_i^{(t)}\,\mathrm{clip}\big(\Delta_i^{(t)},C\big).
\end{aligned}
\end{equation}

and update
$\theta^{(t+1)}=\theta^{(t)}-\eta\,\Delta^{(t)}$.

\textit{Lemma 1:}
Let $\mathcal{M}^{(t)}\subseteq\mathcal{A}^{(t)}$ be the malicious subset and
$\alpha_t=\sum_{i\in\mathcal{M}^{(t)}} w_i^{(t)}$ its total weight. Under Assumption~1,
\begin{equation}
\Big\|\sum_{i\in \mathcal{M}^{(t)}} w_i^{(t)}\,\mathrm{clip}(\Delta_i^{(t)},C)\Big\|_2
\le \alpha_t\,C .
\end{equation}
Thus the bias introduced by malicious clients in each round is linearly bounded by their total weight and the clipping threshold.

Under Assumption~2 and standard smoothness conditions, the expected suboptimality after $T$ rounds admits a bound of the form
\begin{equation}
\mathbb{E}\big[F(\theta^{T})\big]-F^{*}
\;\le\;
O\!\Big(\tfrac{1}{\sqrt{T}}\Big)
+
O\!\Big(\tfrac{\nu^{2}}{\sqrt{T}}\Big)
+
O\!\Big(\tfrac{1}{T}\sum_{t=1}^{T}(\alpha_t C)^{2}\Big),
\end{equation}
where $\nu^{2}$ denotes the variance of the stochastic gradients. As long as Shapley-based weighting keeps $\alpha_t$ small, the method retains standard convergence behavior while limiting adversarial drift.

\vspace{0.4em}
\subsection{Privacy}

Each client first clips its local update and then adds Gaussian noise before broadcasting:
\begin{equation}
\mathcal{M}_{i,t}(D_i)
=
\tilde{\Delta}_i^{(t)}
=
\bar{\Delta}_i^{(t)} + z_{i,t},
\qquad
z_{i,t}\sim\mathcal{N}(0,\sigma_{i,t}^{2}\mathbf{I}),
\end{equation}
where $\|\bar{\Delta}_i^{(t)}\|_2\le C$ by Assumption~1.

\textit{Theorem 1 (Per-Round DP):}
For adjacent datasets $D_i\sim D_i'$ differing in one record, the Gaussian mechanism above guarantees $(\varepsilon_{i,t},\delta_{i,t})$–differential privacy for suitable $\sigma_{i,t}$ chosen as a function of $C$, $\varepsilon_{i,t}$, and $\delta_{i,t}$.

\textit{Theorem 2 (Composition):}
Let $\mathcal{M}_{i,t}$ be $(\varepsilon_{i,t},\delta_{i,t})$–DP for each round $t=1,\ldots,T$.

1) For each client $i$, the sequence
\begin{equation}
\mathcal{M}_{i,[T]}(D_i)
=
\big(\mathcal{M}_{i,1}(D_i),\ldots,\mathcal{M}_{i,T}(D_i)\big)
\end{equation}
satisfies $(\varepsilon_i,\delta_i)$–DP with
$\varepsilon_i=\sum_{t=1}^{T}\varepsilon_{i,t}$,
$\delta_i=\sum_{t=1}^{T}\delta_{i,t}$.

2) If client datasets are disjoint, the joint mechanism
\begin{equation}
\mathcal{M}(D)
=
\big(\mathcal{M}_{1,[T]}(D_1),\ldots,\mathcal{M}_{n,[T]}(D_n)\big)
\end{equation}
satisfies $(\varepsilon,\delta)$–DP with
$\varepsilon=\max_i \varepsilon_i$,
$\delta=\max_i \delta_i$.

\vspace{0.4em}
\subsection{Aggregation Security}

Even in the presence of malicious clients, the proposed aggregation rule guarantees that the global update is bounded and that the deviation from the aggregation is controlled.

\textit{Lemma 2:}
For any $i\in\mathcal{A}^{(t)}$, we have $w_i^{(t)}\ge 0$ and
\begin{equation}
\sum_{i\in\mathcal{A}^{(t)}} w_i^{(t)}
=
\frac{\sum_{i}\psi(\tilde{\Phi}_i^{(t)})}{\sum_{i}\psi(\tilde{\Phi}_i^{(t)})+\varepsilon}
\le 1,
\end{equation}
All aggregation weights are nonnegative and their sum does not exceed~1.

\textit{Lemma 3:}
Under Assumption~1, the global update is uniformly bounded:
\begin{equation}
\|\Delta^{(t)}\|_2
=
\Big\|\sum_{i\in\mathcal{A}^{(t)}} w_i^{(t)}\,\bar{\Delta}_i^{(t)}\Big\|_2
\le C,\qquad \forall t.
\end{equation}

\textit{Lemma 4:}
Let $\mathcal{M}^{(t)}\subseteq\mathcal{A}^{(t)}$ denote malicious clients and
$\alpha_t=\sum_{i\in\mathcal{M}^{(t)}} w_i^{(t)}$. Then
\begin{equation}
\Big\|\sum_{i\in\mathcal{M}^{(t)}} w_i^{(t)}\,\bar{\Delta}_i^{(t)}\Big\|_2
\le \alpha_t C.
\end{equation}
Hence the contribution of malicious updates in each round is bounded by $\alpha_tC$.

\textit{Theorem 3 (Deviation from Honest Aggregation):}
Let $\Delta_H^{(t)}$ be the aggregation result if all clients were honest and $\Delta^{(t)}$ be the actual aggregation under the proposed rule. Then
\begin{equation}
\big\|\Delta^{(t)}-\Delta_H^{(t)}\big\|_2 \le \alpha_t C.
\end{equation}

\textit{Theorem 4 (Shapley-Based Control of Malicious Weight):}
Let $\mathcal{A}^{(t)}=\mathcal{H}^{(t)}\cup\mathcal{M}^{(t)}$ with aggregation weights
\begin{equation}
w_i^{(t)}
=
\frac{\psi(\max(\Phi_i^{(t)},0))}
{\sum_{j\in\mathcal{A}^{(t)}} \psi(\max(\Phi_j^{(t)},0))+\varepsilon},
\qquad
\alpha_t=\sum_{i\in\mathcal{M}^{(t)}} w_i^{(t)}.
\end{equation}
Assume there exists $\tau>0$ such that for any malicious client $i\in\mathcal{M}^{(t)}$,
\begin{equation}
\Pr\!\big(\Phi_i^{(t)}>\tau\big)\le \eta(\tau),
\end{equation}
where $\eta(\tau)$ is non-increasing in $\tau$.
Then, with probability at least $1-|\mathcal{M}^{(t)}|\,\eta(\tau)$,
\begin{equation}
\alpha_t
\le
\frac{|\mathcal{M}^{(t)}|\,\psi(\tau)}
{\sum_{j\in\mathcal{H}^{(t)}} \psi(\max(\Phi_j^{(t)},0))+\varepsilon}.
\end{equation}

This shows that, as long as malicious clients rarely obtain large Shapley scores, their total aggregation weight $\alpha_t$ (and thus their impact on $\Delta^{(t)}$) remains tightly controlled.

\section{EXPERIMENTS}

In this section, we empirically evaluate SI-ChainFL. We first compare it with representative methods on public datasets to validate its performance, and then predict passenger flow at high-speed rail stations using real-world traffic data to assess its effectiveness in practice.

\subsection{Experimental Setup}

\textbf{Dataset:} In our experiments, we will use two datasets. We will use the publicly available datasets as controlled classification benchmarks to verify the convergence and robustness of our proposed method. We will then use two real traffic flow datasets to verify the generalization ability of our method in traffic flow prediction scenarios.

The first series consists of the public datasets MNIST and CIFAR-10. MNIST is a benchmark dataset for handwritten digit recognition, containing 70,000 images across 10 classes (0-9), with 60,000 images in the training set and 10,000 in the test set. It is commonly used to validate the basic performance and training convergence of classification models. CIFAR-10 is a benchmark dataset for natural image classification, containing 60,000 images across 10 classes of animals, with 50,000 images in the training set and 10,000 in the test set. It is commonly used to compare the convergence and performance of models on non-IID data distributions. To simulate the uneven data distribution among clients in federated learning, we first randomly shuffled the dataset, then sampled it according to a log-normal distribution, assigning a different number of samples to each client. Finally, based on the proportions generated by the Dirichlet distribution ($\alpha$), we allocated samples of each class to each client to simulate a federated learning scenario where the data distribution is not independent and identically distributed, and the sample size is imbalanced. In this series of datasets, we used ResNet18 to train local models.

The second series of datasets includes two sets of real traffic flow datasets. The first set is the PeMS dataset\cite{guo2021learning}, which contains highway sensor data collected every 30 seconds by the CARTRANS performance testing system. It includes vehicle speed, traffic flow, and occupancy, with a data granularity of 5 minutes. This experiment uses the standard benchmark datasets PeMS07, extracted and standardized from the PeMS dataset. The time range for PeMS07 is May 1, 2017 to August 31, 2017. The second set of data comes from two sets of real high-speed rail datasets (HSR) from Chengdu Railway Technology Innovation Co., Ltd., all exported from their data platform. The dataset records the arrival and departure passenger volumes of each train at 108 high-speed railway stations from January 1, 2023 to December 31, 2024, spanning 731 days with a data granularity of 1 day. The second dataset is the corresponding weather dataset. We use STGCN for local training on various clients on this series of datasets.

\textbf{Hyperparameter settings:} In all experiments where specific details were not provided, we set the learning rate to 0.01, the number of local iterations to 5, the number of clients in the system to 100, the proportion of malicious clients to 10\%, the number of validation nodes to 10, the optimizer to SGD, and the number of communications to 100. In each round of communication, we randomly selected 70\% of the clients to participate in training. Since each selected client only uploads a clipped and noisy model update once per round, we treat each round as a differential privacy (DP) computation step, resulting in a total of 100 DP steps. We set the concentration parameter $\alpha$ of the Dirichlet distribution to 0.5, the pruning threshold $C$ to 1.0, the Gaussian noise intensity $\mu=\sigma/C$ to 8, and the DP failure probability parameter $\delta_{\mathrm{DP}}$ to $10^{-5}$.

\textbf{Experimental environment:} We used PyTorch 2.8.0 for local model training and conducted experiments on a machine learning device equipped with an RTX 5090 (32GB) and a 25 vCPU Intel(R) Xeon(R) Platinum 8470Q, using CNN as the training network model.

\textbf{Adversarial Behavior:} To evaluate the effectiveness and robustness of the proposed method, we considered two widely used attacks.

\begin{itemize}
    \item \textbf{Free-rider Attack (FR) \cite{xu2020reputation}:} Nodes initiating free-rider attacks are selfish participants who often upload minimal model updates or all-zero vectors, attempting to obtain global model updates without contributing their own data and computing power, and potentially cheating the incentive mechanism. This attack leads to decreased training efficiency and accuracy. Our experiments involved nodes initiating free-rider attacks copying updates from previous rounds as their local model updates and uploading them.

    \item \textbf{Poisoning Attack (PA) \cite{bhagoji2019analyzing}:} Poisoning attacks refer to malicious nodes uploading incorrect model updates, causing a shift in the global model and resulting in a decrease in overall model performance. We simulated the poisoning attack process by mislabeling samples (label flipping).
\end{itemize}

\textbf{Baseline Methods:} In our experiments, we compared SI-ChainFL with the following methods.
\begin{itemize}
    \item \textbf{FedAvg \cite{mcmahan2017communication}:} FedAvg is the most classic federated learning method, in which the central server uses the average of the model parameters obtained after each participant trains locally as the parameters of the global model.
    \item \textbf{FedProx \cite{li2020federated}:} FedProx adds a proximal regularization term to the local objective function to prevent the local model from deviating from the global model. The client minimizes the local loss in each training round, reducing client drift caused by inconsistent training steps.
    \item \textbf{FedEdge \cite{wang2023fededge}:} FedEdge uses edge nodes closer to the terminal as an intermediate layer for aggregation and scheduling, enabling training to be completed more promptly, reducing communication overhead and time, and improving the efficiency of training convergence.
    \item \textbf{FedSage \cite{zhang2021subgraph}:} FedSage approximates cross subgraph aggregation by generating neighbor information locally on the client side and then synchronizing model parameters globally, which can improve the training performance of GNNs without sharing the original graph results.
    \item \textbf{P2FMS \cite{xue2025burst}:} P2FMS improves its sensitivity to heterogeneous burst traffic by analyzing the fluctuation patterns of the global model and local trend changes, and improves the model convergence speed by using an alternating training method.
    \item \textbf{FLTrust \cite{cao2020fltrust}:} FLTrust introduces a trusted root dataset on the central server. The central server uses the gradients trained on this dataset to verify whether the gradients uploaded by each client are consistent. Based on the verification results, suspicious updates are filtered out to ensure the security of the system.
    \item \textbf{RAGA \cite{zuo2025federated}:} RAGA uses the geometric median to aggregate client-uploaded updates during the model aggregation phase and employs the Weiszfeld algorithm to approximate the values, thereby mitigating the impact of anomalous updates.

\end{itemize}

\subsection{Experimental Results}

To evaluate the performance of the proposed SI-ChainFL method, we conducted comprehensive experiments on model accuracy, robustness, efficiency, and ablation studies.

\subsubsection{Accuracy} To evaluate the accuracy of the SI-ChainFL model, we tested the accuracy with varying numbers of iterations and clients, assuming no malicious clients.
\begin{figure}[t]
\centering

\begin{subfigure}{0.49\columnwidth}
  \centering\includegraphics[width=\linewidth]{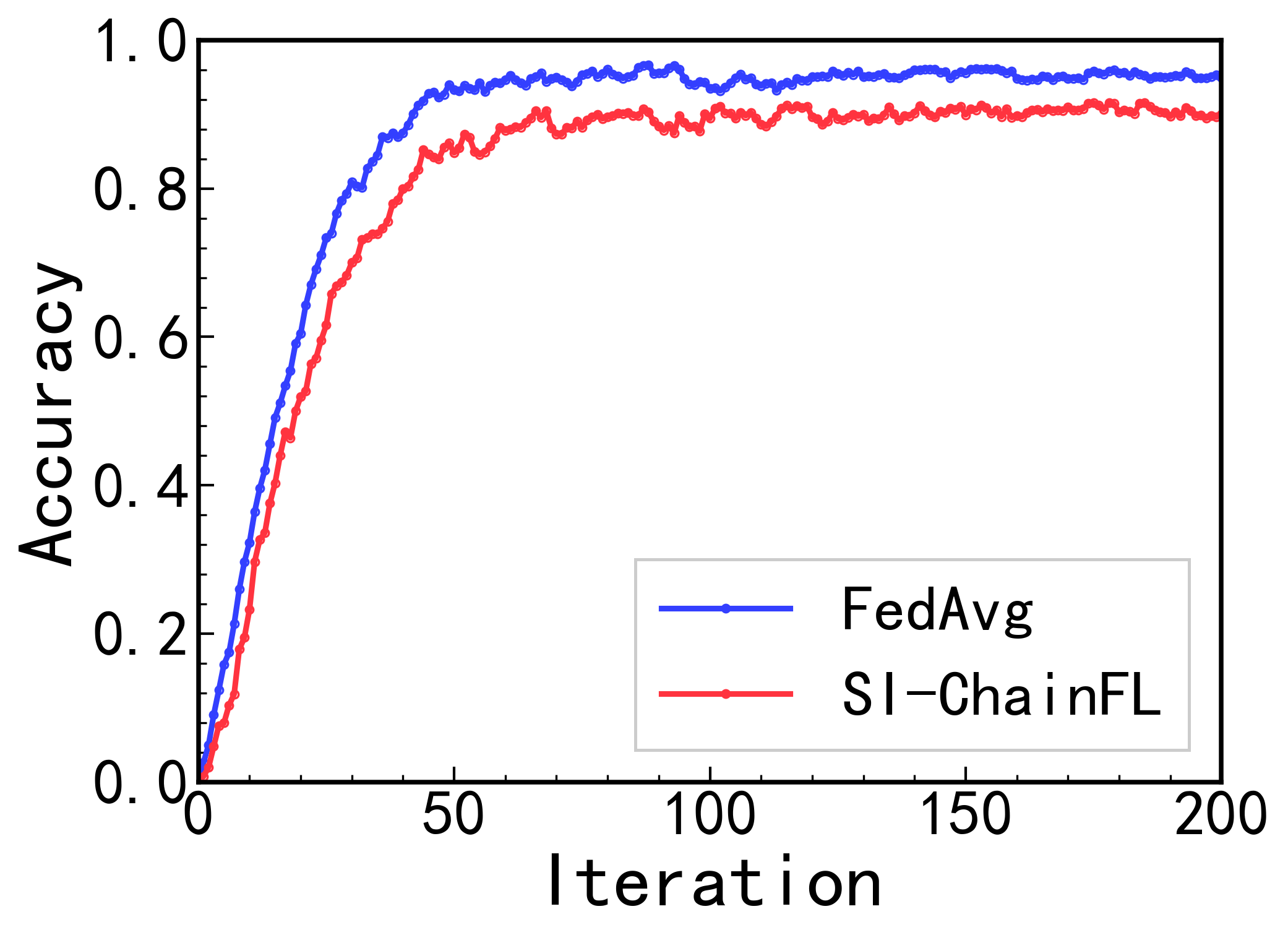}
  \caption{MNIST}
\end{subfigure}\hfill
\begin{subfigure}{0.49\columnwidth}
  \centering\includegraphics[width=\linewidth]{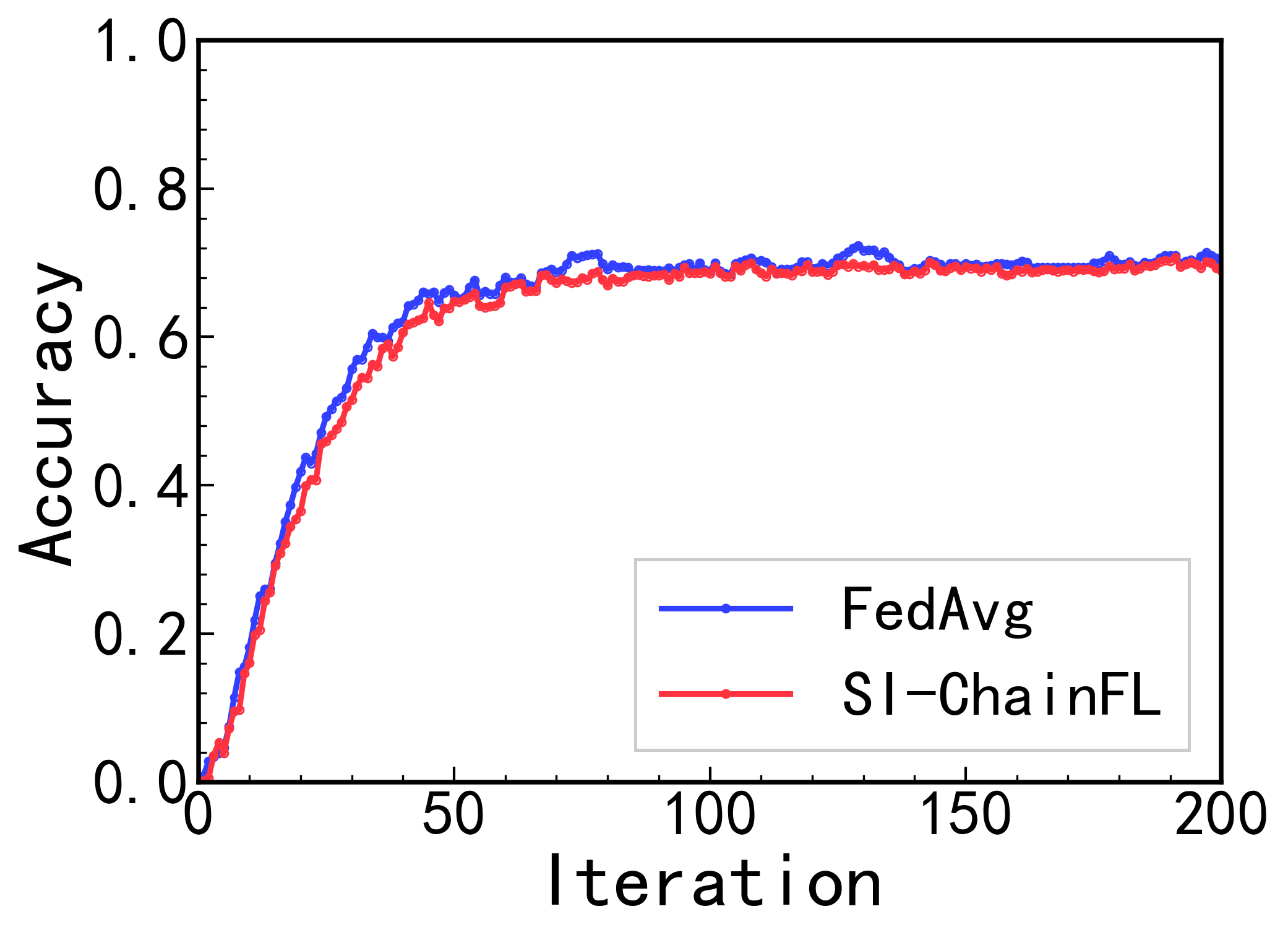}
  \caption{CIFAR-10}
\end{subfigure}

\vspace{2mm}

\begin{subfigure}{0.49\columnwidth}
  \centering\includegraphics[width=\linewidth]{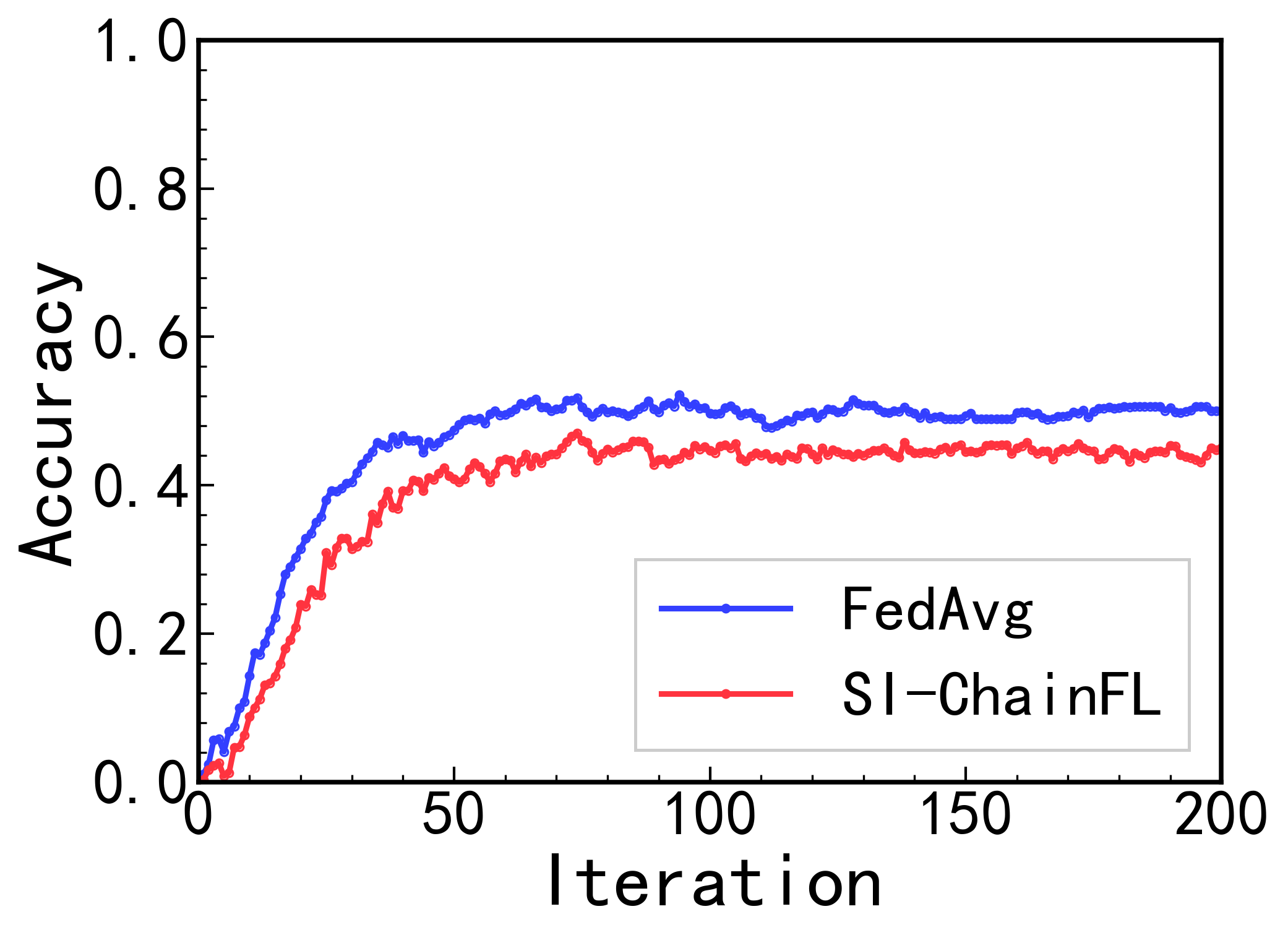}
  \caption{CIFAR-100}
\end{subfigure}\hfill
\begin{subfigure}{0.49\columnwidth}
  \centering\includegraphics[width=\linewidth]{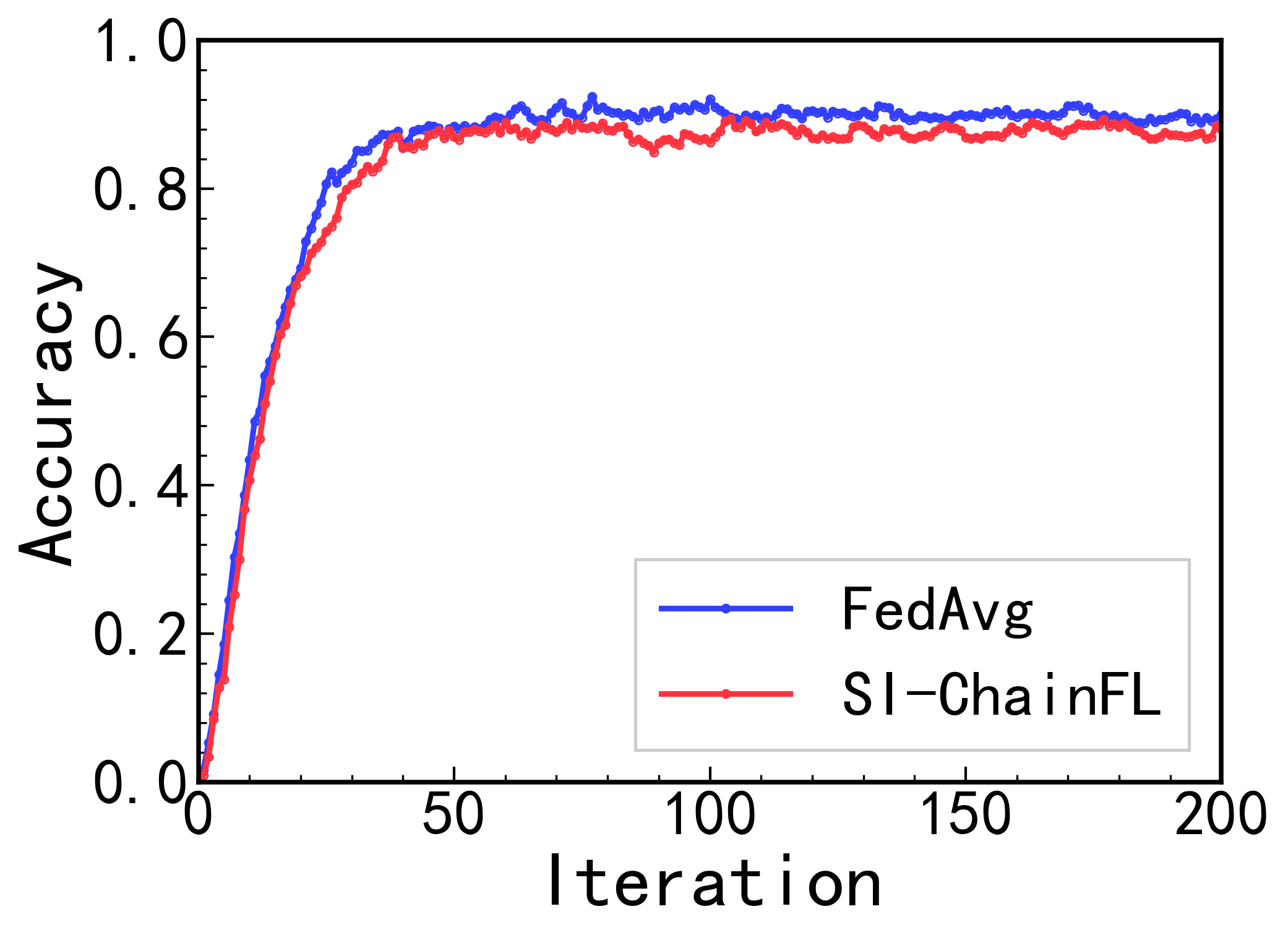}
  \caption{HSR}
\end{subfigure}

\caption{Comparison of convergence performance between SI-ChainFL and FedAvg on MNIST, CIFAR-10, CIFAR-100 and HSR.}
\label{fig:acc_noM}
\end{figure}

\textit{Impact of Iteration Count:} We tested the global model accuracy at different iteration counts and compared it with the mainstream FL method FedAvg. We used a CNN model to test on four datasets: MNIST, CIFAR-10, CIFAR-100, and the High-Speed Rail Network dataset. Detailed experimental results are shown in fig. \ref{fig:acc_noM}. As shown in fig. \ref{fig:acc_noM}, our proposed method has a similar convergence speed to FedAvg. The accuracy of the model slightly decreases on the MNIST and CIFAR-100 datasets because Gaussian noise is added when clients upload their local training updates. However, this decrease in accuracy is slight and manageable. A slight performance degradation is acceptable while ensuring model security. Meanwhile, we observed that the SI-ChainFL model maintains high accuracy on the high-speed rail network dataset, demonstrating its adaptability to such datasets. This is because the SI-ChainFL model improves system safety solely by evaluating node Shapley values to filter nodes and adding Gaussian noise to local updates, without directly modifying the training model.

\begin{figure}[t]
\centering

\begin{subfigure}{0.49\columnwidth}
  \centering\includegraphics[width=\linewidth]{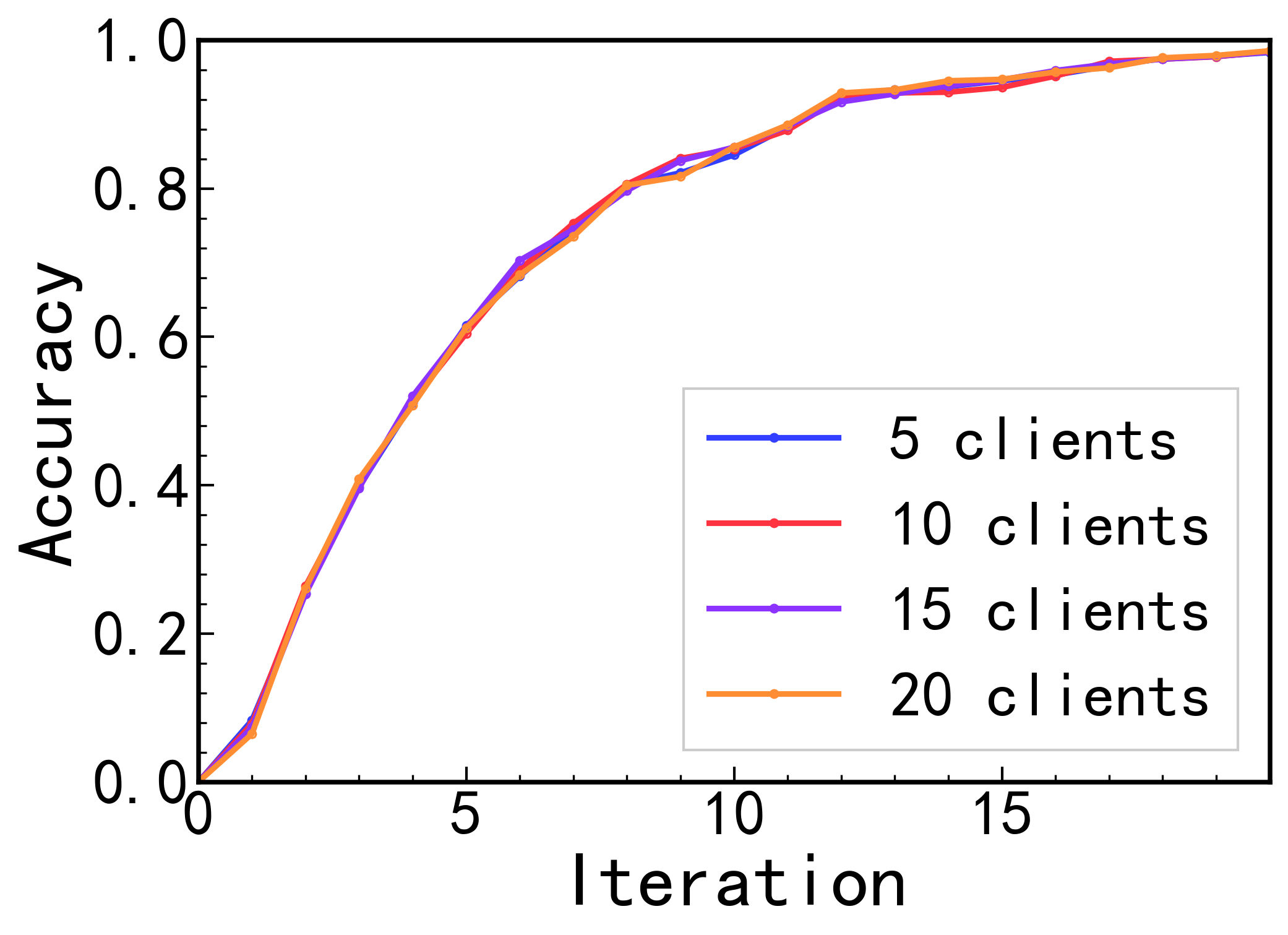}
  \caption{MNIST}
\end{subfigure}\hfill
\begin{subfigure}{0.49\columnwidth}
  \centering\includegraphics[width=\linewidth]{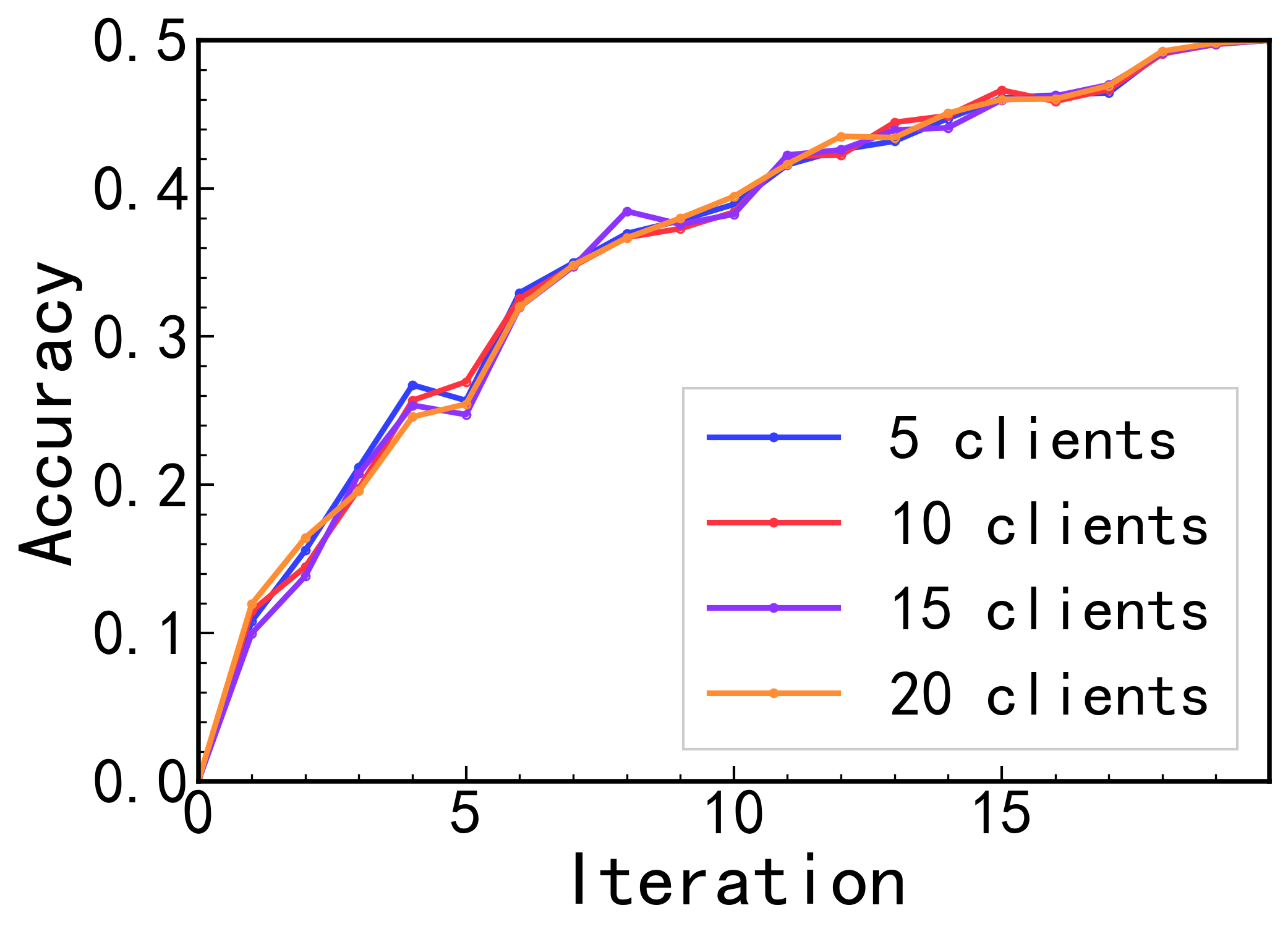}
  \caption{CIFAR-10}
\end{subfigure}

\vspace{2mm}

\begin{subfigure}{0.49\columnwidth}
  \centering\includegraphics[width=\linewidth]{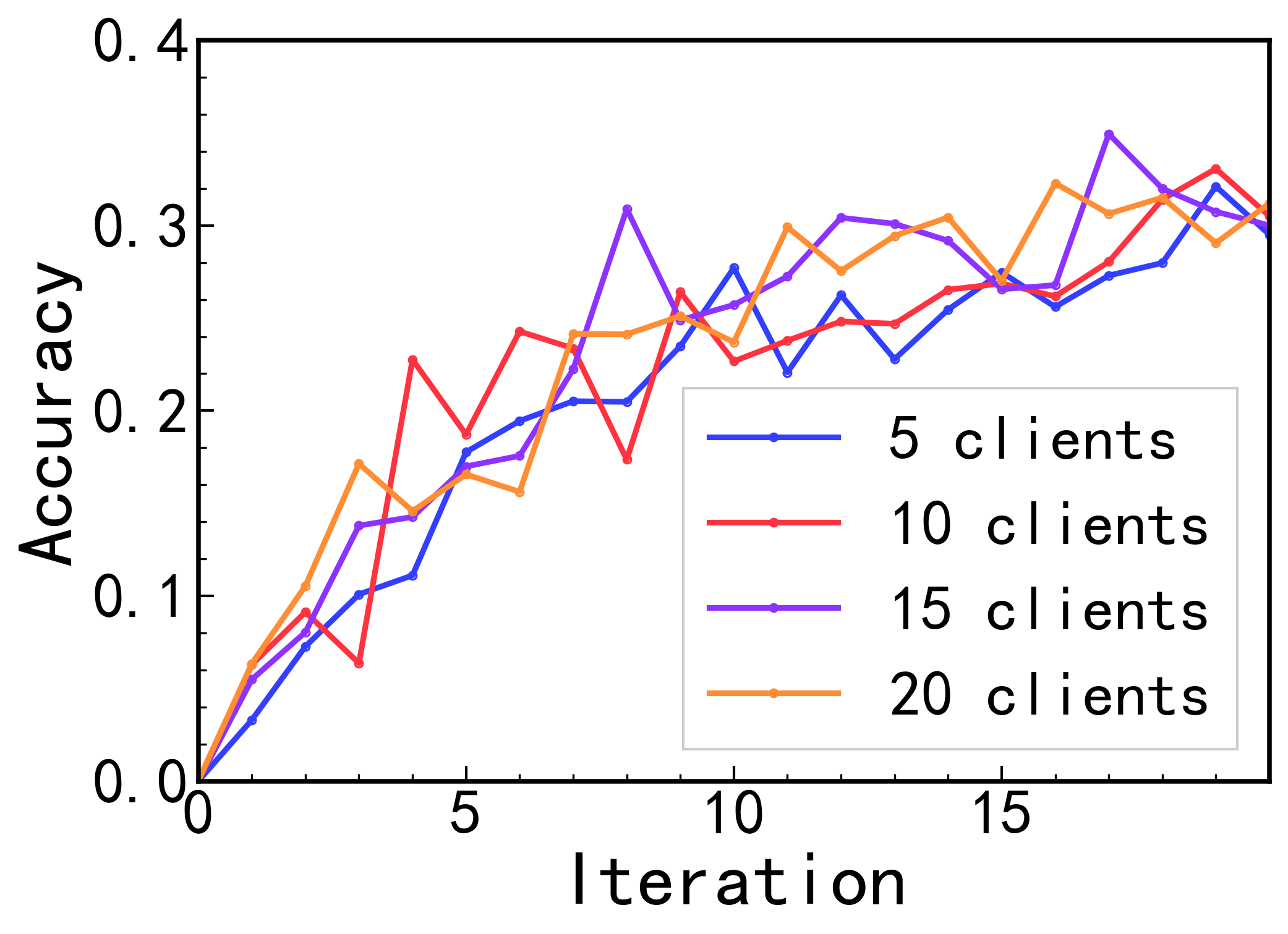}
  \caption{CIFAR-100}
\end{subfigure}\hfill
\begin{subfigure}{0.49\columnwidth}
  \centering\includegraphics[width=\linewidth]{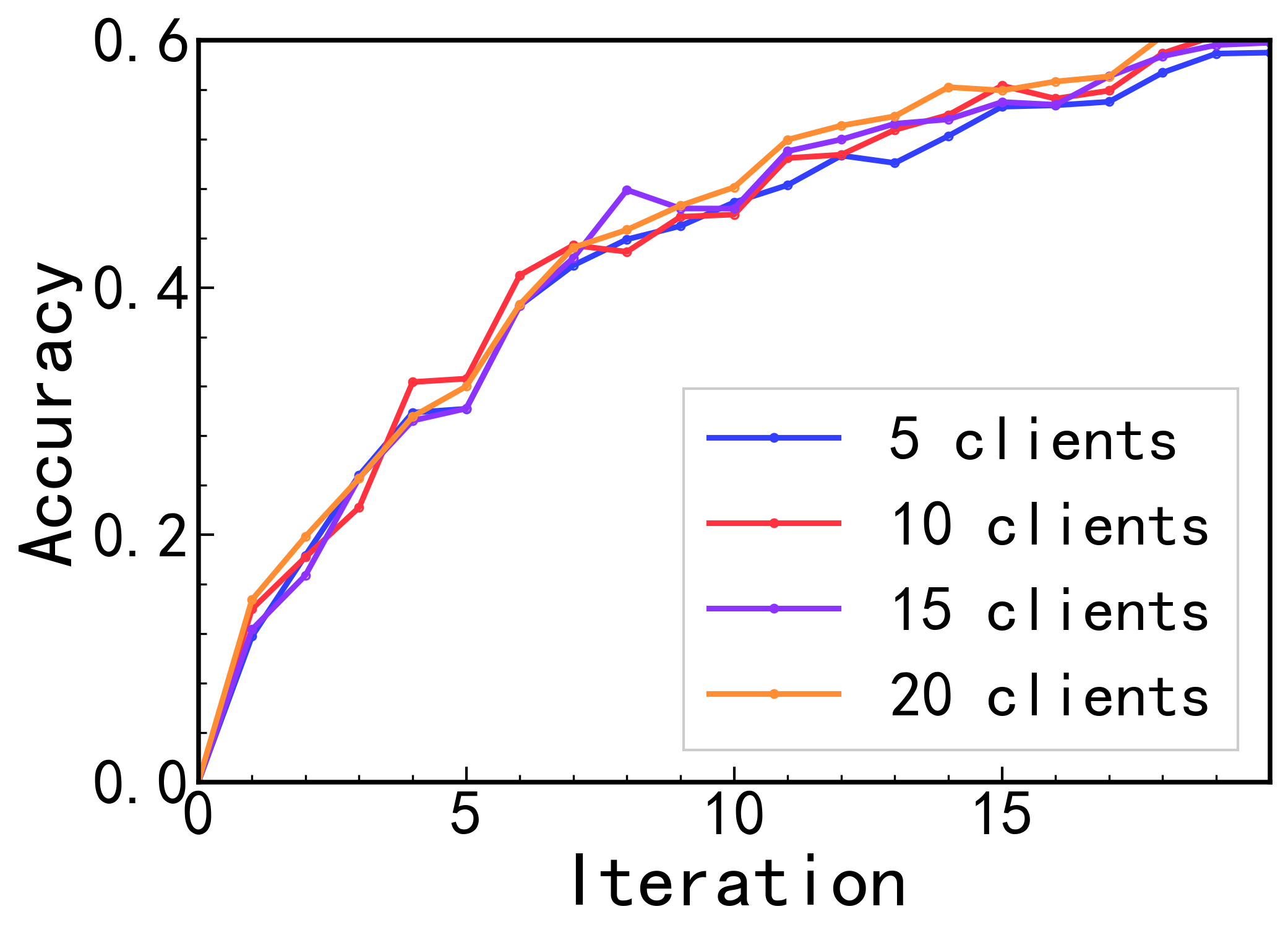}
  \caption{HSR}
\end{subfigure}

\caption{The impact of number of clients on model accuracy.}
\label{fig:acc_noM_nodes}
\end{figure}

\textit{Impact of Client Number:} We conducted experiments to investigate the impact of the number of clients on model performance.  Fig\ref{fig:acc_noM_nodes} shows the accuracy of the test model on four datasets: MNIST, CIFAR-10, CIFAR-100, and the High-Speed Rail Network dataset, with 5, 10, 15 and 20 clients. Our results show that, as can be seen from fig \ref{fig:acc_noM_nodes} (a), (b), and (d), the convergence speed and accuracy of the SI-ChainFL model remain almost constant, with only slight fluctuations in the results shown in fig \ref{fig:acc_noM_nodes} (c). Therefore, we can conclude that different numbers of clients do not affect the performance of the SI-ChainFL model, indicating that our method has good scalability.

\begin{figure}[t]
\centering

\begin{subfigure}{0.49\columnwidth}
  \centering\includegraphics[width=\linewidth]{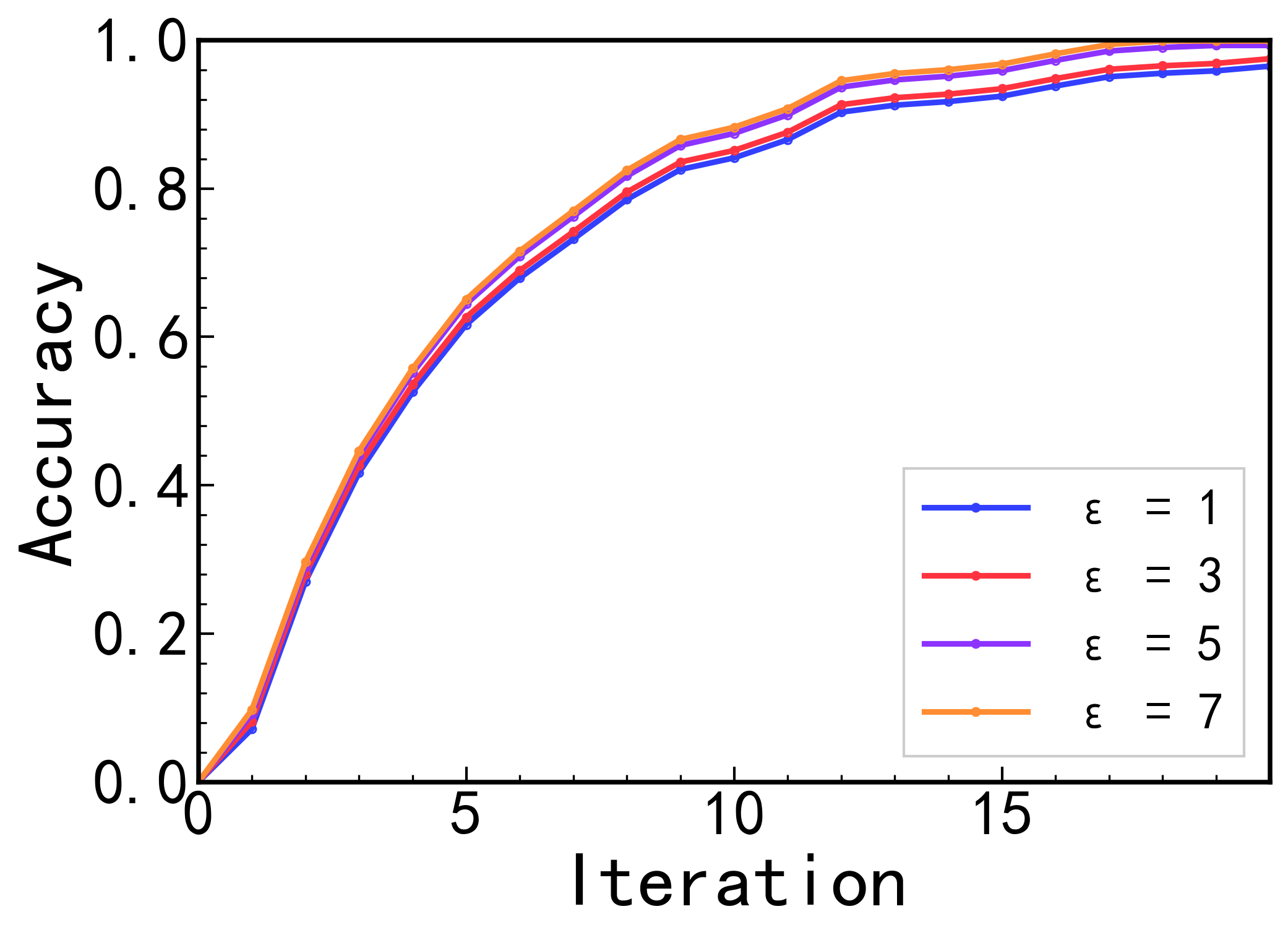}
  \caption{MNIST}
\end{subfigure}\hfill
\begin{subfigure}{0.49\columnwidth}
  \centering\includegraphics[width=\linewidth]{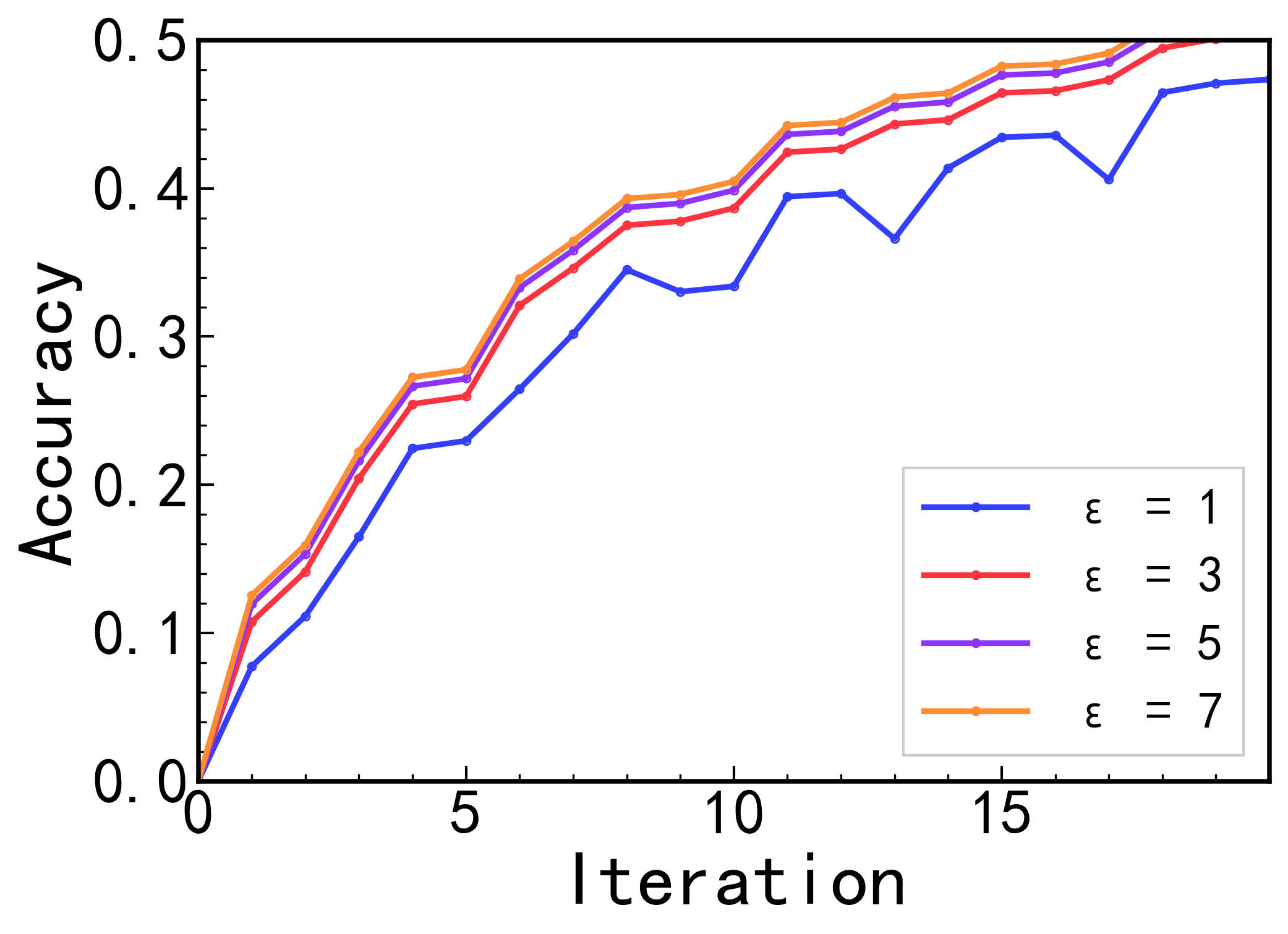}
  \caption{CIFAR-10}
\end{subfigure}

\vspace{2mm}

\begin{subfigure}{0.49\columnwidth}
  \centering\includegraphics[width=\linewidth]{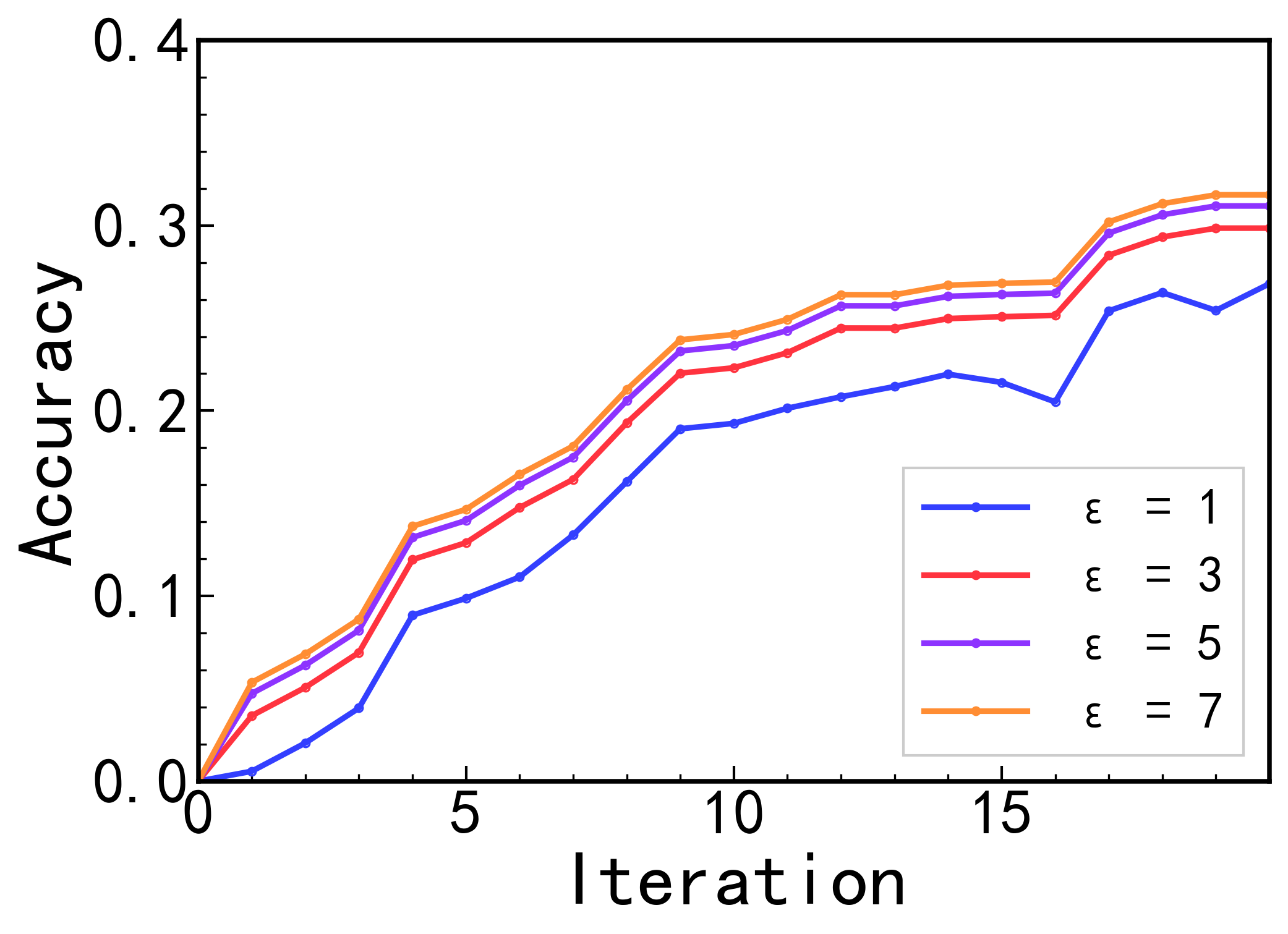}
  \caption{CIFAR-100}
\end{subfigure}\hfill
\begin{subfigure}{0.49\columnwidth}
  \centering\includegraphics[width=\linewidth]{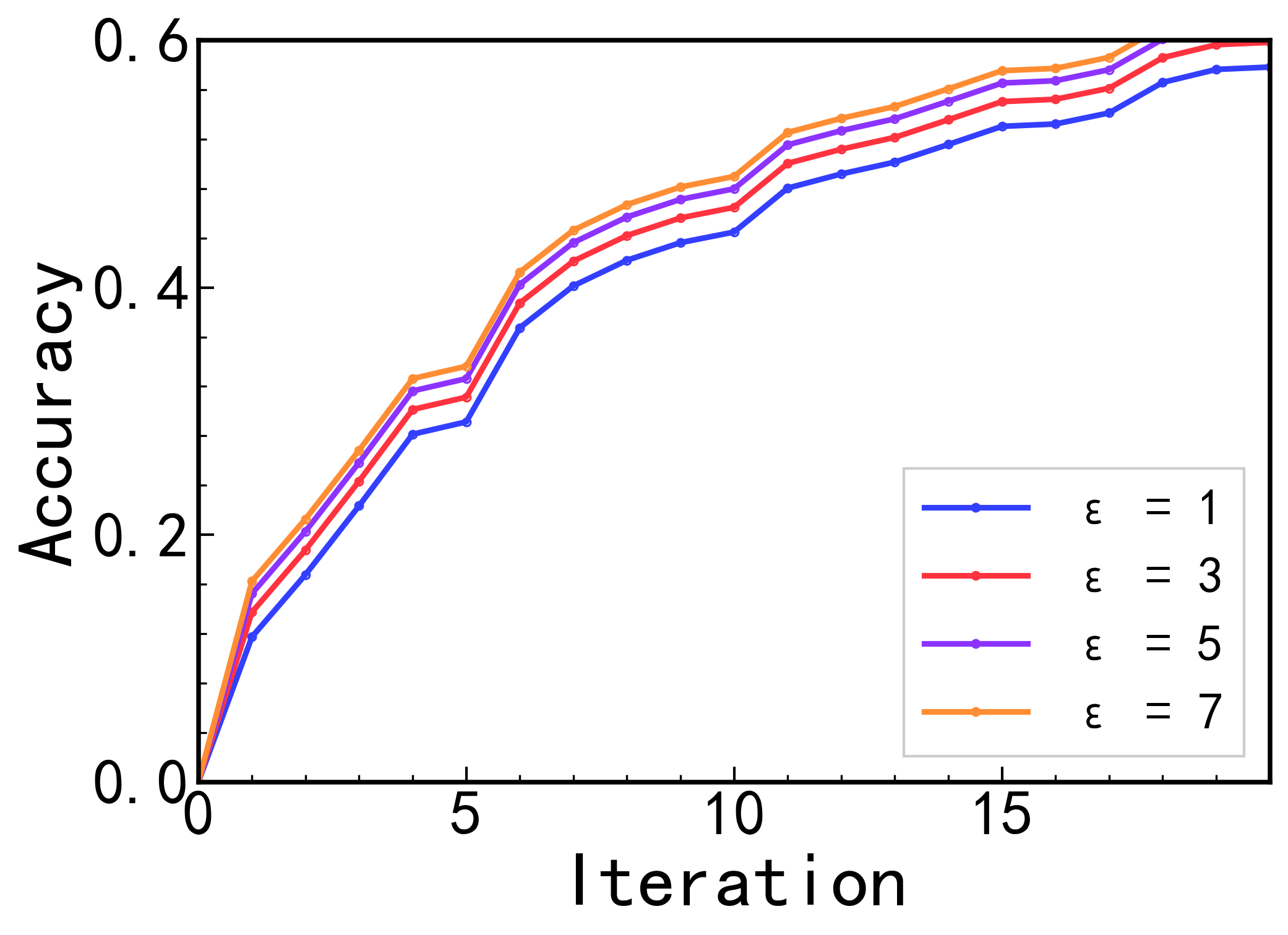}
  \caption{HSR}
\end{subfigure}

\caption{The impact of number of privacy budget on model accuracy.}
\label{fig:acc_privacy_budget}
\end{figure}

\textit{Impact of Privacy Budget:} As shown in Figure \ref{fig:acc_privacy_budget}, within the range of $\epsilon\in{1,3,5,7}$, the model accuracy improves with the increase of the privacy budget. This is because when the privacy budget is low, the noise injected into the local update is stronger, which significantly weakens the effective gradient signal and reduces the model accuracy. On the other hand, in the presence of malicious attacks, a moderate amount of noise may, to some extent, suppress the perturbations caused by malicious updates and improve training robustness; however, excessive noise will overwhelm the effective gradient signal, leading to a decrease in accuracy.

\subsubsection{Robustness}
To evaluate the robustness of the SI-ChainFL model to malicious attacks, we first tested its convergence under two typical attack scenarios. Then, we compared the accuracy and resistance to different numbers of malicious nodes under these two attacks. Next, we tested the impact of the validation dataset size on the model's accuracy.

\begin{figure}[t]
\centering

\begin{subfigure}{0.49\columnwidth}
  \centering\includegraphics[width=\linewidth]{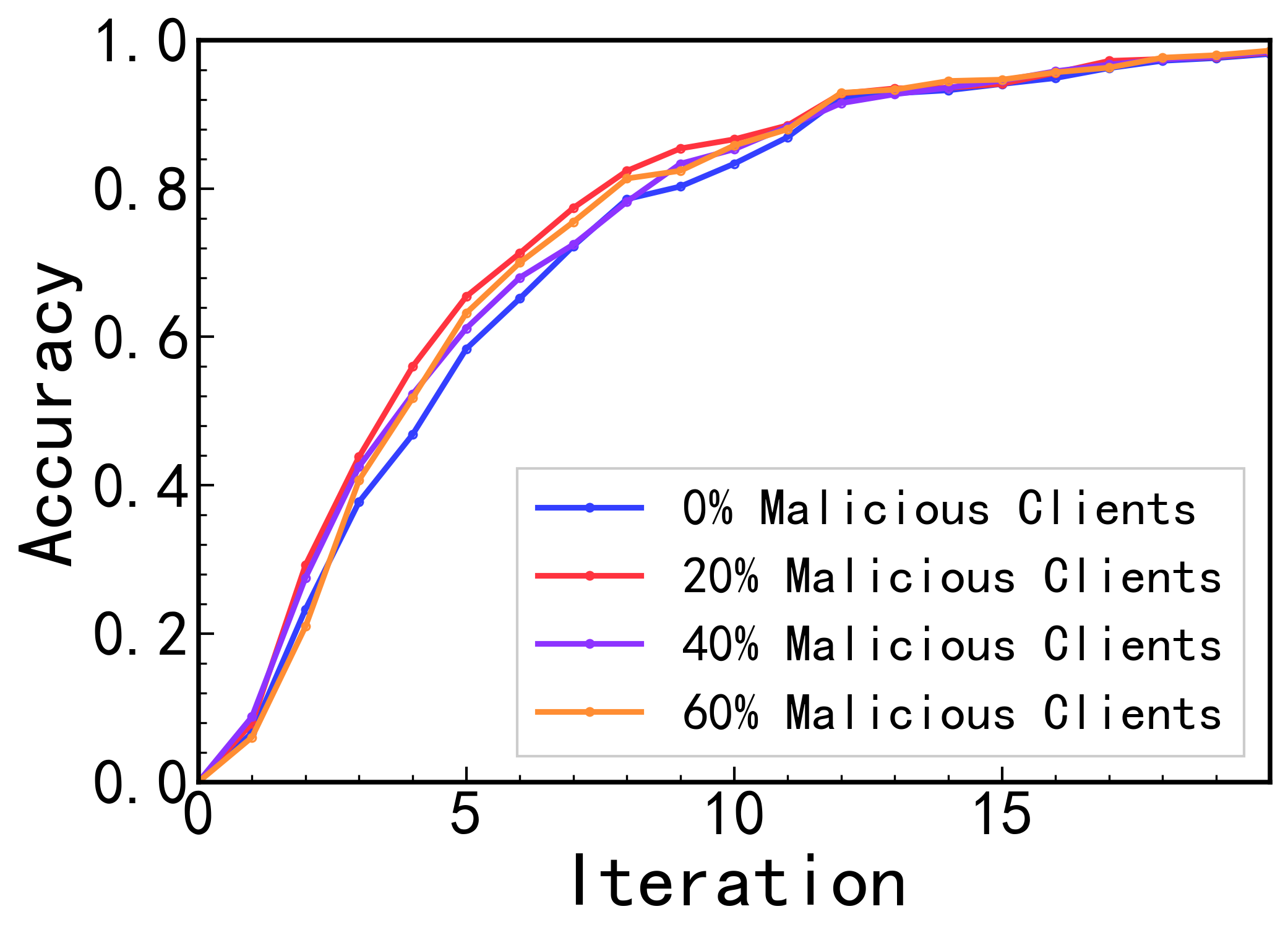}
  \caption{MNIST}
\end{subfigure}\hfill
\begin{subfigure}{0.49\columnwidth}
  \centering\includegraphics[width=\linewidth]{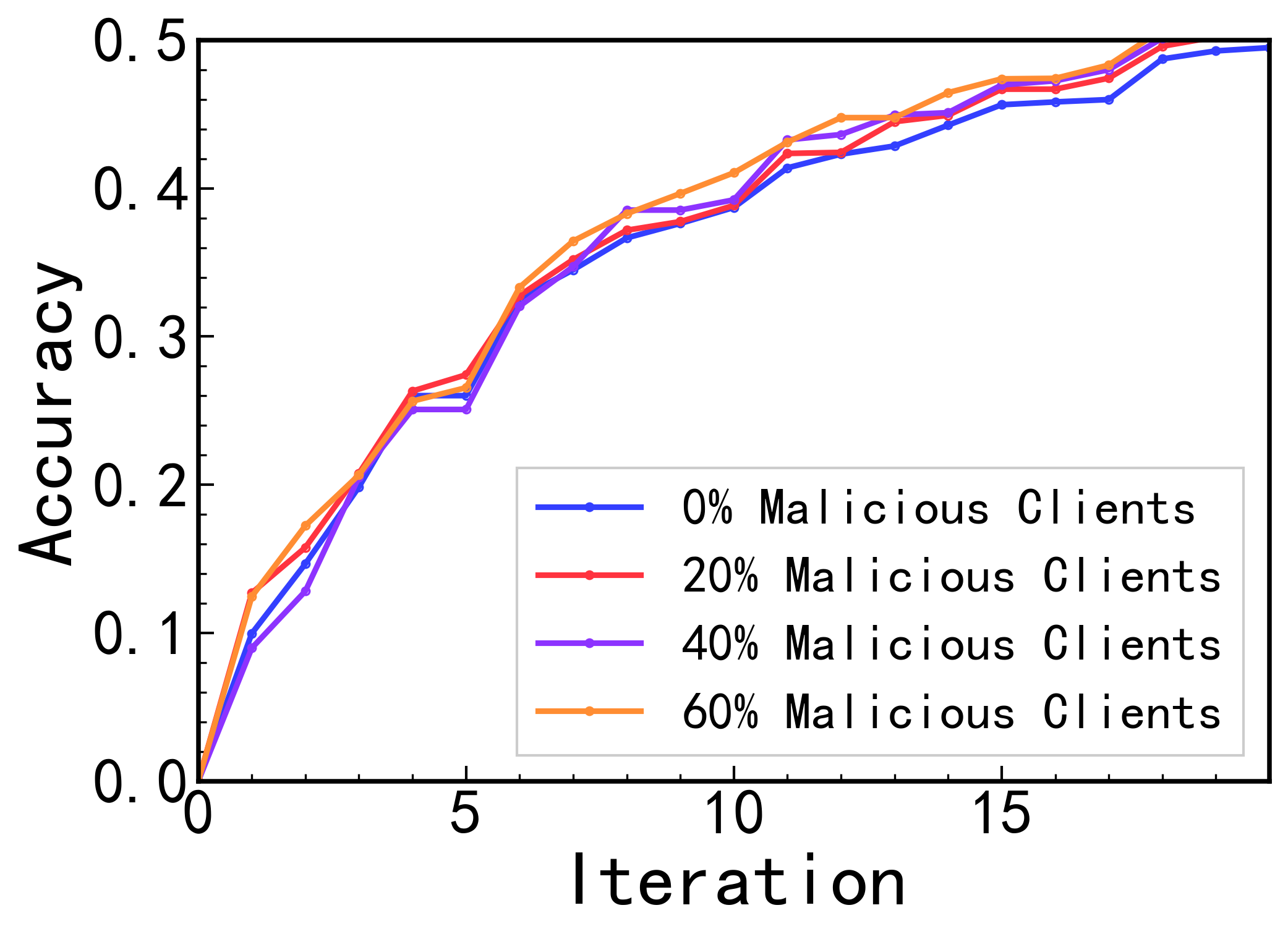}
  \caption{CIFAR-10}
\end{subfigure}

\vspace{2mm}

\begin{subfigure}{0.49\columnwidth}
  \centering\includegraphics[width=\linewidth]{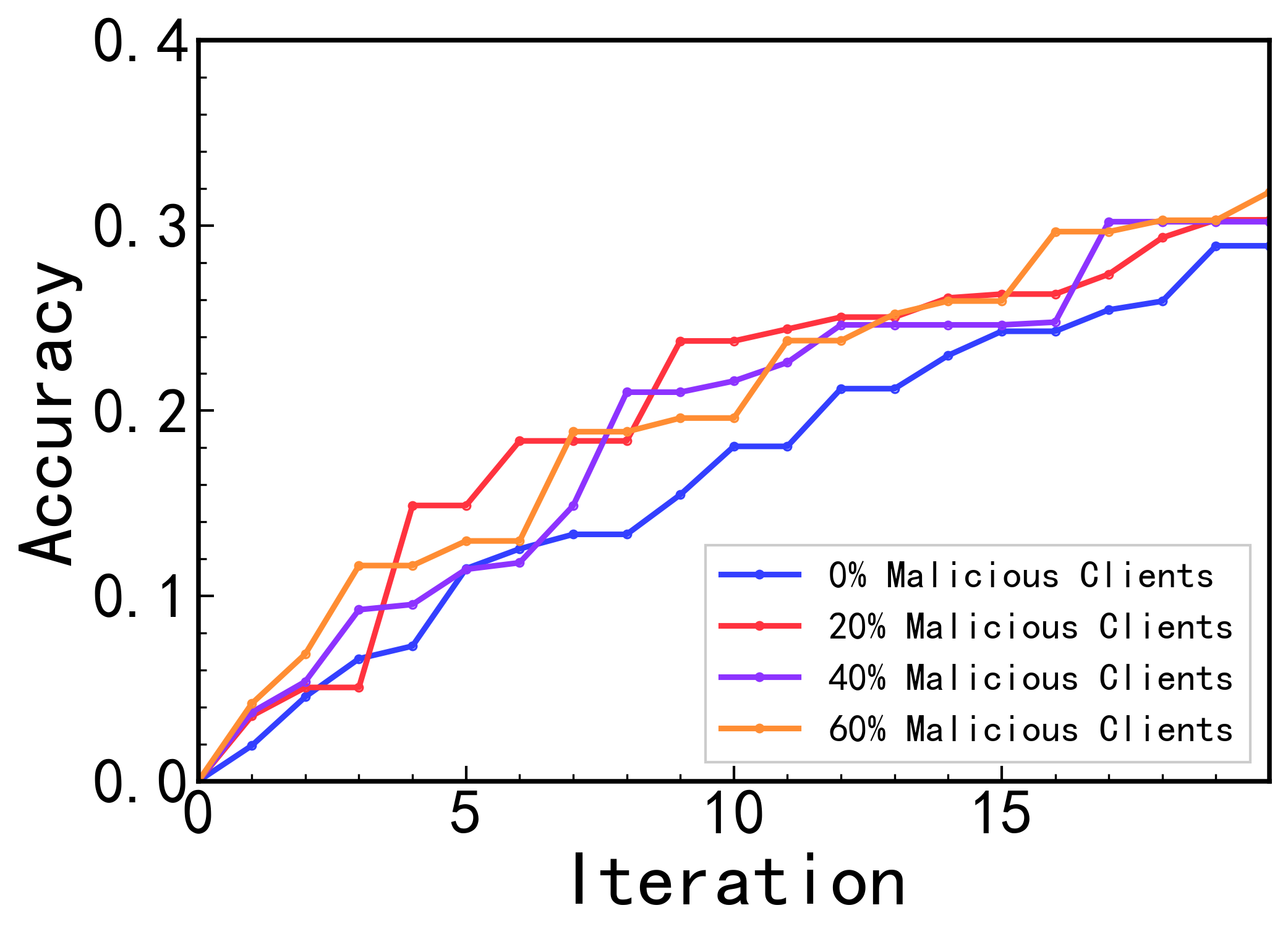}
  \caption{CIFAR-100}
\end{subfigure}\hfill
\begin{subfigure}{0.49\columnwidth}
  \centering\includegraphics[width=\linewidth]{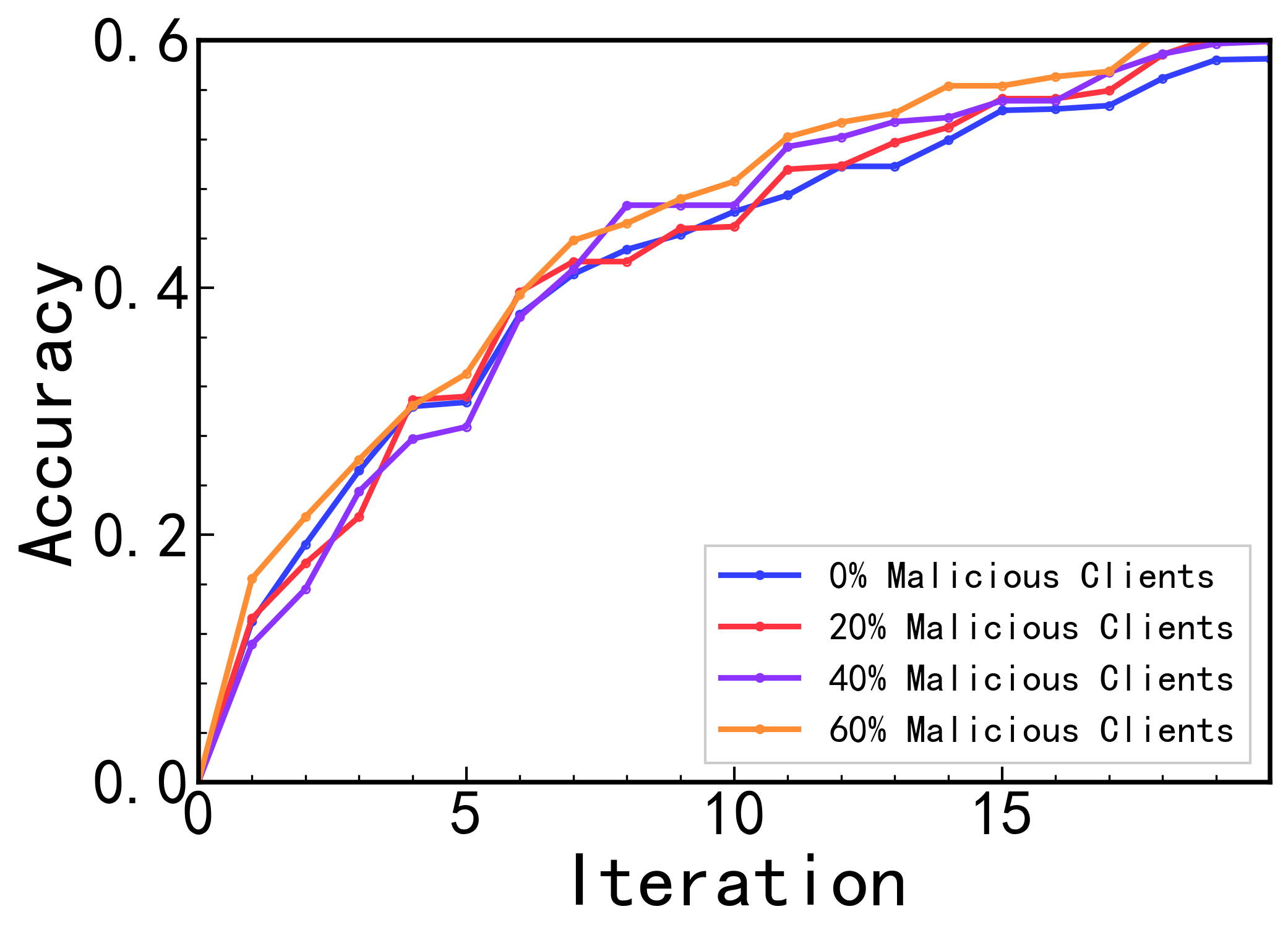}
  \caption{HSR}
\end{subfigure}

\caption{Comparison of model convergence of SI-ChainFL under different numbers of FR attacks.}
\label{fig:acc_M_FR}
\end{figure}

\begin{figure}[t]
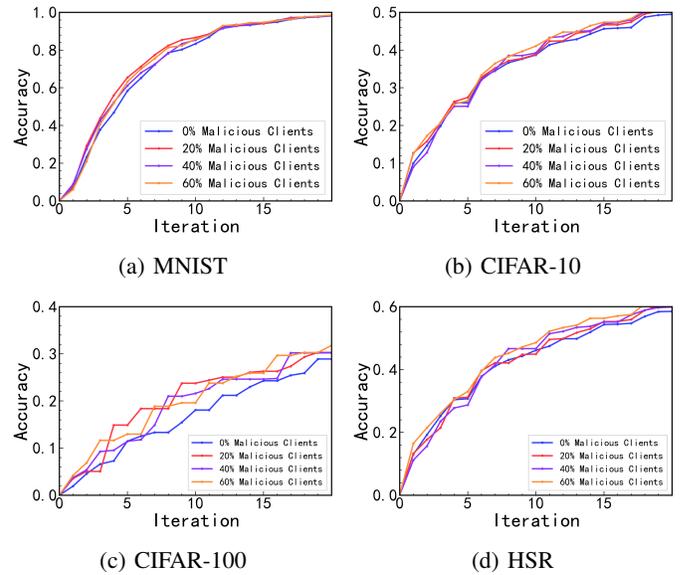

\centering

\begin{subfigure}{0.49\columnwidth}
  \centering\includegraphics[width=\linewidth]{acc_M_FR_01.png}
  \caption{MNIST}
\end{subfigure}\hfill
\begin{subfigure}{0.49\columnwidth}
  \centering\includegraphics[width=\linewidth]{acc_M_FR_02.png}
  \caption{CIFAR-10}
\end{subfigure}

\vspace{2mm}

\begin{subfigure}{0.49\columnwidth}
  \centering\includegraphics[width=\linewidth]{acc_M_FR_03.png}
  \caption{CIFAR-100}
\end{subfigure}\hfill
\begin{subfigure}{0.49\columnwidth}
  \centering\includegraphics[width=\linewidth]{acc_M_FR_04.png}
  \caption{HSR}
\end{subfigure}

\caption{Comparison of model convergence of SI-ChainFL under different numbers of PA attacks.}
\label{fig:acc_M_PA}
\end{figure}

\textit{Convergence Comparison under Malicious Attacks:} To test the performance of the SI-ChainFL model under malicious attacks, we tested the convergence of the global model under free-rider (FR) and poisoning (PA) attacks. Fig \ref{fig:acc_M_FR} and fig \ref{fig:acc_M_PA} illustrate the convergence of the SI-ChainFL model under different numbers of FR and PA attacks, respectively. Fig \ref{fig:acc_M_FR} shows the convergence of the global model's accuracy under FR attacks initiated by 0\%, 20\%, 40\%, and 60\% of clients on four datasets: MNIST, CIFAR-10, CIFAR-100, and the High-Speed Rail Network dataset. Fig \ref{fig:acc_M_PA} shows the convergence of the model under PA attacks initiated by clients. As shown in fig \ref{fig:acc_M_FR}, the impact of FR attacks on model convergence is relatively small as the number of clients launching them increases, especially on the MNIST dataset, where its impact on SI-ChainFL model convergence is almost zero. Fig \ref{fig:acc_M_PA}(d) shows that PA attacks have almost zero impact on the SI-ChainFL model on the high-speed rail network dataset. Although some minor fluctuations may occur on other datasets, the original accuracy of the SI-ChainFL model does not decrease significantly. This result demonstrates the robustness of the SI-ChainFL model in the presence of malicious attacks, further confirming its usability in real-world scenarios where clients may be unreliable or dishonest.

\textit{Accuracy Comparison Under Malicious Attacks:}To evaluate the performance of SI-ChainFL on classification and regression tasks in adversarial environments, we compared the global model accuracy of SI-ChainFL with representative baseline methods in the presence of attacks. In this experiment, we considered free-rider attacks (FR) and poisoning attacks (PA), and reported the global accuracy achieved by different methods on four datasets (MNIST, CIFAR-10, CIFAR-100, and HSR) and two datasets (PeMS07 and HSR). Detailed results are summarized in Table \ref{tab:acc_attacks} and Table \ref{tab:mae_attacks}.

\begin{table*}[t]
\centering
\caption{A Comparison of Global Model Accuracy With other Methods Under Two Types of Attacks}
\label{tab:acc_attacks}
\setlength{\tabcolsep}{4pt}
\renewcommand{\arraystretch}{1.15}
\resizebox{\textwidth}{!}{%
\begin{tabular}{|c|c|c|c|c|c|c|c|c|c|c|}
\hline
\multirow{2}{*}{Dataset} &
\multirow{2}{*}{Pct. of Malicious Clients} &
\multirow{2}{*}{Attack Type} &
\multicolumn{8}{c|}{Accuracy (\%)} \\ \cline{4-11}
& & &
FedAvg & FedProx & FedEdge & FedSage & P2FMS & FLTrust & RAGA & SI-ChainFL \\
\hline

\multirow{6}{*}{MNIST}
& \multirow{2}{*}{10\%}
& FR & 90.44 & 86.70 & 93.16 & 91.80 & \textbf{94.89} & 91.61 & 90.78 & 94.10 \\
&  & PA & 71.28 & 83.65 & 88.28 & \textbf{93.33} & 92.75 & 92.42 & 89.54 & 91.56 \\
\cline{2-11}
& \multirow{2}{*}{50\%}
& FR & 45.63 & 78.12 & 93.08 & 80.88 & 52.74 & 87.12 & 86.48 & \textbf{93.26} \\
&  & PA & 5.16 & 32.46 & \textbf{92.43} & 27.32 & 54.76 & 82.38 & 83.45 & 88.08 \\
\cline{2-11}
& \multirow{2}{*}{90\%}
& FR & 12.13 & 15.29 & 89.23 & 14.52 & 24.25 & 78.88 & 64.79 & \textbf{89.72} \\
&  & PA & 3.35 & 18.69 & 12.33 & 12.04 & 13.43 & 81.33 & 65.81 & \textbf{90.20} \\
\hline

\multirow{6}{*}{CIFAR-10}
& \multirow{2}{*}{10\%}
& FR & 52.56 & 61.33 & 69.48 & 67.28 & \textbf{71.25} & 66.87 & 64.45 & 70.08 \\
&  & PA & 63.83 & 60.22 & 61.56 & \textbf{70.09} & 69.58 & 67.91 & 62.25 & 66.78 \\
\cline{2-11}
& \multirow{2}{*}{50\%}
& FR  & 30.58 & 55.18 & 69.68 & 57.37 & 34.24 & 62.39 & 62.25 & \textbf{70.88} \\
&  & PA & 2.19 & 18.41 & \textbf{66.48} & 15.23 & 24.35 & 58.79 & 61.28 & 63.28 \\
\cline{2-11}
& \multirow{2}{*}{90\%}
& FR & 1.27 & 9.89 & 64.37 & 7.13 & 5.24 & 55.47 & 51.68 & \textbf{65.68} \\
&  & PA & 6.84 & 8.98 & 6.24 & 6.59 & 4.57 & 56.13 & 49.49 & \textbf{65.26} \\
\hline

\multirow{6}{*}{CIFAR-100}
& \multirow{2}{*}{10\%}
& FR & 32.13 & 38.64 & 44.29 & 43.14 & 44.56 & 42.77 & 42.56 & \textbf{45.06} \\
&  & PA & 39.42 & 37.16 & 38.34 & \textbf{45.05} & 43.65 & 43.66 & 42.79 & 44.26 \\
\cline{2-11}
& \multirow{2}{*}{50\%}
& FR & 24.76 & 34.84 & \textbf{44.66} & 35.65 & 32.25 & 39.26 & 42.55 & 43.83 \\
&  & PA & 1.86 & 13.99 & 41.22 & 11.68 & 25.56 & 36.36 & 38.56 & \textbf{42.19} \\
\cline{2-11}
& \multirow{2}{*}{90\%}
& FR & 1.18 & 7.67 & 40.89 & 5.48 & 15.56 & 34.71 & 32.45 & \textbf{41.56} \\
&  & PA & 4.96 & 6.35 & 4.66 & 4.85 & 12.65 & 33.97 & 28.58 & \textbf{42.68} \\
\hline

\multirow{6}{*}{HSR}
& \multirow{2}{*}{10\%}
& FR & 69.09 & 82.20 & 92.21 & 91.04 & \textbf{94.85} & 90.89 & 91.49 & 93.33 \\
&  & PA & 89.29 & 85.49 & 87.14 & \textbf{92.70} & 92.59 & 91.72 & 90.49 & 90.89 \\
\cline{2-11}
& \multirow{2}{*}{50\%}
& FR & 43.00 & 76.32 & 92.18 & 79.21 & 53.25 & 85.85 & 82.62 & \textbf{92.52} \\
&  & PA & 3.93  & 29.77 & \textbf{91.69} & 24.14 & 14.56 & 80.65 & 76.69 & 86.61 \\
\cline{2-11}
& \multirow{2}{*}{90\%}
& FR & 10.64 & 15.39 & 87.91 & 11.18 & 15.56 & 77.10 & 65.83 & \textbf{88.63} \\
&  & PA & 2.00 & 13.65 & 9.93 & 10.43 & 12.56 & 79.99 & 75.26 & \textbf{89.38} \\
\hline

\end{tabular}}
\end{table*}

As shown in tab. \ref{tab:acc_attacks}, SI-ChainFL maintains high accuracy in both attack scenarios, and its performance does not drop sharply with the increase in the number of malicious clients. In contrast, most baseline models struggle to maintain satisfactory accuracy as the proportion of malicious clients increases. Specifically, when the proportion of malicious clients reaches 50\%, the performance of FedAvg, FedProx, and FedSage drops significantly, indicating that their security decreases drastically with the increase in the number of attackers. FLTrust also experiences some minor drops in accuracy. In contrast, while FedEdge's accuracy also decreases under FR attacks, it still achieves a level comparable to SI-ChainFL; however, its performance drops rapidly under heavy PA attacks. As shown in Table \ref{tab:mae_attacks}, SI-ChainFL maintains relatively stable performance in traffic flow prediction regression tasks even when attacked. In conclusion, these results demonstrate that SI-ChainFL is highly robust to adversarial behavior and effectively mitigates the negative impact of malicious clients on the global model. Therefore, we conclude that SI-ChainFL achieves its intended goal of enhancing system security.

\begin{table*}[t]
\centering
\caption{A Comparison of Global Model MAE With other Methods Under Two Types of Attacks}
\label{tab:mae_attacks}
\setlength{\tabcolsep}{4pt}
\renewcommand{\arraystretch}{1.15}
\resizebox{\textwidth}{!}{%
\begin{tabular}{|c|c|c|c|c|c|c|c|c|c|c|}
\hline
\multirow{2}{*}{Dataset} &
\multirow{2}{*}{Pct. of Malicious Clients} &
\multirow{2}{*}{Attack Type} &
\multicolumn{8}{c|}{MAE} \\ \cline{4-11}
& & &
FedAvg & FedProx & FedEdge & FedSage & P2FMS & FLTrust & RAGA & SI-ChainFL \\
\hline

\multirow{6}{*}{PeMS07}
& \multirow{2}{*}{10\%}
& FR & 0.3187 & 0.2274 & 0.1046 & 0.1083 & \textbf{0.1015} & 0.1129 & 0.1017 & 0.1068 \\
&  & PA & 0.1052 & 0.2061 & 0.2318 & \textbf{0.1027} & 0.1088 & 0.1194 & 0.1140 & 0.1176 \\
\cline{2-11}
& \multirow{2}{*}{50\%}
& FR & 0.6421 & 0.3096 & 0.1284 & 0.2079 & 0.5275 & 0.2157 & 0.2413 & \textbf{0.1019} \\
&  & PA & 0.9346  & 0.8067 & 0.1138 & 0.8340 & 0.9182 & 0.2291 & 0.3015 & \textbf{0.1059} \\
\cline{2-11}
& \multirow{2}{*}{90\%}
& FR & 0.9483 & 0.9074 & 0.2198 & 0.9365 & 0.9149 & 0.3246 & 0.4028 & \textbf{0.1056} \\
&  & PA & 0.9671 & 0.9313 & 0.9187 & 0.9042 & 0.9430 & 0.3379 & 0.3092 & \textbf{0.1014} \\
\hline

\multirow{6}{*}{HSR}
& \multirow{2}{*}{10\%}
& FR & 0.3315 & 0.2148 & 0.1061 & \textbf{0.1026} & 0.1079 & 0.1182 & 0.1153 & 0.1034 \\
&  & PA & 0.1117 & 0.2395 & 0.2054 & 0.1042 & \textbf{0.1011} & 0.1098 & 0.1169 & 0.1136 \\
\cline{2-11}
& \multirow{2}{*}{50\%}
& FR & 0.6812 & 0.3257 & \textbf{0.1018} & 0.2941 & 0.5438 & 0.2366 & 0.2072 & 0.1091 \\
&  & PA & 0.9724  & 0.8583 & 0.3087 & 0.8039 & 0.9075 & 0.2033 & 0.3427 & \textbf{0.1488} \\
\cline{2-11}
& \multirow{2}{*}{90\%}
& FR & 0.9468 & 0.9216 & 0.2571 & 0.9177 & 0.9024 & 0.3038 & 0.4915 & \textbf{0.1023} \\
&  & PA & 0.9639 & 0.9057 & 0.9491 & 0.9382 & 0.9228 & 0.3696 & 0.3157 & \textbf{0.1065} \\
\hline

\end{tabular}}
\end{table*}

\begin{figure}[t]
\centering
\begin{subfigure}{0.49\columnwidth}
  \centering
  \includegraphics[width=\linewidth]{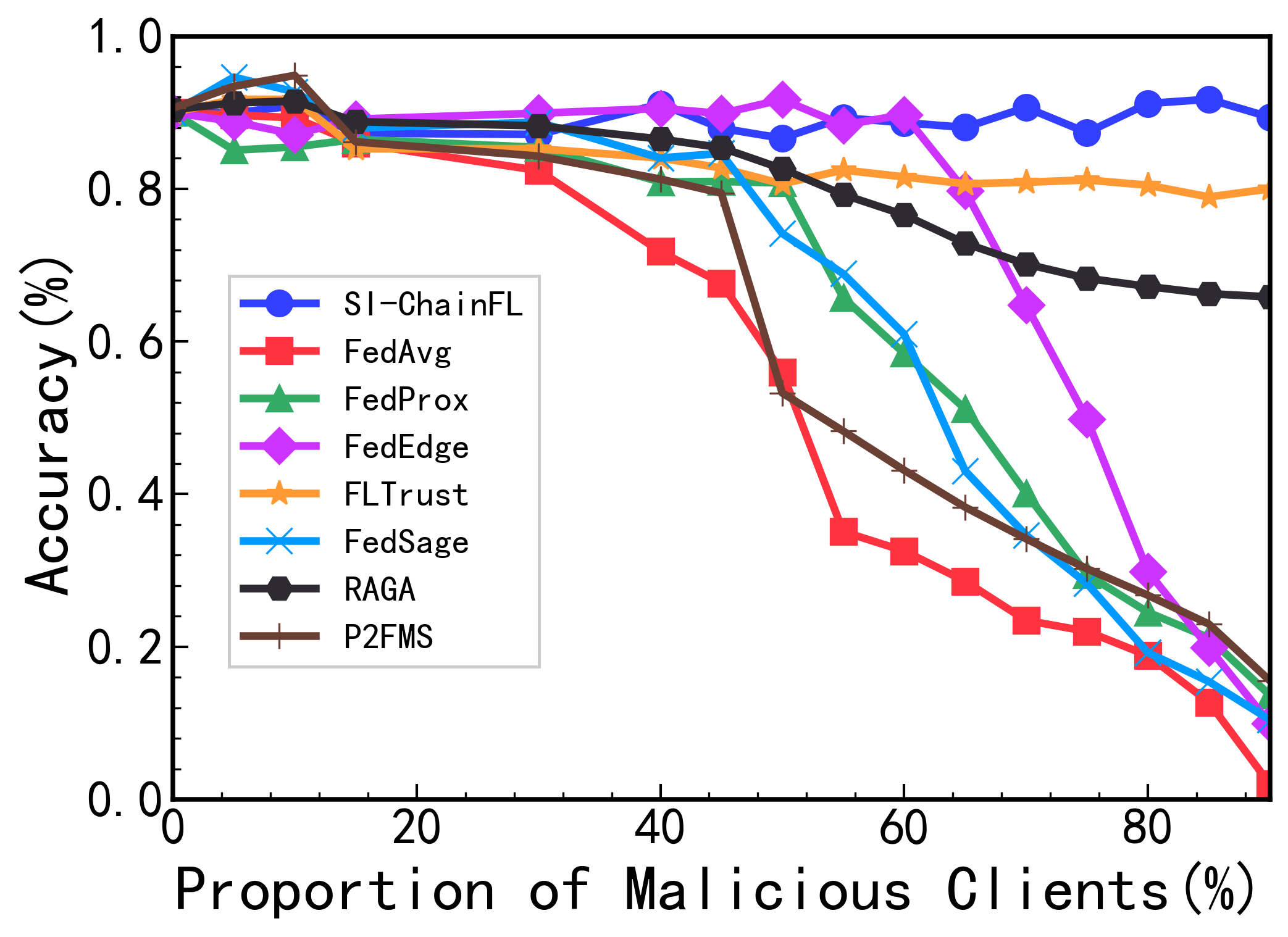}
  \caption{FR (Accuracy)}
\end{subfigure}\hfill
\begin{subfigure}{0.49\columnwidth}
  \centering
  \includegraphics[width=\linewidth]{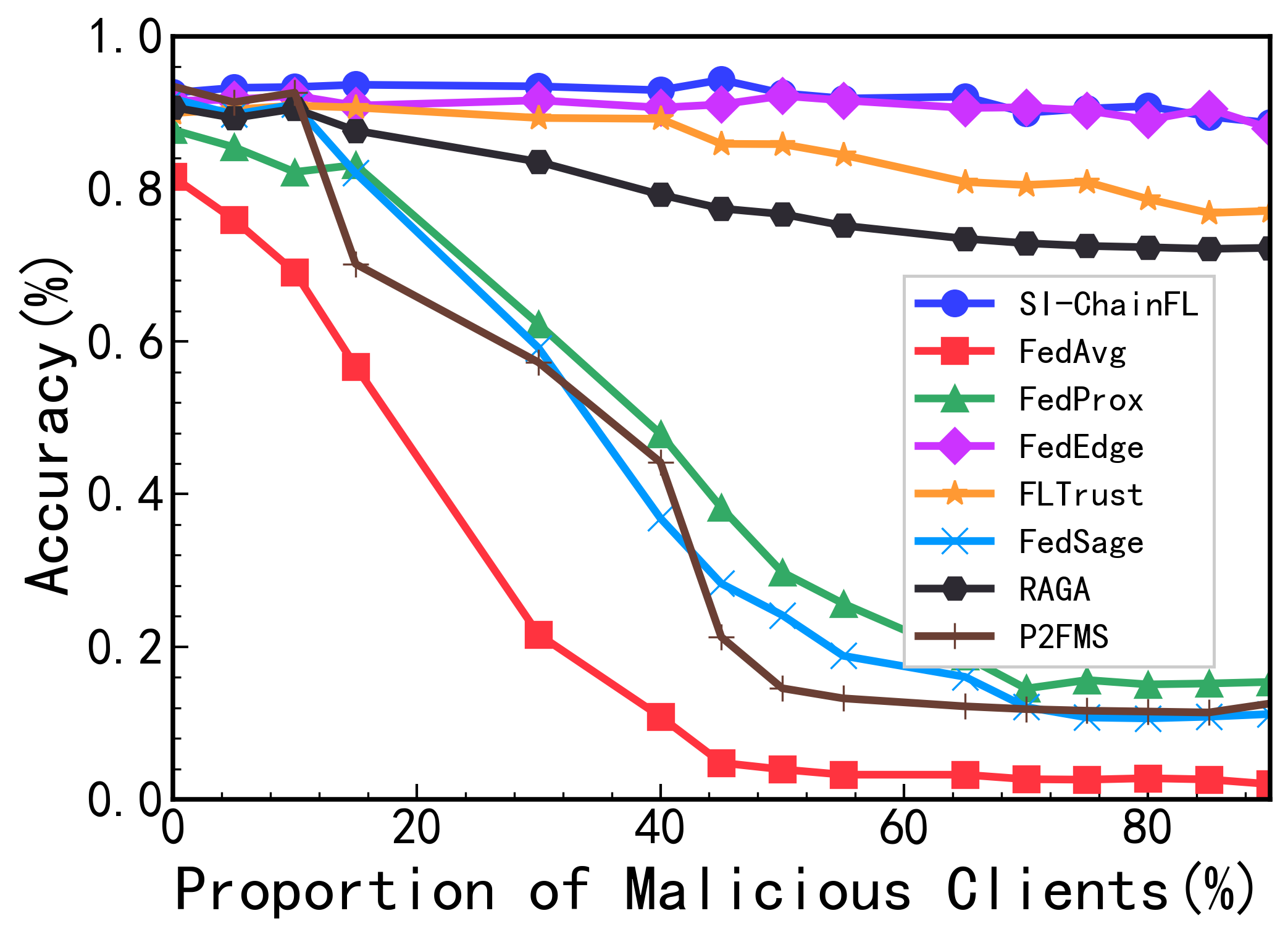}
  \caption{PA (Accuracy)}
\end{subfigure}

\caption{Impact of the number of malicious clients on model accuracy in the HSR dataset.}
\label{fig:acc_M_0_100}
\end{figure}

\begin{figure}[t]
\centering
\begin{subfigure}{0.49\columnwidth}
  \centering
  \includegraphics[width=\linewidth]{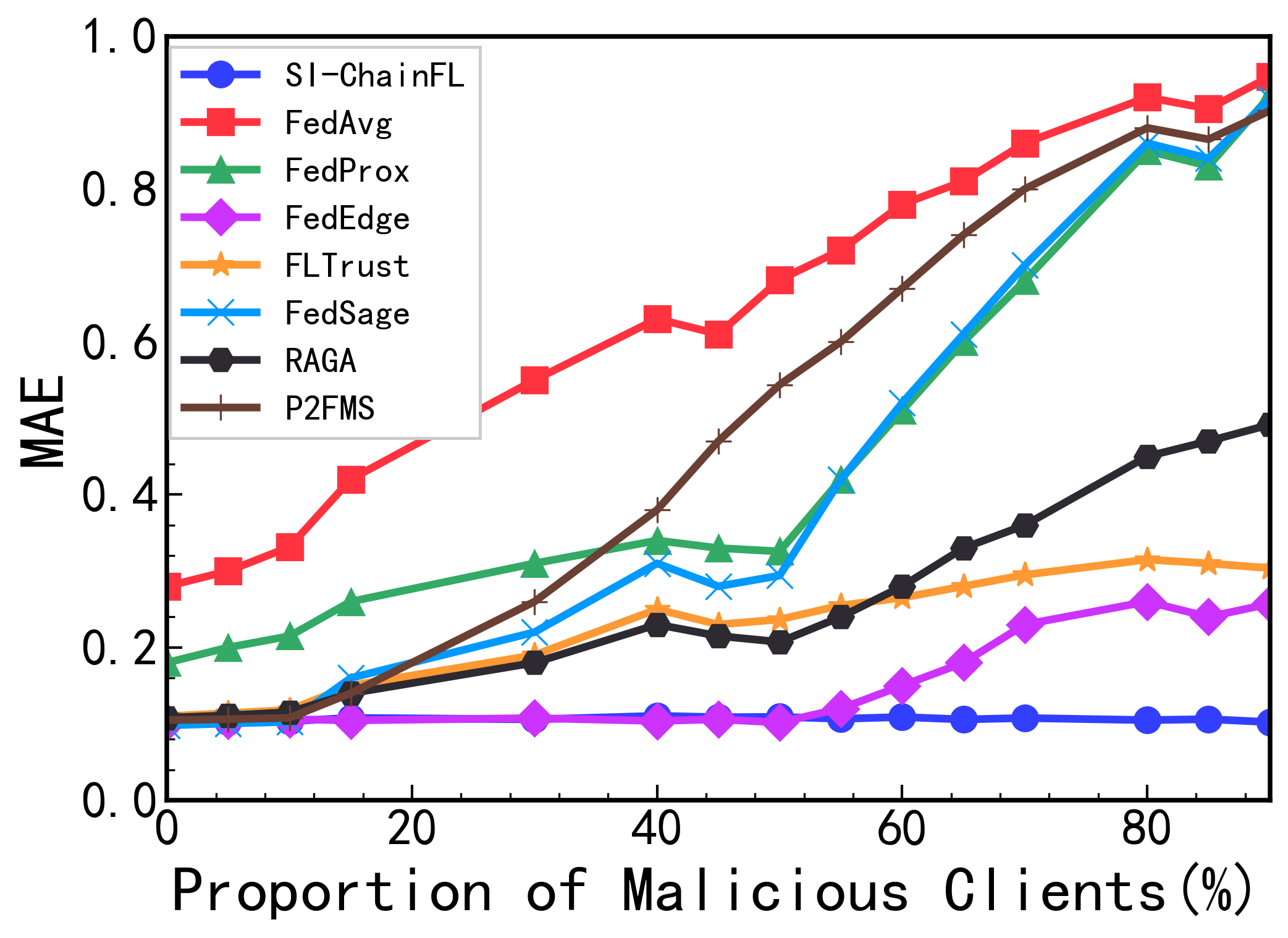}
  \caption{FR (MAE)}
\end{subfigure}\hfill
\begin{subfigure}{0.49\columnwidth}
  \centering
  \includegraphics[width=\linewidth]{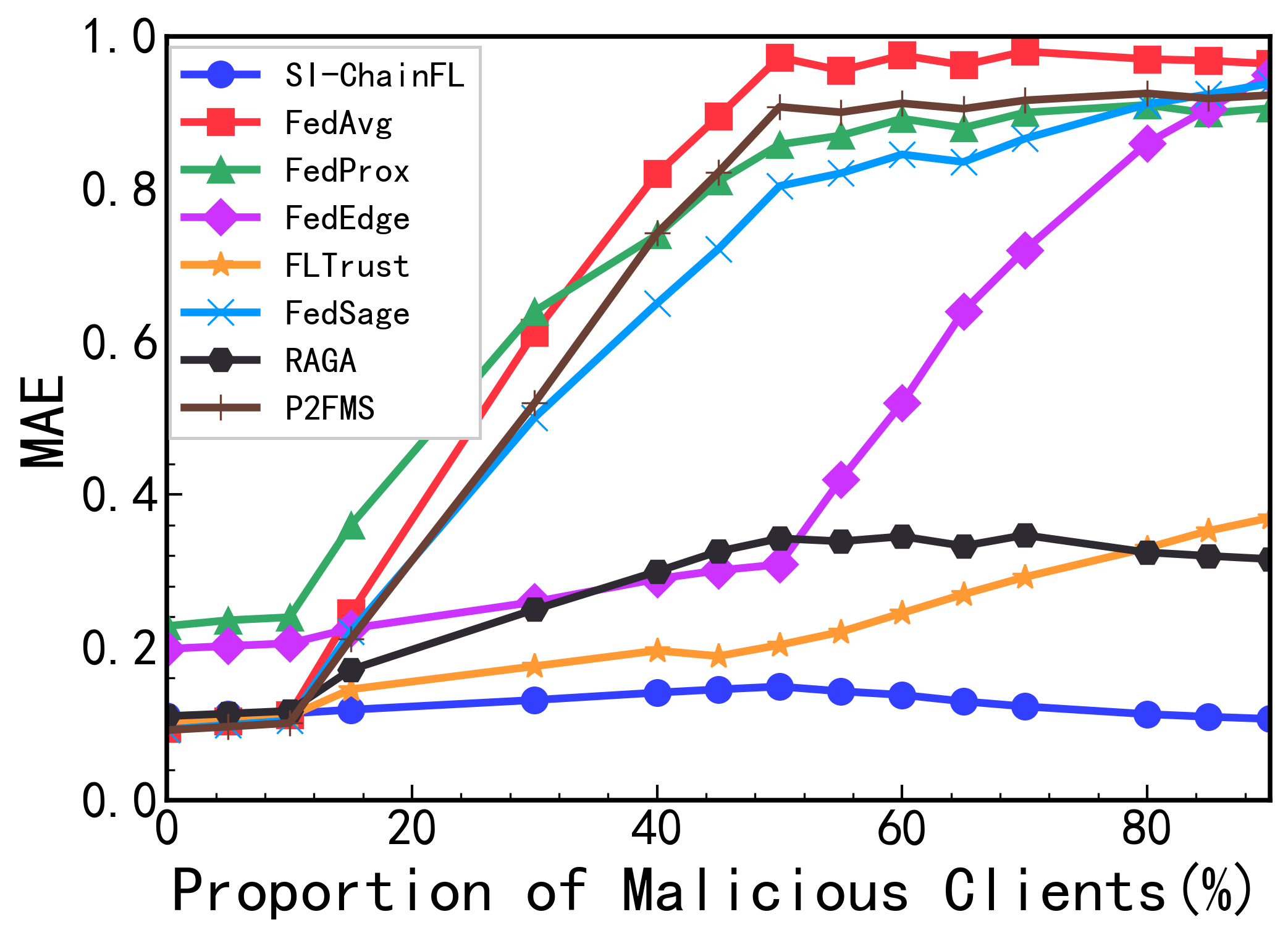}
  \caption{PA (MAE)}
\end{subfigure}

\caption{Impact of the number of malicious clients on model MAE in the HSR dataset.}
\label{fig:mae_M_0_100}
\end{figure}

\textit{Comparison of the impact of different methods on the number of malicious clients:} To further test the robustness of the SI-ChainFL model against different numbers of malicious clients, we analyzed the performance of the SI-ChainFL model and other federated learning methods. Specifically, as the proportion of malicious clients increased from 0\% to 90\%, we tested the accuracy and mean absolute error (MAE) of the SI-ChainFL model and other methods in predicting sudden surges in passenger flow on the high-speed rail dataset under FR and PA attacks. As can be seen from Figure \ref{fig:acc_M_0_100}, the accuracy of the SI-ChainFL model did not significantly decrease or increase with the increase in the proportion of malicious clients, demonstrating high robustness under both types of attacks. In contrast, the global model accuracy of FedAvg, FedProx, FedSage, and P2FMS all decreased significantly with the increase in the proportion of malicious clients. FedEdge maintained high accuracy under FR attacks, but its accuracy dropped sharply when subjected to PA attacks with a malicious client proportion reaching 60\%. While FLTrust's accuracy didn't significantly decrease under FR and PA attacks, it remained below normal levels. RAGA maintained good performance when the proportion of malicious clients reached 10\% and 50\%, but its performance dropped significantly when the proportion increased to 90\%. As shown in the figure \ref{fig:mae_M_0_100}, meanwhile, the MAE of the SI-ChainFL model remained consistently low, while FedAvg, FedProx, FedSage, and P2FMS experienced sharp increases in MAE. RAGA and FLTrust saw slight increases. FedEdge maintained a low MAE under FR attacks, but its MAE increased sharply when the proportion of clients launching PA attacks exceeded 50\%. Even when the proportion of malicious clients launching both attacks reached 90\%, the SI-ChainFL model still maintained good performance. Therefore, we can conclude that the SI-ChainFL model is highly robust to FR and PA attacks.

\begin{figure}[t]
\centering

\begin{subfigure}{0.49\columnwidth}
  \centering\includegraphics[width=\linewidth]{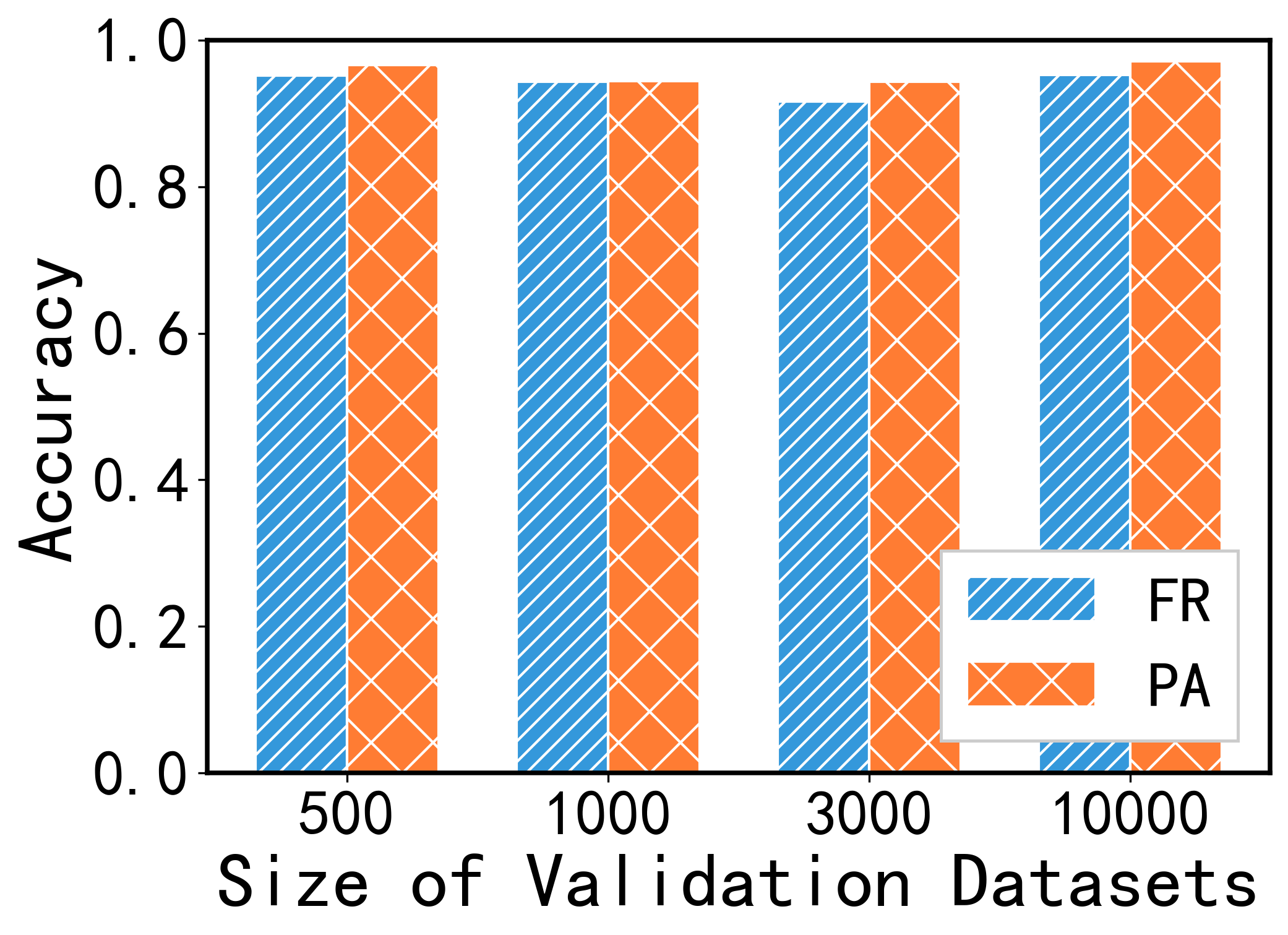}
  \caption{MNIST}
\end{subfigure}\hfill
\begin{subfigure}{0.49\columnwidth}
  \centering\includegraphics[width=\linewidth]{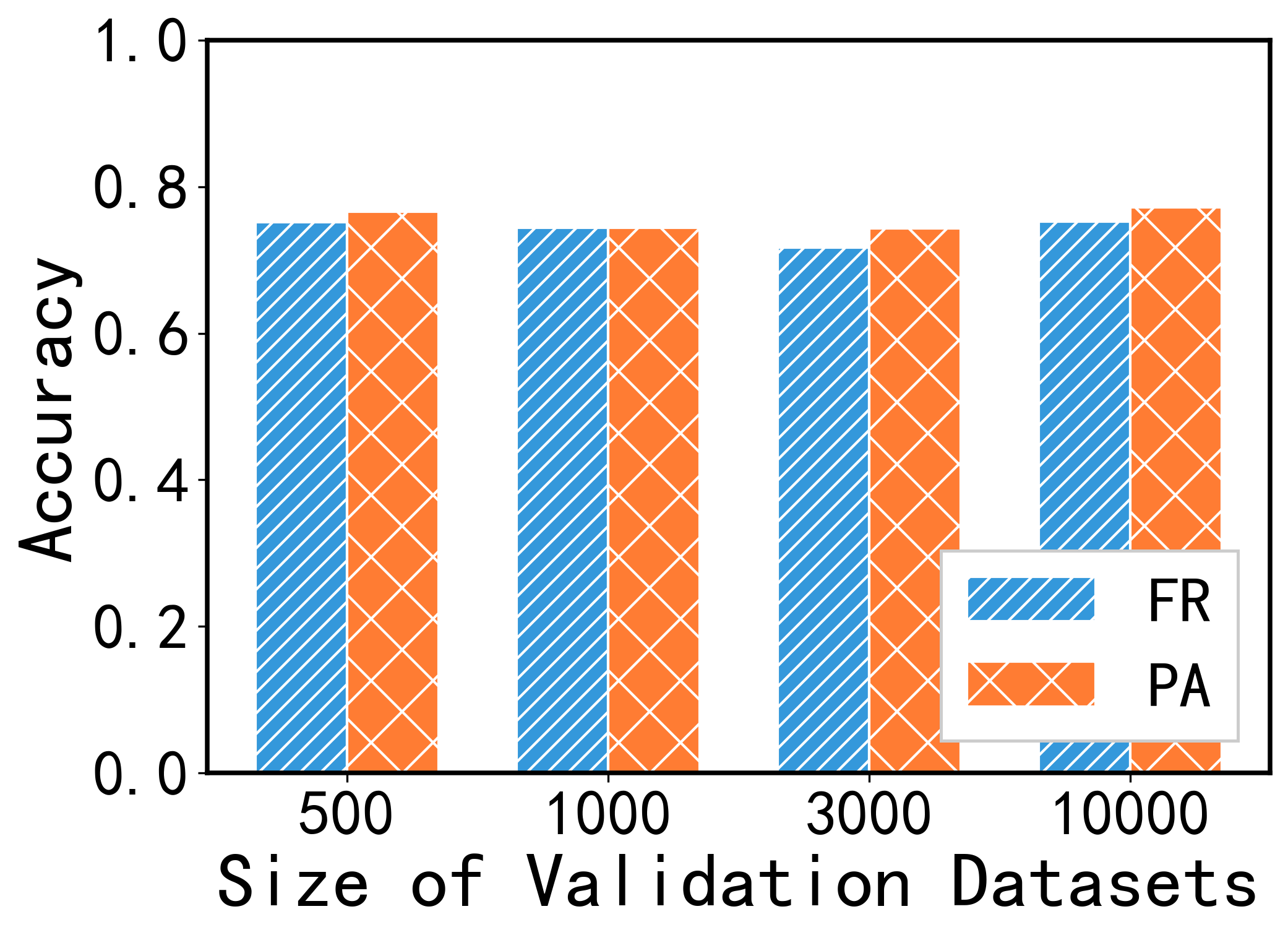}
  \caption{CIFAR-10}
\end{subfigure}

\vspace{2mm}

\begin{subfigure}{0.49\columnwidth}
  \centering\includegraphics[width=\linewidth]{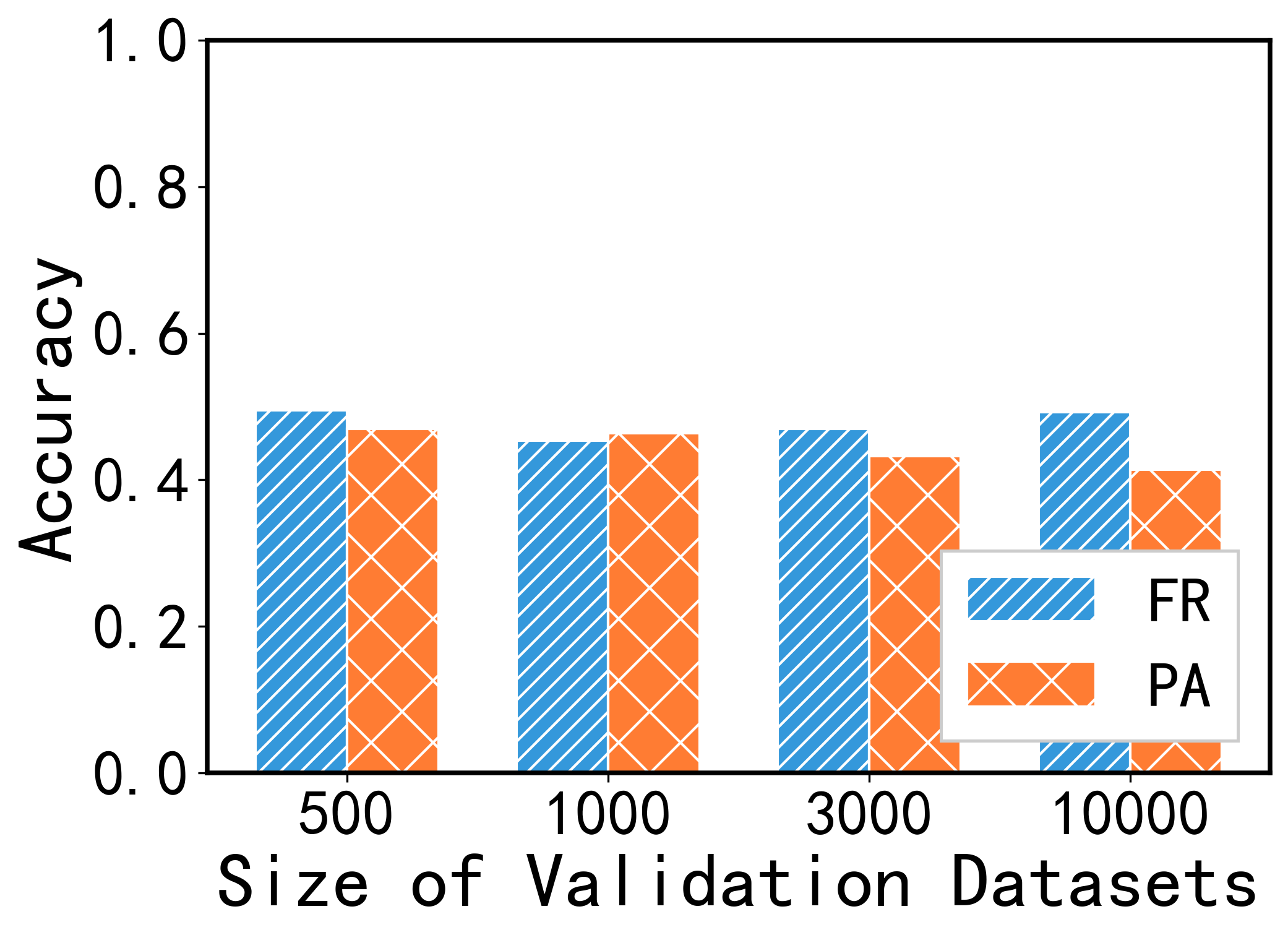}
  \caption{CIFAR-100}
\end{subfigure}\hfill
\begin{subfigure}{0.49\columnwidth}
  \centering\includegraphics[width=\linewidth]{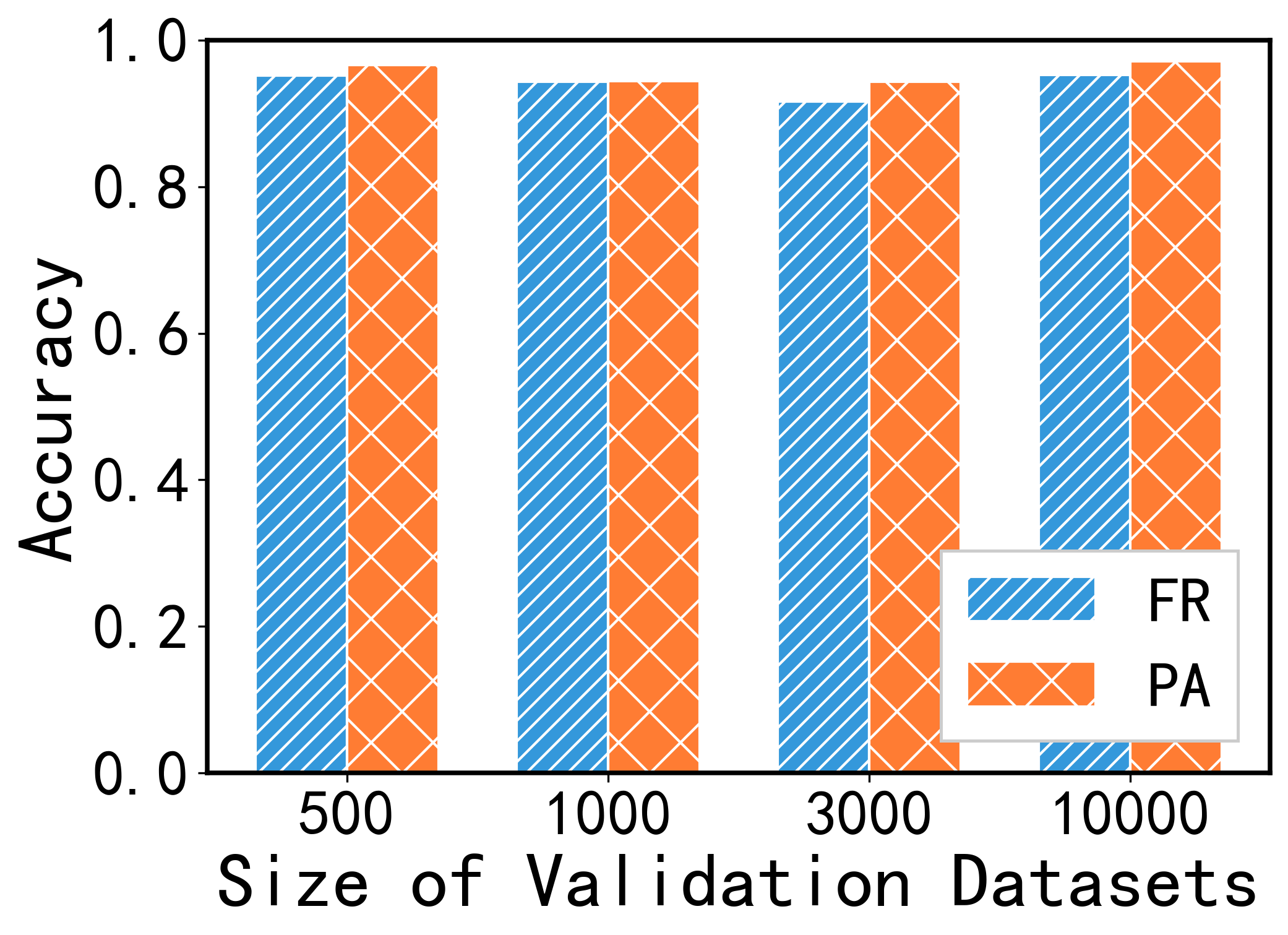}
  \caption{HSR}
\end{subfigure}

\caption{Impact of evaluation datasets size on the model accuracy.}
\label{fig:acc_datasize}
\end{figure}

\textit{Impact of validation dataset size on model accuracy:} To evaluate the impact of validation dataset size on model performance, we tested the accuracy of the SI-ChainFL model on validation sets of different sizes: 500, 1000, 3000, and 10000. As shown in fig. \ref{fig:acc_datasize}, the size of the validation dataset has a relatively small impact on model performance, especially on the MNIST and high-speed rail network datasets. The model accuracy hardly changes with the size of the validation dataset, indicating that the performance of the SI-ChainFL model is not dependent on the size of the validation dataset and exhibits a certain degree of stability. In other words, to protect the privacy of the data used for verification without compromising security, we will use a small scale validation set for model training.

\subsubsection{Efficiency}
To evaluate the efficiency of the SI-ChainFL model, we tested its training time and compared it with other similar methods, finally analyzing its time complexity.

\begin{figure}[h!]
    \centering
    \hspace*{-0.6cm} 
    \includegraphics[width=0.5\textwidth]{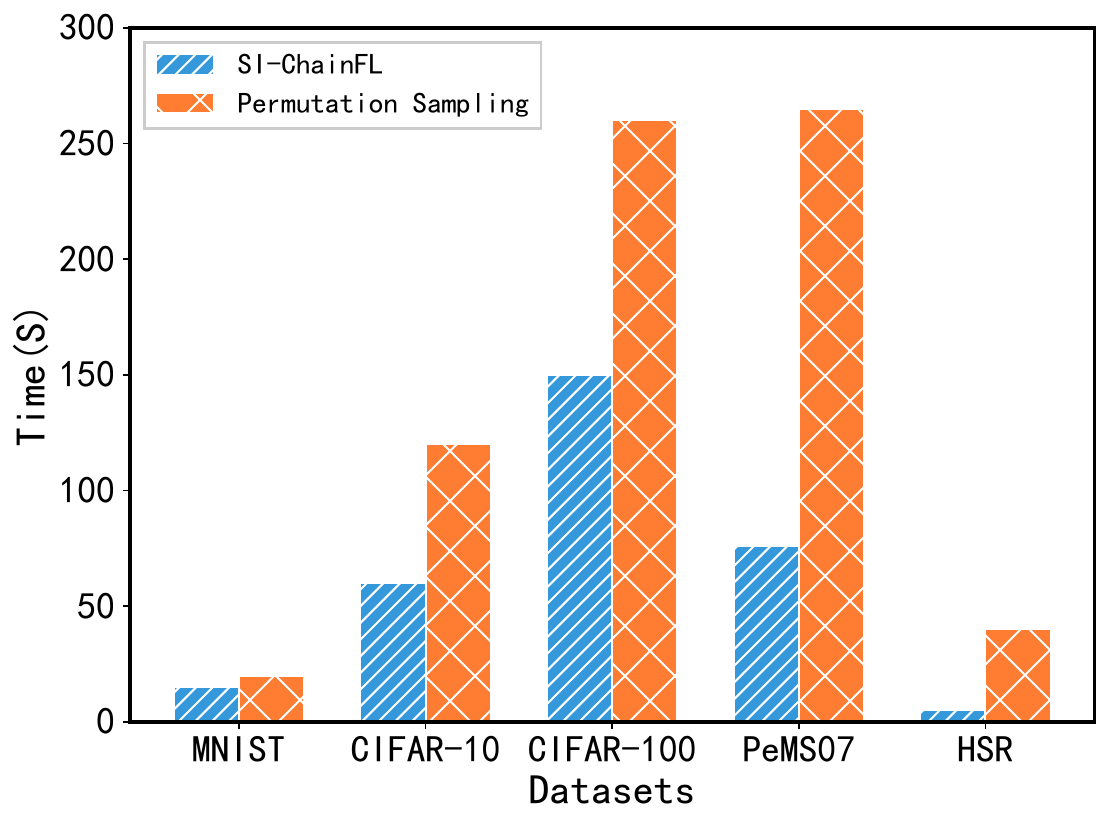}
    \caption{Comparison of Shapley value calculation time.}
    \label{fig:time_shapley}
\end{figure}

\textit{Comparison of Shapley value calculation time:} To test the training efficiency of the SI-ChainFL model, we tested its Shapley value computation time and compared it with the random sampling Shapley value computation method. Figure \ref{fig:time_shapley} shows that the SI-ChainFL model effectively reduces the Shapley value computation time. On the high-speed rail dataset, its computation time is lower than that of the random sampling method. On the CIFAR-10 and CIFAR-100 datasets, the computation time of the SI-ChainFL model is approximately half that of the random sampling method. Specifically, on the high-speed rail dataset, its computation time is only one-eighth that of the random sampling method, significantly reducing training time. The improved computational efficiency is due to the fact that the SI-ChainFL model only computes Shapley values for specific sparse data, thus significantly reducing the computational burden of Shapley value computation. Therefore, the SI-ChainFL model has lower computational overhead and higher efficiency.

\textit{Computational Complexity Analysis:} To further evaluate the efficiency of the SI-ChainFL model, we analyze its computational complexity and compare it with other solutions. SI-ChainFL reduces computational overhead from two aspects: the number of samples and the number of participating clients. First, it partitions the dataset according to scenarios based on fixed rare events, reducing the size of the validation dataset from $|D_{val}|$ to $R(M+H)$. Second, it retains only clients that contribute significantly to rare positive examples and merges the remaining clients, reducing the number of participating clients from $n_t$ to $K+1$. The time complexity is reduced from $O(K_{\pi} n_t C_v)$ for random permutation to $O\!\left(K_{\pi}(K+1)\,R(M+H)\,f\right)$. Therefore, the time complexity of our SI-ChainFL model's Shapley value calculation method is much smaller than that of small-computation methods such as random sampling. This means that SI-ChainFL requires only a small amount of computation and can meet the needs of scenarios with limited computing resources. In rare event scenarios where $K \ll n_t$ and $M+H \ll \lvert D_{\mathrm{val}} \rvert$, the computational cost of the SI-ChainFL model is significantly reduced.

\subsubsection{Ablation Experiments}
To evaluate the effectiveness of the two main components of the SI-ChainFL model: the contribution-aware Shapley value incentive method (SI) and the blockchain-based global model aggregation method (Chain-FL). we tested the model's accuracy and MAE under two different levels of attacks on the HSR dataset. The experimental results are shown in Table \ref{tab:acc_acp}. As can be seen from Table \ref{tab:acc_acp}, even under attacks, our model maintains good performance, indicating that the contribution-aware Shapley value incentive method effectively reduces the impact of malicious clients on the global model performance by quantifying the client's contribution across multiple dimensions. When the proportion of malicious nodes is 10\%, the model effectively handles FR and PA attacks even without using any method, or only using the SI method without further blockchain-based model aggregation. This is because the method only removes malicious clients by evaluating their effectiveness, thereby improving the CNN model; therefore, when the proportion of malicious clients is low, the performance difference between the base model and the SI-ChainFL model is not significant. The contribution-aware Shapley value incentive method has already screened out honest clients with high-quality data to participate in global model training, initially ensuring the model's security in the event of malicious attacks. The blockchain-based global model aggregation method uses the client Shapley values calculated in the previous step for voting, further ensuring the security and fairness of the client election for global model training when the proportion of malicious clients is high.

\begin{table}[t]
\centering
\caption{Accuracy (\%) and MAE under different number of malicious clients and attack types.}
\label{tab:acc_acp}
\setlength{\tabcolsep}{8pt}
\renewcommand{\arraystretch}{1.15}

\resizebox{0.95\columnwidth}{!}{%
\begin{tabular}{|c|c|ccc|ccc|}
\hline
\multirow{2}{*}{Pct. of Malicious Clients} &
\multirow{2}{*}{Attack Type} &
\multicolumn{3}{c|}{Accuracy (\%)} &
\multicolumn{3}{c|}{MAE} \\ \cline{3-8}
& & None & SI & SI-ChainFL & None & SI & SI-ChainFL \\
\hline
\multirow{2}{*}{10\%}
& FR & 65.83 & 84.49 & 93.33 & 0.3623 & 0.1341 & 0.1034 \\
& PA & 74.95 & 71.93 & 90.89 & 0.2834 & 0.1436 & 0.1136 \\ 
\hline
\multirow{2}{*}{50\%}
& FR & 39.03 & 64.95 & 92.52 & 0.5654 & 0.282
4& 0.1091 \\
& PA & 15.86 & 55.83 & 86.61 & 0.9832 & 0.2813 & 0.1488 \\
\hline
\multirow{2}{*}{90\%}
& FR & 10.50 & 60.29 & 88.63 & 0.8617 & 0.3824 & 0.1023 \\
& PA & 2.23 & 54.37 & 89.38 & 0.9854 & 0.3964 & 0.1065 \\
\hline
\end{tabular}%
}
\end{table}

\section{CONCLUSION}

In this paper, we propose SI-ChainFL, a secure and efficient federated learning framework for contribution aware data sharing. SI-ChainFL combines a multi objective Shapley value contribution metric with blockchain based secure aggregation, improving the usefulness of local updates while limiting free riding, poisoning attacks, and computation cost. The Shapley formulation jointly considers rare event prediction utility, data diversity, data quality, and timeliness, and clients are grouped by their influence on rare positive samples to further reduce computation. The resulting Shapley scores are used to elect validator nodes and to decide which clients join global aggregation and receive model updates. We provide theoretical security analysis and extensive experiments showing that SI-ChainFL effectively resists free riding and poisoning. Future work will apply graph federated learning to high speed rail mesh data and evaluate SI-ChainFL in more dynamic environments and broader intelligent transportation tasks.

\bibliographystyle{IEEEtran}  
\bibliography{main}            


\begin{IEEEbiography}[{\includegraphics[width=1in,height=1.25in,clip,keepaspectratio]{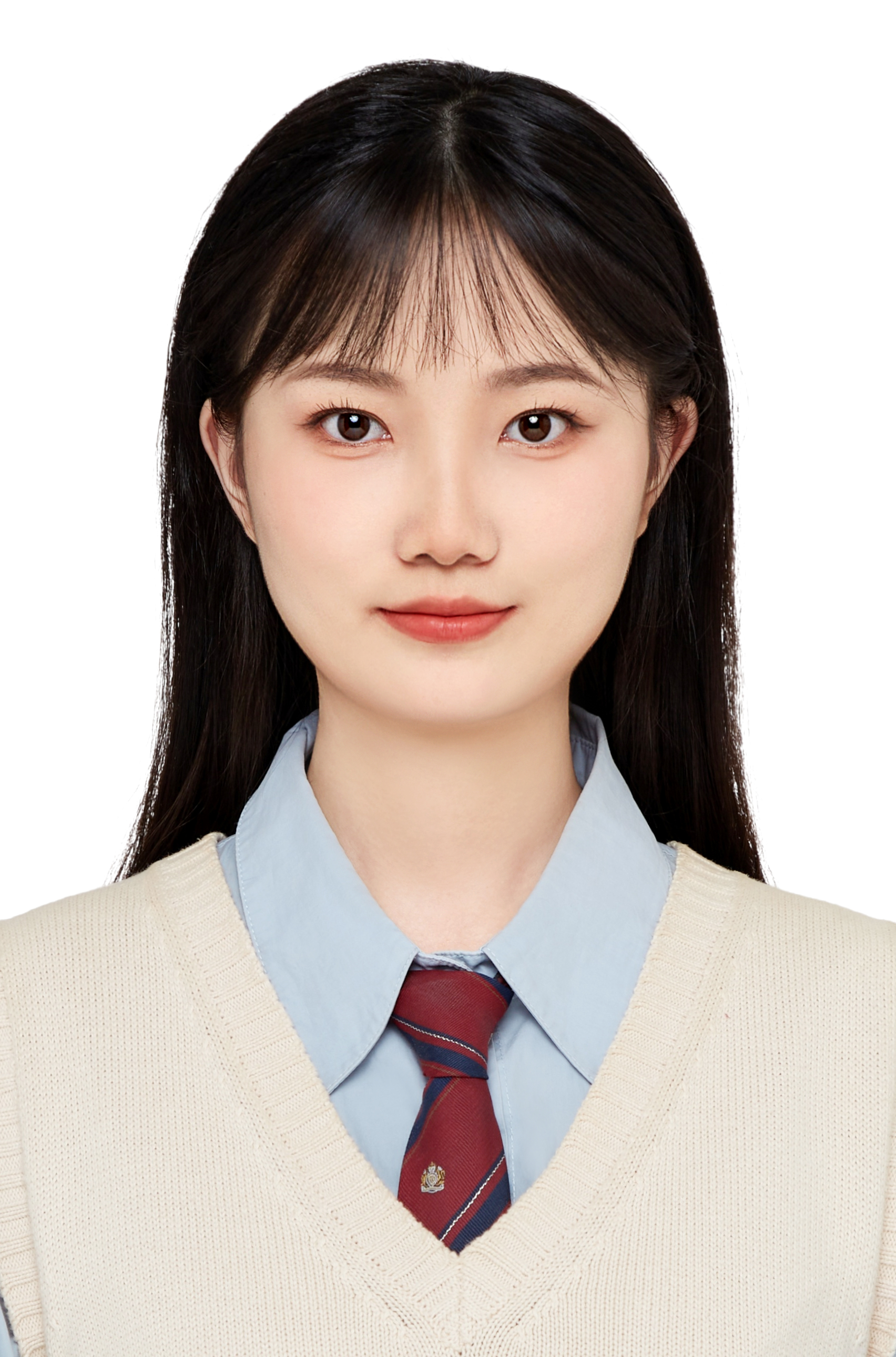}}]{Mingjie Zhao} (Student Member, IEEE) received the M.E. degree in software engineering from Sichuan University in 2023. She is a Ph.D. candidate in the school of Compute Science and technology at Sichuan University. She is a Visiting Student with the ENeS Lab, Muroran Institution of Technology, Muroran, Japan, supported by the China Scholarship Council Program from October 2024 to April 2027.

Her research interests include data sharing, federated learning, blockchain, and intelligent transportation systems.
\end{IEEEbiography}


\begin{IEEEbiography}[{\includegraphics[width=1in,height=1.25in,clip,keepaspectratio]{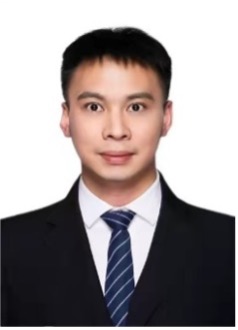}}]{Cheng Dai}%
(Member, IEEE) received the Ph.D. degree in signal and information processing from the School of Information and Communication Engineering, University of Electronic Science and Technology
of China, Chengdu, China, in 2021.

He is currently an Associate Professor with the College of Computer Science, Sichuan University, Chengdu. His research interests include intelligent transportation system, tensor decomposition and traffic 
\end{IEEEbiography}


\begin{IEEEbiography}[{\includegraphics[width=1in,height=1.25in,clip,keepaspectratio]{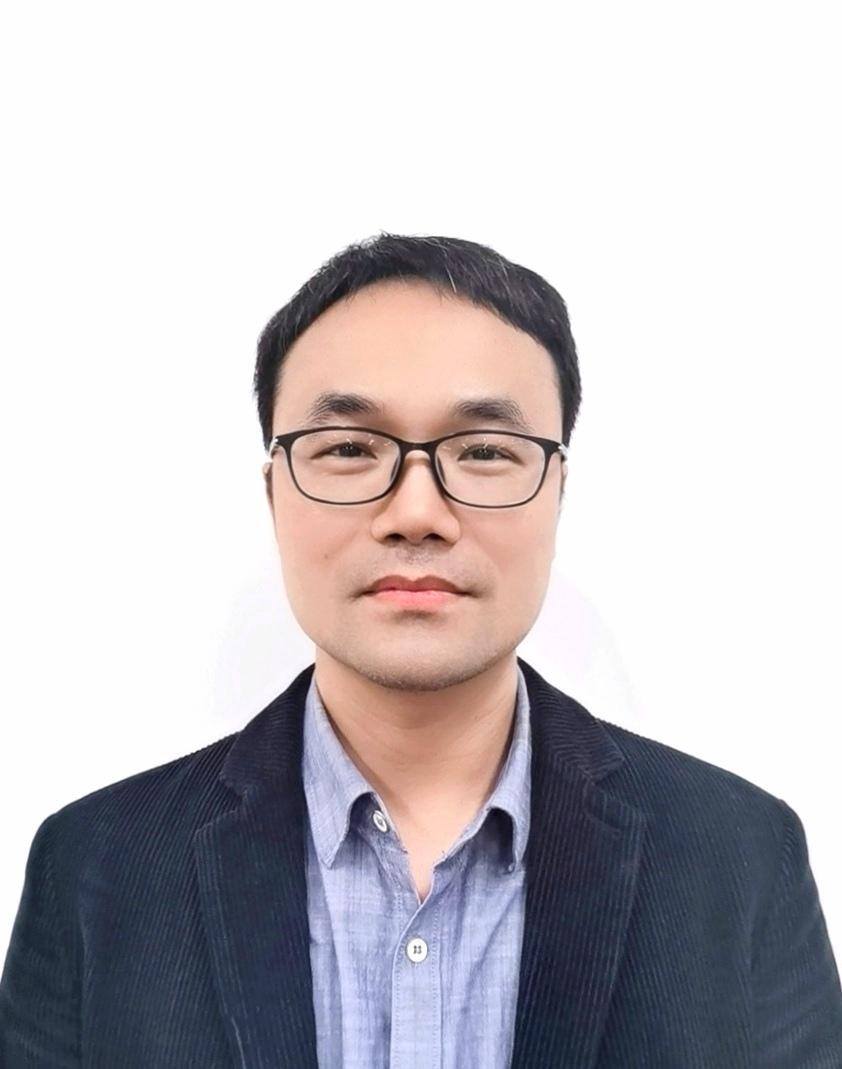}}]{Fei Cheng}%
received the M.S. degree from Sichuan Normal University, in 2009. Currently, he is working toward the Ph.D. degree with the College of
Computer Science, Sichaun University. 

His research interests include big data sharing and big data privacy protection.
\end{IEEEbiography}
\vspace{-15mm}
\begin{IEEEbiography}[{\includegraphics[width=1in,height=1.25in,clip,keepaspectratio]{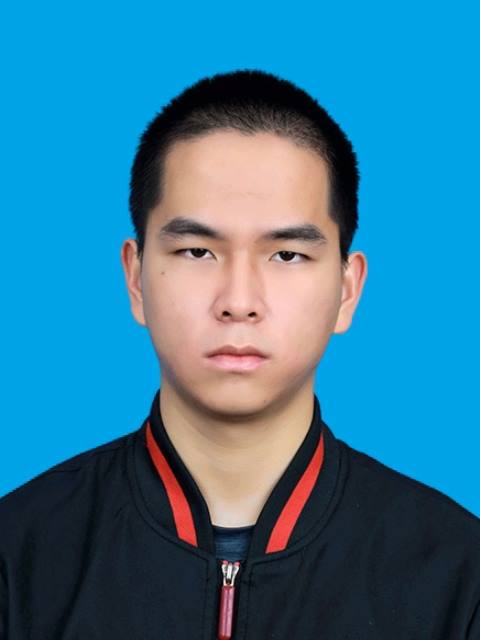}}]{Kui Ye}%
received the B.E. degree in software engineering from Sichuan University, in 2021. Currently, he is working toward the Ph.D. degree with the College of Computer Science, Sichuan University.

His research interests include quantum computation, quantum logic design, and computer-aided design of integrated circuits and systems.
\end{IEEEbiography}
\vspace{-10mm}
\begin{IEEEbiography}[{\includegraphics[width=1in,height=1.25in,clip,keepaspectratio]{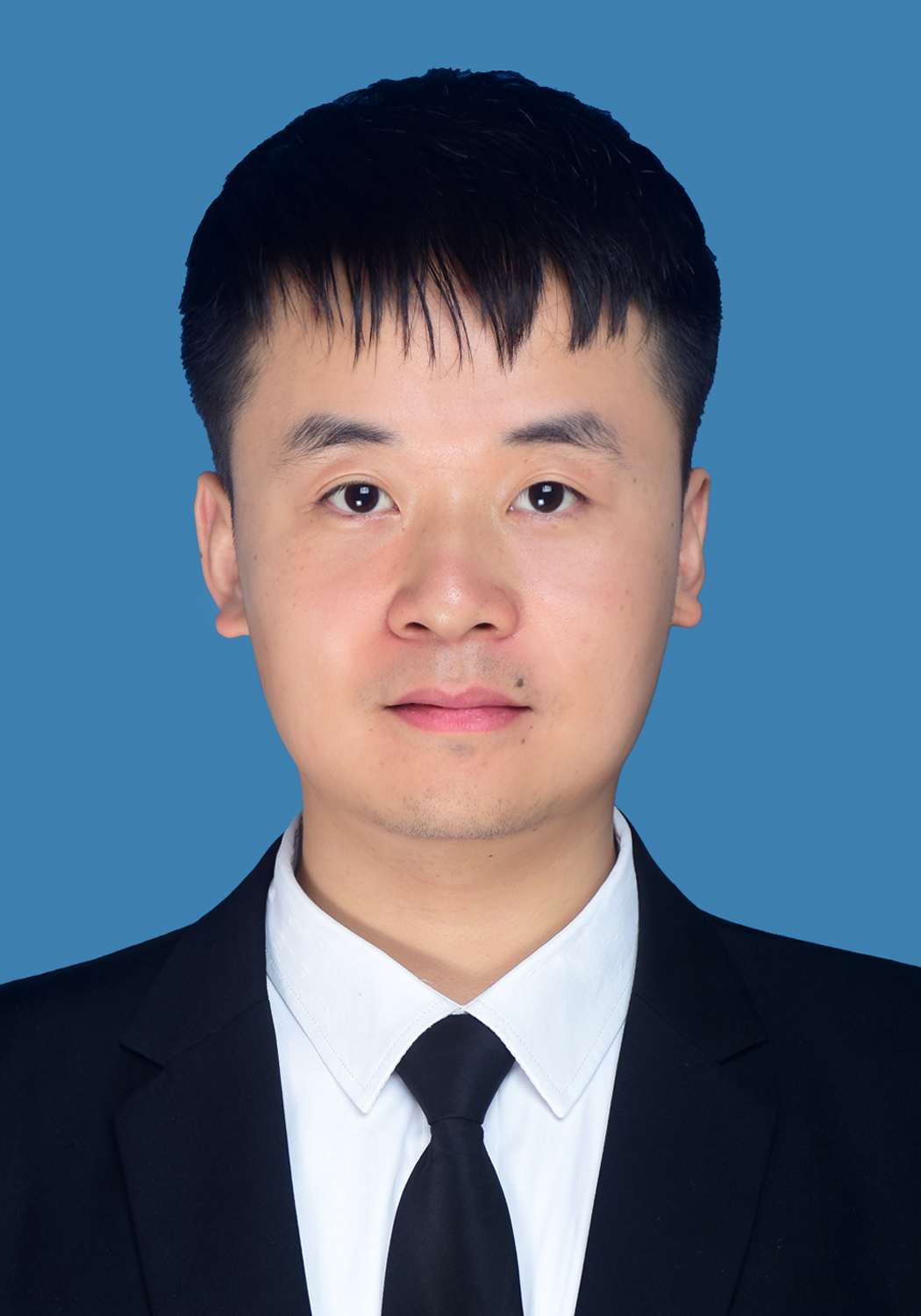}}]{Xin Chen}%
is a Ph.D. candidate at the College of Computer Science, Sichuan University. 

His research focuses on time series forecasting and large language models, with interests in frequency-domain analysis, control theory–inspired intelligent prediction, and efficient adaptation of large models for sequential understanding tasks.
\end{IEEEbiography}
\vspace{-2mm}
\begin{IEEEbiography}[{\includegraphics[width=1in,height=1.25in,clip,keepaspectratio]{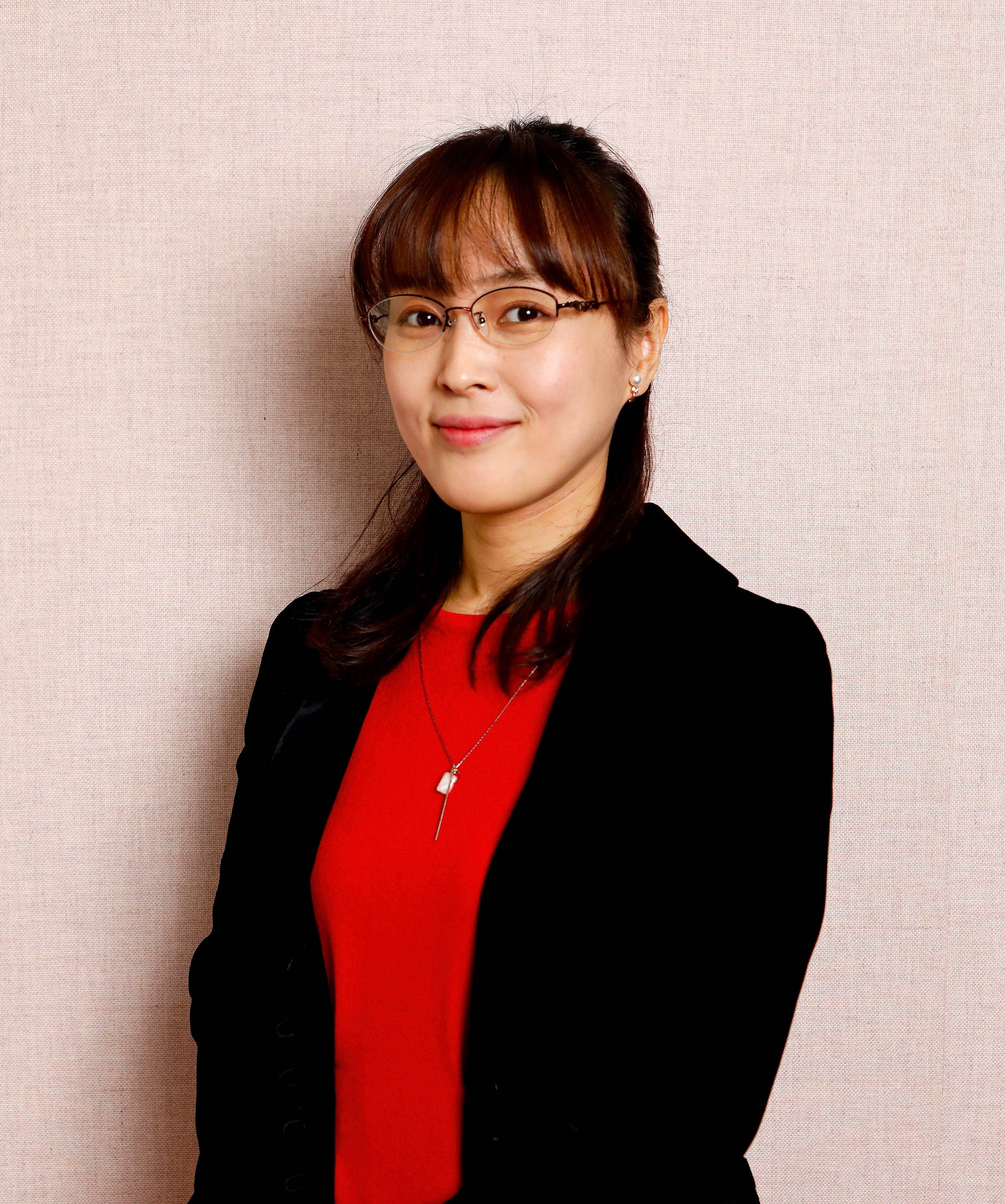}}]{Kaoru Ota}%
(Member, IEEE) received her B.S. and Ph.D. degrees from the University of Aizu, Japan, in 2006 and 2012, respectively, and her M.S. degree from Oklahoma State University, USA, in 2008. She is a Distinguished Professor at the Graduate School of Information Sciences, Tohoku University, Japan, and a Professor at the Center for Computer Science (CCS), Muroran Institute of Technology, Japan, where she served as the founding director. She was also a Ministry of Education, Culture, Sports, Science and Technology (MEXT) Excellent Young Researcher. Kaoru is the recipient of IEEE TCSC Early Career Award 2017, The 13th IEEE ComSoc Asia-Pacific Young Researcher Award 2018, 2020 N2Women: Rising Stars in Computer Networking and Communications, 2020 KDDI Foundation Encouragement Award, and 2021 IEEE Sapporo Young Professionals Best Researcher Award, The Young Scientists’ Award from MEXT in 2023. She is Clarivate Analytics 2019, 2021, 2022 Highly Cited Researcher (Web of Science) and is selected as JST-PRESTO researcher in 2021, Fellow of EAJ in 2022, and Fellow of AAIA in 2025.
\end{IEEEbiography}

\begin{IEEEbiography}[{\includegraphics[width=1in,height=1.25in,clip,keepaspectratio]{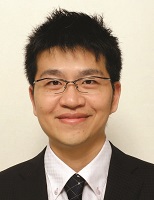}}]{Mianxiong Dong}%
 (Senior Member, IEEE) received B.S., M.S. and Ph.D. in Computer Science and Engineering from The University of Aizu, Japan. He is the Vice President and Professor of Muroran Institute of Technology, Japan. He was a JSPS Research Fellow with School of Computer Science and Engineering, The University of Aizu, Japan and was a visiting scholar with BBCR group at the University of Waterloo, Canada supported by JSPS Excellent Young Researcher Overseas Visit Program from April 2010 to August 2011. Dr. Dong was selected as a Foreigner Research Fellow (a total of 3 recipients all over Japan) by NEC C\&C Foundation in 2011. He is the recipient of The 12th IEEE ComSoc Asia-Pacific Young Researcher Award 2017, Funai Research Award 2018, NISTEP Researcher 2018 (one of only 11 people in Japan) in recognition of significant contributions in science and technology, The Young Scientists’ Award from MEXT in 2021, SUEMATSU-Yasuharu Award from IEICE in 2021, IEEE TCSC Middle Career Award in 2021. He is Clarivate Analytics 2019, 2021, 2022, 2023, 2025 Highly Cited Researcher (Web of Science) and Foreign Fellow of EAJ.
\end{IEEEbiography}
\vspace{-60mm}
\begin{IEEEbiography}[{\includegraphics[width=1in,height=1.25in,clip,keepaspectratio]{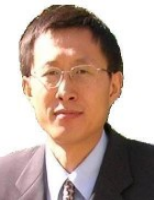}}]{Bing Guo}%
 received the B.S. degree in computer science from Beijing Institute of Technology, China, in 1991, and the M.S. and Ph.D. degrees in computer science from the University of Electronic Science and Technology of China, China, in 1999 and 2002, respectively. Currently, he is a Professor with the College of Computer Science, Sichuan University, China. 
 
His research interests include embedded systems, big data management, and blockchain.
\end{IEEEbiography}

\end{document}